\documentclass[sigconf]{acmart}

\usepackage{url}
\usepackage{graphicx}
\AtBeginDocument{%
  \providecommand\BibTeX{{%
    \normalfont B\kern-0.5em{\scshape i\kern-0.25em b}\kern-0.8em\TeX}}}

\setcopyright{acmcopyright}
\copyrightyear{2024}
\acmYear{2024}
\acmDOI{XXXXXXX.XXXXXXX}

\acmConference[CSCW '24]{Conference on Computer-Supported Cooperative Work \& Social Computing}{November 9--13, 2024}{San José, Costa Rica}
\acmPrice{15.00}
\acmISBN{978-1-4503-XXXX-X/18/06}

\usepackage{xcolor}
\usepackage[normalem]{ulem}
\newif\ifdraft

\newcommand{\newt}[1]{\ifdraft\textcolor{black}{#1}\else\textcolor{black}{#1}\fi}
\newcommand{\newtext}[1]{\ifdraft\textcolor{blue}{#1}\else\textcolor{black}{#1}\fi} %
\newcommand{\strike}[1]{\ifdraft\textcolor{red}{\sout{#1}}\fi}

\begin{document}

\title[OSINT Research Studios]{OSINT Research Studios: A Flexible Crowdsourcing Framework to Scale Up Open Source Intelligence Investigations}

\author{Anirban Mukhopadhyay}
\orcid{0009-0003-0925-1084}
\affiliation{%
  \institution{Department of Computer Science, Virginia Tech}
  \city{Blacksburg}
  \state{VA}
  \country{USA}
}
\email{anirban@vt.edu}

\author{Sukrit Venkatagiri}
\orcid{0000-0002-3888-7693}
\affiliation{%
  \institution{Department of Computer Science, Swarthmore College}
  \city{Swarthmore}
  \state{PA}
  \country{USA}
}
\email{sukrit@swarthmore.edu}

\author{Kurt Luther}
\orcid{0000-0003-1809-6269}
\affiliation{%
  \institution{Department of Computer Science, Virginia Tech}
  \city{Arlington}
  \state{VA}
  \country{USA}
}
\email{kluther@vt.edu}

\renewcommand{\shortauthors}{Mukhopadhyay, et al.}

\begin{abstract}
Open Source Intelligence (OSINT) investigations, which rely entirely on publicly available data such as social media, play an increasingly important role in solving crimes and holding governments accountable. The growing volume of data and complex nature of tasks, however, means there is a pressing need to scale and speed up OSINT investigations. Expert-led crowdsourcing approaches show promise, but tend to either focus on narrow tasks or domains, or require resource-intense, long-term relationships between expert investigators and crowds. We address this gap by providing a flexible framework that enables investigators across domains to enlist crowdsourced support for discovery and verification of OSINT. We use a design-based research (DBR) approach to develop OSINT Research Studios (ORS), a sociotechnical system in which novice crowds are trained to support professional investigators with complex OSINT investigations. Through our qualitative evaluation, we found that ORS facilitates ethical and effective OSINT investigations across multiple domains. 
We also discuss broader implications of expert--crowd collaboration and opportunities for future work.
\end{abstract}

\ccsdesc[500]{Human-centered computing~Collaborative and social computing systems and tools}

\ccsdesc[500]{Human-centered computing~Empirical studies in HCI}

\keywords{OSINT, open source intelligence, design-based research, social media investigation, collaboration, crowdsourcing}

\maketitle

\section{Introduction}

Open Source Intelligence (OSINT) involves the use of publicly available information to generate intelligence that addresses a particular need \cite{williams_defining_2018}. OSINT analysis is increasingly used by journalists \cite{higgins2021we}, human rights activists \cite{noauthor_digital_2020,kermode_objects_2020}, and law enforcement \cite{department_of_homeland_security_ufouoles_2010,esteban_borges_securitytrails_2019}, among other professions. For example, OSINT is used to verify breaking news and combat disinformation, monitor international weapons development, locate suspected terrorists and victims of human trafficking, and document war crimes \cite{y77, belghith_compete_2022, kermode_objects_2020,y13}. These investigations are widely recognized for their ability to use data sources like social media, satellite imagery, flight tracking information, and metadata from smartphones and IoT devices to conduct investigations \cite{y77,y48}. 

There is a pressing need to scale and speed up OSINT investigations, especially those focused on time-sensitive topics such as documenting war crimes or addressing disinformation. Apart from time pressure, investigators find it difficult to manage the growing volume of data that they must process \cite{glassman2012intelligence,noauthor_russia_2022}. The ephemerality of online information \cite{noauthor_digital_2020} and prevalence of misleading or misrepresented content \cite{noauthor_digital_2020,hulnick_downside_2002} pose additional challenges. Many investigators may also lack the data science and software development skills required to fully utilize OSINT tools and techniques \cite{haughey2020misinformationbeat}, presenting another barrier to the adoption of OSINT.

There are two common approaches to scale and speed up such investigations: automation \cite{fletcher2018measuring,shao_hoaxy_2016,hassan_claimbuster_2017} and crowdsourcing \cite{Matatov2018DejaVu,noauthor_whatsapp_2021,diakopoulos_towards_2021,pennycook_fighting_2019}. OSINT investigations require creatively leveraging multiple tools and techniques \cite{belghith_compete_2022}. Thus, software tools cannot fully automate and scale up the complex sensemaking involved in investigative work \cite{li2018crowdia}. Computational methods have produced high volumes of unverifiable information, which has low utility for experts \cite{micallef2022true}. OSINT analysts also aim to minimize dependency on custom-built tools, as they can easily become obsolete  \cite{muller_gatekeeper_2021}.

Crowdsourcing provides a second, more flexible way to augment investigators' complex sensemaking efforts. However, unfettered access to information as with OSINT investigations have resulted in ``bottom-up'' crowdsourced investigations that exhibit biased results \cite{allen2022birds}, doxxing \cite{nhan2017digilantism}, and even sabotaging of ongoing investigations \cite{trottier2017digital}. Expert supervision has sometimes resulted in more successful investigations in terms of both process (ethical, safe, and privacy-protecting) and outcomes (results). 
For example, CrowdSolve \cite{venkatagiri2021crowdsolve} described an effort in which law enforcement officials and experts supervised a crowd of 250 true crime enthusiasts in investigating two cold cases in a co-located, weekend-long event where information was tightly controlled rather than publicly available or open source. The Human Rights Center (HRC) at the University of California, Berkeley trains law students to partner with professional human rights investigators on OSINT projects lasting months or longer \cite{HRC_Berkeley}. GroundTruth \cite{venkatagiri2019groundtruth} brought together the complementary strengths of experts and an online, novice crowd for performing image geolocation, an important but very specific type of OSINT task. 
While these diverse projects show promise, they tend to either focus on narrow tasks or domains, or require resource-intense, long-term relationships between expert investigators and crowds. There is a need for a more flexible approach that enables investigators across a diverse set of domains to enlist crowdsourced support for a wide variety of OSINT investigation tasks.

In this paper, we use a design-based research (DBR) approach \cite{cobb2003design} to develop OSINT Research Studios (ORS), a sociotechnical framework in which novice crowds are trained and provided with scaffolding to support professional investigators with complex OSINT investigations. We developed ORS in a semester-long class with 30 trained students serving as the crowd. Using an industry-standard OSINT model \cite{williams_defining_2018}, we developed five types of macrotasks, i.e., discovery and verification techniques for OSINT. We recruited five OSINT experts who worked as journalists, fact-checkers, in law enforcement, and as human rights investigators to divide their investigations into one or more of the five macrotask types. We evaluated ORS through a semester-long deployment of the model and five study sessions where experts and the crowd conducted real-world investigations. 

Our qualitative evaluation revealed that the macrotasks were relevant to expert work practices. Experts said that the sessions were productive, and mentioned strengths like speed, safety, quality and quantity of submissions, and the crowd's adaptability in response to expert feedback. The crowd enjoyed working with experts and felt that they successfully applied their OSINT skills. We also discuss experts' involvement during the sessions, the crowd's perceptions of their contributions, and how the type of tasks impacted the crowd's experience. 

Our paper makes the following three contributions: 
\begin{enumerate}
    \item We make a conceptual contribution by developing training modules for crowdsourcing diverse and complex OSINT tasks.
    \item Using design-based research (DBR), we present OSINT Research Studios (ORS), a sociotechnical framework that enables collaboration between investigators and a trained crowd. Experts from multiple domains can delegate OSINT tasks to a trained crowd and generate leads for their investigation. 
    \item We evaluate ORS through multiple deployments during the OSINT lab course. We find ORS is effective for scaling and speeding up expert-led OSINT investigations and discuss the key design elements that make it successful.
\end{enumerate}
\section{Background and Related Work}
\subsection{OSINT Investigations}
OSINT investigations must deal with large amounts of digital content, including from social media platforms, search engines, online databases, among other sources. These investigations frequently employ a diverse set of tools and techniques to analyze digital traces, augmenting investigations in domains like journalism \cite{higgins2021we}, law enforcement \cite{department_of_homeland_security_ufouoles_2010,esteban_borges_securitytrails_2019}, and human rights advocacy \cite{noauthor_digital_2020,kermode_objects_2020}. Numerous applications are possible because of the vast quantities of detailed data available online alongside tools used to gather and analyze this data\cite{mckeown_investigating_2014}.	

OSINT can be divided into four stages: information discovery, verification, archival and reporting \cite{williams_defining_2018}. Researchers have developed systems to support each of these processes. CrowdTangle \cite{fletcher2018measuring} and Algorithm Tips \cite{diakopoulos_towards_2021} help in automating discovery; Hoaxy \cite{shao_hoaxy_2016} and DejaVu \cite{Matatov2018DejaVu} provide structure for verification tasks, such as verifying Twitter bot networks and accounts; The Web Archive Workbench \cite{hswe_web_2009} supports digital archiving; and Hunchly \cite{hunchly} is used for documenting and reporting. The number of open source OSINT tools is growing quickly with citizen journalism organizations like Bellingcat \cite{bellingcat_investigation_team_diversifying_2015} collaborating with and funding software developers and data scientists \cite{bellingcat2015calltoarms}. To map this space, aggregators like OSINT Explorer collate these resources and provide guides on which OSINT tools to use \cite{abdullah_osint_2021}. However, most current tools only support \textit{individual} steps of the OSINT process. These tools are also frequently rendered obsolete by changes in the underlying information architecture controlled by large online platforms (e.g., social media platforms, search engines). In contrast, we leverage the adaptability of a trained crowd to approach multiple tasks using a conceptual approach rather than relying on specific or customized tools. 

OSINT investigations are carried out by both novices and experts. The OSINT community has grown rapidly in recent years due to its low barrier to entry and gained attention with influential investigations of the Ukraine-Russia conflict \cite{gutierrez2015digital} and the storming of the U.S. Capitol in January 2021 \cite{splc_2021_jan6}. Analysts in the OSINT community volunteer for organizations like Bellingcat \cite{bellingcat}, Trace Labs~\cite{cox2018tracelabs}, and the Syrian Archive~\cite{syrian_archive}. These organizations run collaborative projects to analyze open information from global events including wars, missing person investigations \cite{tracelabs_2022_twitter}, human rights violations \cite{amnesty_international_syria_2020}, and election irregularities \cite{schumacher-matos_election_2012}. 
There have been successful instances of this form of collaboration, but there is little structure to how they are performed \cite{cochrane2022citizen}. Lack of intelligence training and coherence in preparing policy options for decision-makers among OSINT enthusiasts emerged as challenges for collaboration between the OSINT community and law enforcement agencies \cite{cochrane2022citizen}. Our work addresses this gap in structuring crowdsourced OSINT investigations and augmenting expert work practices. 

OSINT investigations can have wide variation in scope and depth. Existing collaborations with OSINT analysts for investigative journalism and human rights advocacy predominantly tackle long-running investigations that require a deep understanding of the context and its evolution \cite{muller_gatekeeper_2021,HRC_Berkeley}. These investigations require high levels of involvement, communication and training. Instead, our work here seeks to support rapid, more targeted tasks, scaling up and speeding up larger, more complex investigations led by experts. 

Finally, OSINT, though popular in its application across many domains, has not received much research attention in the form of frameworks and systems that help to scale up and speed up these investigations \cite{aliprandi_caper_nodate,belghith_compete_2022, iorga2021early}. Our work contributes to the development of OSINT Research Studios (ORS), a collaborative crowdsourcing framework that can support experts with multiple steps within complex OSINT investigations. The work also demonstrates that OSINT can be a valuable domain of study for the CSCW and HCI community.

\subsection{\newt{Collaborative Sensemaking and} Crowdsourced Investigations}

\subsubsection{\newt{Collaborative sensemaking}}
\newt{Investigations are a type of sensemaking task, as they involve collecting and analyzing large amounts of information to reach a conclusion \cite{venkatagiri2019groundtruth,dailey_dispersants_2015,alharthi_2021}. OSINT investigations, when broken down into the steps of the OSINT cycle \cite{williams_defining_2018, belghith_compete_2022} --- discover, verify, preserve and publish --- follow the sensemaking process closely. Collaborations also play a crucial role in facilitating sensemaking by dividing tasks related to discovery and verification and incorporating diverse perspectives during data analysis \cite{fisher2012distributed}.} 

\newt{Previous studies on supporting collaborative sensemaking have primarily focused on co-located teams working synchronously \cite{vogt2011colocated,venkatagiri2021crowdsolve}, distributed crowd working asynchronously \cite{dailey2014crowdsourcerers,li2018crowdia,fisher2012distributed}, and even a distributed crowd working synchronously \cite{venkatagiri2023cosint}. In contrast, this paper studies a new, hybrid setting, with remote experts collaborating synchronously with co-located crowds distributed across two locations. We present a semester-long deployment with 30 university students in a classroom setting, accommodating experts from across the United States through remote participation.} 

\newt{Extensive research has studied the development of collaborative systems to assist fact-checkers and journalists. Various collaborative tools have been designed specifically for fact-checking news articles \cite{sethi2018extinguishing}, videos \cite{carneiro2019deb8}, and visual disinformation \cite{Matatov2018DejaVu, venkatagiri2019groundtruth}. Additionally, systems like Newstrition \cite{news} and Checkdesk \cite{checkdesk} apply crowdsourcing approaches to verify information. The Datavoidant tool \cite{flores2022datavoidant} facilitated human-AI collaboration to empower journalists in addressing data voids through information discovery and verification. The CAPER tool \cite{aliprandi_caper_nodate} aids law enforcement agents in collaborating for sensemaking tasks to prevent organized crime. Garcia et al. \cite{garcia2018quotidian, garcia2021data} studied the exploration of social media data for human rights investigations and public safety. Most closely related to our work, Venkatagiri et al. \cite{venkatagiri2023cosint} enabled a trained crowd to be effective in debunking online misinformation using OSINT techniques by introducing collaboration in a competitive Capture-the-Flag environment. We contribute to this line of work by developing a flexible expert-crowd collaborative framework where investigators can enlist crowdsourced support for a wide variety of tasks within broader, more complex OSINT investigations.}

\newt{Studies investigating the effectiveness of crowdsourced sensemaking and fact-checking are highly relevant to our work. For instance, Arif et al. \cite{arif_closer_2017} and Dailey et al. \cite{dailey2014crowdsourcerers} demonstrated that distributed crowds can effectively debunk rumors online, while Saeed et al. \cite{saeed2022crowdsourced} revealed that crowdsourced fact-checking on Twitter often performs as well as professional fact-checkers. Experimental investigations into the efficacy of crowdsourced fact-checking by Pennycook and Rand \cite{pennycook_fighting_2019} and Allen et al. \cite{allen2021scaling} found that crowdsourced trustworthiness ratings can distinguish between authentic and fake news sources. However, Godel et al. \cite{godel2021moderating} discovered that real-time crowdsourced veracity ratings performed worse than those generated by professional fact-checkers. While prior crowdsourcing approaches predominantly focused on studying online crowds operating independently without expert supervision, we show that a trained crowd can augment investigations of online information when led by experts.}

\subsubsection{Crowdsourced investigations}
CSCW literature presents three types of crowdsourced investigations: top-down \cite{poelman_as_2012,alcaidinho_mobile_2017}, bottom-up \cite{huang_connected_2015,erete_engaging_2015}, and hybrid investigations \cite{venkatagiri2021crowdsolve}. Bottom-up investigations are driven by non-professional crowds and tend to move through the sharing, validation and analysis stages of an investigation in an online setting. Investigations have been studied in the context of collective sensemaking on social media during crisis events \cite{dailey2014crowdsourcerers}, e.g., analysis of photos related to 2013 Boston Marathon bombings \cite{huang_connected_2015}, as well as correcting online information on social media \cite{arif_closer_2017}. Though potentially effective, these investigations have resulted in harmful behavior like misidenfications \cite{kornfield2021wrong}, doxxing \cite{nhan2017digilantism}, and perpetuating conspiracy theories \cite{metaxas2017infamous}. 

Other related investigations demonstrate that crowds can collaborate under the guidance of experts to augment investigations \cite{venkatagiri2019groundtruth,li2018crowdia,venkatagiri2021crowdsolve}. Among hybrid investigations, GroundTruth \cite{venkatagiri2019groundtruth} demonstrated crowd-augmented expert work using a novice crowd to reduce the search area for geolocating images, an important OSINT task. Venkatagiri et al.\cite{venkatagiri2021crowdsolve} described an expert-led crowdsourcing model through CrowdSolve, which is characterized by experts (law enforcement officers in this case) leading investigations by providing resources, training and feedback; and the crowd performing analytical tasks to generate leads. Our work has two key differences compared with CrowdSolve: 1) here OSINT investigations are performed without the use of restricted information and led by experts from \textit{multiple} domains; and 2) the hybrid sessions we studied are mediated by online collaboration and not restricted to working within a co-located setting involving physical paper case files. Our work provides additional flexibility to the concept of expert-led crowdsourcing \cite{venkatagiri2021crowdsolve} for diverse and complex OSINT investigations involving experts from multiple domains. 

\subsection{Decomposition and Training for Crowdsourcing Complex Work}
OSINT analysis involves complex problem-solving tasks. One common approach to crowdsourcing is to decompose complex tasks into smaller subtasks that are easier to handle \cite{dailey2014crowdsourcerers,papoutsaki_crowdsourcing_2015,chan_solvent:_2018,kittur_standing_2014}. Among related complex tasks, entire sensemaking processes have been decomposed into microtasks to solve fictional murder mysteries and terrorist plots \cite{li2018crowdia,li2019dropping} and generating text content from journalism to how-to guides \cite{bernstein2010soylent,hahn2016knowledge,beckett2017wikitribune}. \newt{While microtask-based crowdsourcing offers scalability and efficiency by distributing small tasks among a large crowd, it may not be suitable for complex, creative OSINT investigations that require retaining contextual information. Macrotasks, on the other hand, allow for deeper engagement, contextual comprehension, and complex problem-solving \cite{doroudi_toward_2016}.} In our work, we contribute a decomposition of the OSINT tasks of discovery and verification into macrotasks. We describe the steps and skills required for five types of macrotasks that are transferable across multiple domains. 

Even with task decomposition, OSINT exhibits \textit{complex problem-solving tasks}~\cite{doroudi_toward_2016} that suggests multiple possible strategies to the crowd; workers can arrive at solutions, but not without relevant skills \cite{doroudi_toward_2016}. Decomposition also creates added coordination challenges that can be addressed through appropriate workflows \cite{tausczik_distributed_2018}. We investigate how to crowdsource complex OSINT tasks, which are relatively less studied in crowdsourcing. 

Previous studies have developed effective ways of training crowdworkers for complex crowdsourcing tasks \cite{gong_social_2019,kim2018hit,noronha2011platemate}. Training provides more agency to the crowd and enables them to perform investigations without rigid roles and constrained workflows --- strategies which have been effective for other types of creative and complex work \cite{retelny2017noworkflow, alharthi_2021}. There are multiple ways to conduct training \cite{zhu2014reviewing,oleson2013evaluating,mitra2015comparing}. First, crowdworkers can gain experience by solving problems that are relevant to the task \cite{agapie2015crowdsourcing}. Second, reviewing expert and peer solutions can improve crowd performance \cite{suzuki_atelier:_2016}. Third, crowdworkers can improve their performance based on self-assessment and expert feedback \cite{dow2012shepherding}. Doroudi et al. \cite{doroudi_toward_2016} compared multiple such training strategies for a complex web search problem, which is closely related to OSINT tasks that involve consulting multiple sources of information and arriving at a conclusion \cite{aula2008complex}. They found that all training strategies improved performance compared to the no training condition. Training through expert examples improved crowdworker accuracy the most. Wang et al. \cite{wang_exploring_2018} explored trained based on analytical thinking skills for historical analysis, that also closely relates to our strategy of imparting skills to help the crowd approach challenging OSINT tasks instead of relying on any particular tool. They found that crowdworkers developed domain expertise and performed at least as well as other training strategies mentioned above. Unlike prior work, we develop a semester-long training process focused on OSINT tasks.

Previous examples show that only minimal active training is provided to volunteers within real-world OSINT investigations \cite{cochrane2022citizen, osintethics2020hanham}. In our work, we contribute an OSINT crowd training module that combines multiple training strategies. Training includes demonstrations for acquiring OSINT skills, honing them during practice and expert sessions, as well as feedback and self-evaluations. We also illuminate how training can be effective within crowdsourced OSINT investigations.

\section{Design Challenges and Opportunities for Crowdsourcing Complex OSINT Investigations}
\label{sec:design-challenges}
Based on prior work, we discuss the four main types of challenges in designing crowdsourcing solutions to support OSINT investigations.

As previously mentioned, OSINT investigators are overwhelmed due to the large volume of information involved and the complex nature of investigative tasks \cite{noauthor_russia_2022}. Effectively scaling up OSINT investigations can help investigators improve the speed and/or accuracy of their work. As automation by itself is difficult to achieve for highly contextual and nuanced tasks, crowdsourcing provides a viable approach. Crowdsourcing can leverage humans' creative and sensemaking capabilities to augment ongoing investigations~\cite{venkatagiri2019groundtruth,venkatagiri2021crowdsolve}. However, there are major challenges faced by investigators from domains like journalism, human rights investigations, and law enforcement who seek crowdsourcing support for the analysis of open online information \cite{cochrane2022citizen}. We identify four design \textit{challenges} based on prior work and OSINT investigation reports, detailed below, \newt{along with \textit{design goals} for a sociotechnical framework to address these challenges.}

\subsection{Delegation of Complex OSINT Tasks}
Previous studies have shown that investigators spend most of their time in the discovery and verification phases of OSINT \cite{micallef2022true}. \newt{The increasing volume of digital data online presents a significant challenge for investigators to manage effectively \cite{glassman2012intelligence,noauthor_russia_2022}. Additionally, the transient nature of open online information \cite{noauthor_digital_2020} and the widespread presence of deepfakes, mis-, and disinformation \cite{noauthor_digital_2020,hulnick_downside_2002} further complicate the tasks of discovery and verification, posing challenges for successful investigations.} These tasks belong to the class of problems known as \textit{complex problem-solving tasks} that have a number of potential strategies and are difficult to solve without acquiring relevant skills~\cite{doroudi_toward_2016}. Experts do not have a way to crowdsource such tasks using existing workflows.

\subsubsection{Discovery}
\paragraph{Challenges:} Micallef et al. \cite{micallef2022true} specify that monitoring social media for interesting content and contextualizing it is one of the most time-consuming parts of fact-checking. Law enforcement involves collecting and analyzing open source information for corroboration as well, but in a way that is admissible in court. The focus is on referring to sources that are genuine and ethically collecting the information \cite{ghioni2022open}. Human rights investigations are long-running and highlight transparency in data collection. The rapid pace of generation, multiple platforms, and recycling of content make the collection process more challenging \cite{micallef2022true}. The strategy to discover relevant information needs to adapt based on newly found information and pivot around them.

\paragraph{Design Goals:} Investigators need to discover relevant information by understanding the evolution of a topic, gathering relevant hashtags, and identifying actors who spread the information \cite{micallef2022true,FirstDraft_discovery}. These topics can be broad like COVID misinformation, anti-vaccine protests, conspiracy theories, and many more. Or they can be more specific topics like a calamity, investigating claims by a local government representative, or reporting on local phenomena like crime and illegal activities. These activities require mining information tied to a particular geographical location and time range. Investigators also need to look for potential mis/disinformation around hot topics.

\subsubsection{Verification} 
\paragraph{Challenges:} Verification of online information is crucial in human rights, law enforcement investigations and fact-checking. Human rights investigations involve documenting events within specific regions, often utilizing Geospatial Information Systems (GIS) and performing geolocation \cite{unver_digital_2018}. However, the lack of metadata and geotags for social media content poses challenges in geolocation efforts \cite{unver_digital_2018}. To ensure the credibility of information, various professionals such as law enforcement officers, journalists, and fact-checkers examine the background of the account that generated the content, while being mindful of bot accounts and coordinated campaigns that propagate disinformation \cite{wardle2014verifying,bellingcat_verification}. Image analysis becomes a vital component of the verification process in journalism, addressing the difficult task of identifying manipulated or fake visual information \cite{micallef2022true}. In addition, fact-checking plays a crucial role by gathering trusted information to assess the veracity of claims \cite{micallef2022true}. However, the complexity of these tasks, combined with limitations in time, personnel, and analysis skills among experts, presents challenges in conducting successful investigations \cite{haughey2020misinformationbeat}.

\paragraph{Design Goals:} Investigators need to verify any information that is collected from publicly available sources for their use. According to Wardle~\cite{wardle2014verifying}, there are four main elements that need to be verified: 1) Provenance --- is it original or has it been used before in a different context? 2) Source --- what is the background of the account that created the content? Is it a bot? 3) Time --- When was it created? 4) Location --- Where is the place shown in it? These task, combined with fact-checking and image analysis, are essential crowd skills that meet the verification requirements of investigators.		
\subsection{Safety, Privacy, and Other Ethical Considerations} 
\paragraph{Challenges:} Previous instances of crowdsourced OSINT investigations have led to biased results \cite{allen2022birds}, misidentifications \cite{kornfield2021wrong}, and vigilante behavior like doxxing \cite{nhan2017digilantism}. The high-profile failures are perhaps more prominent than successful investigations \cite{huang_connected_2015}. Investigators are also wary of leaks and sabotaging ongoing investigations during collaboration with crowds \cite{trottier2017digital}. Investigators across domains benefit from careful consideration of ethics, safety and privacy in their work. 

\paragraph{Design Goals:} OSINT comes with its own ethos: prioritizing transparency, avoiding subterfuge, and limiting investigations to passive reconnaissance \cite{belghith_compete_2022}. This ethos can be leveraged to enable more successful investigations and reduce ethical mishaps. Crowdsourced investigators can operationalize these values, e.g., 1) Transparency: documenting the process of investigation for reproducibility; 2) Avoiding subterfuge: prohibiting forms of hacking and impersonation to gather private information; and 3) Passive reconnaissance: ensuring investigators view the information but don’t engage. \newtext{We must ensure that experts are actively overseeing and guiding the investigation. Experts should be responsible for validating information and making final decisions based on the crowd’s input.}

\subsection{Organizational Overheads of Synchronous Collaboration} 
\paragraph{Challenges:} Collaboration and communication are key elements of OSINT investigations \cite{belghith_compete_2022}. Investigators find it hard to collaborate with a crowd in real time. Generally, a single expert has to work with a group of volunteers and the overheads of coordination and communication impact the effectiveness of the investigation \cite{venkatagiri2021crowdsolve}. Information silos and duplication of effort within the crowd are known issues with both competitive and collaborative OSINT investigations \cite{venkatagiri2021crowdsolve}.

\paragraph{Design Goals:} Both expert and crowd investigators need technological solutions to orchestrate resources that allow them to document and present their findings. Information has to be easily accessible and open to all participants during investigations. Having a robust infrastructure can mitigate risks of data loss and usability issues. 

\subsection {Maintaining the quality of investigation}
\paragraph{Challenges:} Investigators need to insist on highest standards to establish the legitimacy of their reports and be confident in facing public scrutiny \cite{muller_gatekeeper_2021}. In previous microtask-based crowdsourcing models for investigative work, experts cannot make interventions or provide feedback to improve the quality of crowd results \cite{venkatagiri2019groundtruth,li2018crowdia}. The success of the particular task is overly dependent on the initial design and how results from the microtasks are aggregated. 

\paragraph{Design Goals:} Based on previous crowdsourcing studies for complex work, feedback from experts and self-evaluation can be helpful in improving the results \cite{dow2012shepherding}. The crowd needs to improve their work based on expert feedback. Crowd submissions need to meet these characteristics: 1) Relevant: the claim is relevant to the topic they selected; 2) Specific: the claim can be attributed to a specific statement or piece of content, such as a tweet, photo, video, or quote in an article; and 3) Verifiable: the claim has the potential to be verified, so it must be a factual statement (i.e., not an opinion) that can be shown to be true or false. Relevance and verifiability have been used earlier in crowdsourced OSINT capture-the-flag events for scoring submissions \cite{TraceLabs_search}. 
\section{A design-based research approach to expert-led OSINT investigations}

\subsection{Our Approach}
Collaboration and crowdsourcing have been embraced by investigators as a way to increase the scale and speed of their work~\cite{fisher2012distributed}. Previous studies \cite{oleson2013evaluating,mitra2015comparing,kim2018hit,noronha2011platemate} have shown that training workers to prepare them with sophisticated domain knowledge can be effective to complete complex tasks. Based on the challenges faced by investigators from multiple domains while dealing with publicly available information, we argue that collaboration between the investigators and a crowd trained in OSINT can be helpful. The first part of the problem deals with developing a comprehensive training module that can enable the crowd to perform common OSINT tasks used by investigators, understand the ethical considerations, apply relevant skills in different contexts, and present information that meets the expert requirements. The second part involves getting the trained crowd to collaborate with experts synchronously.  

In this work, we sought to address the challenges we identified in Section~\ref{sec:design-challenges} through the development of OSINT Research Studios (ORS), where we empower a group of students to apply OSINT analysis to augment real-world expert investigations. This group of students serves as the crowd in our crowdsourcing framework for the study. Previous research has shown that a crowd of students can effectively test new forms of crowdsourcing and generate recommendations for the process \cite{xu2015classroom,williams2016axis}. \newtext{Students have worked carefully and safely on real-world OSINT investigations for cyber vulnerability assessments and human rights as part of experiential learning previously \cite{MITCybersecurityClinic, HRC_Berkeley}.} Here, we design a sociotechnical framework where experts can get valuable crowdsourced support on diverse, complex OSINT tasks by addressing the design challenges. 

\subsection{Methods}
We take a design-based research (DBR) approach to develop the ORS model for overcoming the challenges with current crowdsourced OSINT investigations. DBR \cite{zhang2017agile,cobb2003design,easterday2014design} is characterized by iterative cycles of design and evaluation to develop insights into learning experiences. DBR also provides a way for researchers to simultaneously iterate on and study complex models \cite{zhang2017agile,easterday2014design,plomp2013educational}. We leverage that to evaluate the collaborative framework as it evolves with deployments in a classroom setting. There are constraints of experimentation in such education settings where the learning goals of students and their experience with the study sessions are also important considerations. DBR allows us to quickly identify failures, make changes to the design and evaluate the resulting system. We can evaluate the performance of the trained crowd across multiple deployments without comparing it against baselines. 

Iteration is an essential part of the DBR process and we use reflection assignments after each study session to capture the crowd’s feedback on three experiences: 1) coordination with expert; 2) teamwork; and 3) overall difficulty and enjoyability of the session. We gather insights into experts' experience through post-session interviews. We iterate based on these perspectives and observations to finalize how tasks can be assigned and carried out by crowd teams. Feedback from practice and study sessions is useful for optimizing team structure and communication between the investigator and the crowd. \newtext{We highlight the iterations for each of the challenges in Section~\ref{subsec:design-arguments}.} We then evaluated the ORS model through a case study of the OSINT lab course, a semester-long university course taught by the third author.

Our overall approach is also heavily inspired by Agile Research Studios \cite{zhang2017agile}. This work explored a classroom-based approach to collaborative, real-world HCI research with students. We adapt this approach for the context of OSINT investigations.

\subsubsection{Data collection and analysis}
As a part of the OSINT lab course, we conducted five study sessions where the trained crowd worked with investigators to augment their ongoing investigations. Data was not collected for the first, practice session\newt{, a pilot study} aimed at developing coordination between the crowd and expert and iterating on the script for the interview study. The first study session was conducted during week~7 of the semester, and the final one during week~15, with the other deployments roughly every two weeks. 

\newt{We collected qualitative and quantitative data from the study sessions to understand the experts' and crowd's attitudes and performance.} The different data sources were: 1) observations during the study sessions \newtext {by the first and third authors}; 2) reflection surveys submitted by the crowd;  3) spreadsheets containing crowd submissions and expert feedback; 4) semi-structured interviews with each expert after sessions; and 5) separate focus group interviews with members of the crowd. \newtext {We present the reflection survey form details, a snapshot of a spreadsheet containing crowd submissions and corresponding expert feedback, and interview scripts for experts and the crowd in Appendix~\ref{appendix}.} We had pre-session meetings with three of the five investigators to decide on investigation topics and break down the investigation into tasks. We worked asynchronously through email with the other two experts. 

We conducted a total of ten interviews with 14 students and five investigators, consisting of five semi-structured interviews with investigators and five focus group interviews with students. \newt{The first author transcribed all recorded interviews. In collaboration with the rest of the research team, the first author conducted a deductive thematic analysis \cite{braun2006thematic} of the transcripts. The themes were informed by prior work (Section~\ref{sec:design-challenges}) and aligned with our interview guide. They were aimed at capturing the user experience and evaluating our design arguments \newtext{(Section~\ref{subsec:design-arguments})}. We extended the thematic analysis to include the crowd's reflection survey responses.}

\newt{For experts, we identified the following themes: 1) the usefulness and effectiveness of OSINT macrotasks; \newtext {2)} criteria for successful completion of tasks; 3) planning versus actual activities during sessions; \newtext {4)} interaction and communication with the crowd; 5) assessment of submitted information in terms of quality and quantity; 6) comparison of trained crowd's effectiveness with the general crowd; 7) crowd's self-evaluation of submissions; 8) suggestions for improvements; and 9) willingness to work with the crowd again. For crowdworkers we identified the following themes: 1) team formation and evolution of teamwork; 2) usefulness of practice sessions; 3) perceived change of performance over time; 4) tools and techniques used in tasks; 5) challenges faced with tasks; \newtext {6)} positive and negative aspects of sessions; \newtext {7)} use of self-evaluation; and 8) suggestions for improvement. \strike {As a research team, we iteratively coded transcripts based on these themes, further refined themes through discussion, compared similarities and differences across codes and themes, and identified higher-level themes to organize our findings.}} \newtext{We took multiple steps to analyze the transcripts and organize the findings. First, we coded each transcript based on the established themes. After this, we engaged in detailed discussions to refine these themes. We sought insights into each design argument and goal, deepening our understanding of the collaboration process. We compared the similarities and differences across codes and themes to form higher-level themes. These themes helped organize our results and are finally presented in our findings (Section~\ref{sec:findings}).} 

\subsubsection{Participants recruitment and demographics} 
This study was approved by our university’s IRB. The first set of participants were students in the course, who were junior and senior students in the Computer Science department of two universities (U1 and U2). There were a total of 30 students in the course, 20 from U1 and 10 from U2. The first author recruited the students during an in-class lecture and their participation was voluntary. The consenting participants received \$20 after completing a post-course completion interview. 

Eighteen out of 20 students from U1 consented to data collection. Two out of those 18 students identified as female while others identified as male. Ten out of those 18 students, all of them identifying as male, participated in focus group interviews conducted by the first author. Eight out of the 10 students at U2 consented to the study and four students participated in a focus group interview. All students from U2 identified as male. 
We refer to the 14 crowdworkers who participated in the focus group interviews as CW1 -- CW14.

Our study consisted of six investigators as expert participants. The first pilot session, which we omitted from our data analysis, was conducted by a journalist recruited through an ad on Upwork and was compensated with \$100. \newtext{ We estimated a total time commitment of 2 hours and 15 minutes for the investigators. There were three phases: a 30-minute pre-session training meeting, a 75-minute study session, and a 30-minute post-session interview. We decided on a rate of \$45/hour based on the hourly rates of freelance journalists on Upwork and the compensation reported in previous studies \cite{micallef2022true, venkatagiri2019groundtruth, belghith_compete_2022}} Details about the next set of five experts (referenced as E1 -- E5; we use the terms ``expert'' and ``investigator'' interchangeably) who led the study sessions, including the topics and tasks are presented in Table~\ref{tab:sessions}. These investigators were invited to participate in the study through email and social media advertisements. All of the expert participants were based in North America and identified as male, reflecting the demographic trends in OSINT described above. 
There was a wide range in experts' professional experience, from 3--5 years to 11+ years. Two out of the five investigators did not have previous experience with crowdsourcing. Only one of the investigators accepted the offered compensation of \$100 for their participation. 

\subsubsection{Limitations} We did not set explicit learning goals and tests to measure performance of the crowd for each task/deployment. Instead, our evaluation is based on expert feedback and the session’s utility for augmenting ongoing OSINT investigations. \newtext{To reduce the potential for biased results in crowd feedback, as they are part of a for-credit course, students' participation in the study was completely voluntary and interviews were conducted after the conclusion of the course. Another constraint was the organization of single sessions for each domain, which does not fully explore the potential of different investigations within each domain. In future work, replicating sessions for each domain could offer more varied insights and contribute to a more robust and comprehensive understanding of the crowdsourcing framework.}  

\strike{Bringing the gender gap of the OSINT field into light \cite{bellingcat_investigation_team_diversifying_2015}, we were unfortunately only able to recruit male participants.}
\newtext{Another limitation of our study was that most of the crowd participants and all of the expert participants were male. Despite efforts by the instructor and universities to recruit more women, these predominantly male participant demographics are representative of broader gender gaps identified both within the university departments and the broader OSINT~\cite{bellingcat_investigation_team_diversifying_2015} and Computer Science~\cite{varma2010so} fields.
This limitation may introduce gender bias into the research findings, as the perspectives and experiences of female participants are not represented. Future work should focus on approaches to broaden gender diversity in the OSINT field and courses such as this one.}

\begin{table*}[]
\caption{Details about experts, topics, and tasks for our study sessions}
\resizebox{2\columnwidth}{!}{
\begin{tabular}{cllll}
\hline
\textbf{\begin{tabular}[c]{@{}c@{}}Session /\\ Participant\\  Identifier\end{tabular}} &
  \multicolumn{1}{c}{\textbf{Expert profession}} &
  \multicolumn{1}{c}{\textbf{\begin{tabular}[c]{@{}c@{}}Years of \\ Experience\end{tabular}}} &
  \multicolumn{1}{c}{\textbf{Topic}} &
  \multicolumn{1}{c}{\textbf{Tasks Involved}} \\ \hline
1 / E1 &
  Fact-checker &
  5+ years &
  \newt{Verify origins and content of video about Dr. Anthony Fauci} &
  \begin{tabular}[c]{@{}l@{}}Verification: Source analysis, \\ fact-checking, image analysis\end{tabular} \\ \midrule 
2 / E2 &
  Law enforcement officer &
  11+ years &
  \begin{tabular}[c]{@{}l@{}}Collect social media images and videos for a \\ particular mountain range within certain dates \\ that \newt{contain people or vehicles}\end{tabular} &
  \begin{tabular}[c]{@{}l@{}}Discovery;\\ Verification: Geolocation, \\ source analysis\end{tabular} \\ \midrule 
3 / E3 &
  Human rights investigator &
  3--5 years &
  \begin{tabular}[c]{@{}l@{}}\newt{Identify whereabouts of leader in Ukraine} \\ \newt{prior to their death}; \\ \newt{Collect evidence from social media of European} \\ \newt{immigrants being used as political pawns}\end{tabular} &
  \begin{tabular}[c]{@{}l@{}}Discovery;\\ Verification: Geolocation, \\ source analysis, image analysis\end{tabular} \\ \midrule 
4 / E4 &
  Investigative journalist &
  3--5 years &
  \begin{tabular}[c]{@{}l@{}}\newt{Identify discourse around anti-vaccine protests occurring} \\ \newt{throughout Europe} \newt{as well as groups involved}\end{tabular} &
  \begin{tabular}[c]{@{}l@{}}Discovery;\\ Verification: Geolocation, \\ source analysis\end{tabular} \\ \midrule 
5 / E5 &
  Local news journalist &
  3--5 years &
  \begin{tabular}[c]{@{}l@{}} \newt{Identify local U.S. politician's public appearances}; \\ Report on the discourse around deer hunting \\ within local city limits\end{tabular} &
  \begin{tabular}[c]{@{}l@{}}Discovery;\\ Verification: fact-checking, \\ geolocation\end{tabular} \\ \hline
\label{tab:sessions}
\end{tabular}}
\end{table*}

\subsection{OSINT Lab Course}
\label{sec:osint-lab-course}
The OSINT lab course was taught simultaneously with students from two universities in the Fall of 2021. Students in the program learned about the OSINT lab course through word of mouth, recruitment emails, and course catalogs. They received credit for completing the course but were not required to stay in the course. The collaboration with investigators from multiple domains was designed to provide an authentic learning experience. 

Participants spent the first half of a semester learning about the entire OSINT analysis process. The training covered technical and ethical aspects of OSINT investigations. The first two weeks were spent on introducing the field of OSINT with examples of impactful investigations and its key elements of a culture of transparency, an adversarial mindset, and collaboration among individuals. Each subsequent week focused on one of the identified OSINT macrotasks. The third author demonstrated tools and techniques associated with that task. Details about these skills are provided in Table~\ref{tab:tasks}. All the authors participated in designing and implementing practice sessions where students formed teams and solved demo tasks that required the application of relevant OSINT skills. Based on previous work demonstrating the benefit of goal setting in training \cite{rechkemmer_motivating_2020}, we asked each team to submit at least 3 high-quality submissions during practice sessions.

The crowd conducted real-world investigations based on expert prompts during the second half of the semester. \newt{After recruiting each expert, we scheduled a 30-minute pre-session training meeting a few days before the session. First, we presented a short slide deck that gave an overview of the study, our expectations for the expert, and described the five crowd macrotasks in detail. Second, the expert brainstormed an appropriate investigation topic and we discussed how to decompose it and map it onto one or more macrotasks. Third, we answered any questions the expert had about the forthcoming session.} 

The investigations varied in terms of topics and contexts as they were based on real ongoing investigations of experts. One common thread was the use of publicly available information (i.e., OSINT), predominantly social media content, and involved a combination of discovery and verification tasks. OSINT investigations ranged from fact-checking videos, documenting human rights violations, finding traces of homicide suspects, and investigating the whereabouts of public servants.

\begin{figure*}[]
\includegraphics[width=17cm]{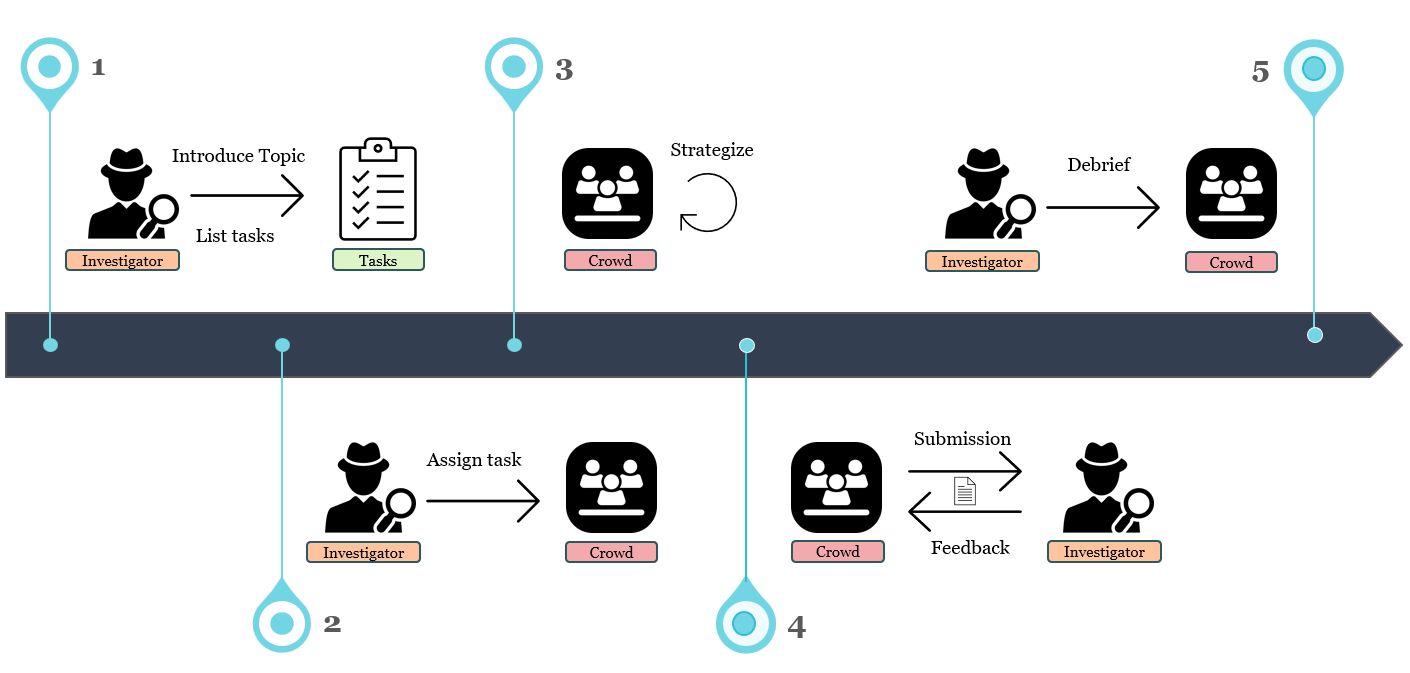}
\caption{\newt{Phases of our study. ORS connects experts with a trained crowd to perform real-world OSINT investigations described in Section~\ref{sec:osint-lab-course}. The study sessions facilitate synchronous collaboration through 4 chronological phases: (1) The investigator presents the investigation topic and lists tasks that relate to the 5 OSINT macrotasks (presented in Table~\ref{tab:tasks}). (2) The authors collaborate with the investigator to assign tasks to teams and provide links for submission. (3) The crowd strategizes the division of work within their teams. (4) The crowd conducts investigations for the rest of the session, submitting their findings through a tailored Google Form. Responses are aggregated in a spreadsheet, allowing real-time expert feedback and guidance. (5) The expert debriefs the crowd by sharing their insights and high-level feedback.}}
\label{fig:study_session}
\Description{}
\end{figure*}

There were five investigation sessions. Each session was 75 minutes long and conducted during the class. Experts joined remotely through a video call and the crowd had the option of either joining remotely or being co-located in a classroom. The student crowd consisted of eight teams of three to four members for every session. Each session had four phases. First, the investigator gave a short presentation on what they wanted the crowd to investigate. Second, the authors worked with the expert to assign teams to different subtasks and pointed to resources for submission. Third, the crowd spent the next five minutes strategizing the division of labor within each team. Fourth, the crowd investigated for an hour. Teams made submissions to a Google Form tailored for each session. A spreadsheet aggregated these responses and experts reviewed them in real-time to provide feedback and guidance to the crowd. \newt{Finally, at the conclusion, the expert engaged in a debriefing process, discussing what they learned and giving some high-level feedback to the crowd. An overview of the structure of the investigation sessions and the roles of experts and crowds is presented in Figure~\ref{fig:study_session}.}

\begin{table*}[]
\caption{Discovery and verification tasks for OSINT investigations}
\resizebox{2\columnwidth}{!}{\begin{tabular}{@{}lll@{}}
\toprule
\textbf{Task}                   & \textbf{Skills}                                                                                                                                   & \textbf{Tools and techniques}                                                                                                                                                                                                                                                                   \\ \midrule
Discovery \cite{FirstDraft_discovery}                      & \begin{tabular}[c]{@{}l@{}}Gather hashtags, identify\\ interesting social media  \\ accounts, and tailor search for \\ global and local events \cite{FirstDraft_discovery}\end{tabular} & \begin{tabular}[c]{@{}l@{}}Advanced searches on social media platforms \\ based on keywords and location; \\ Follow digital trail to identify key actors and \\ their associates; CrowdTangle \cite{fletcher2018measuring};\\ Archive information through archive.is\cite{archive_is}\end{tabular}                              \\ \midrule
Verification: Source analysis  \cite{FirstDraft_verification} & \begin{tabular}[c]{@{}l@{}}Look at provenance of content; \\ research the person, \\ organization (or bot) behind \\ posts \cite{verification_handbook}\end{tabular}              & \begin{tabular}[c]{@{}l@{}}Reverse image search through engines like \\ Yandex\cite{yandex}, TinEye \cite{tineye}, Google\cite{google_image}; \\ Search on social media platforms including \\ closed ones like Reddit and Telegram;\\Hoaxy\cite{shao_hoaxy_2016}, BotSentinel\cite{bot_sentinel} to analyze sources\end{tabular} \\ \midrule
Verification: Fact-checking  \cite{bellingcat_verification}   & \begin{tabular}[c]{@{}l@{}}Find prior fact-checks or\\ conduct original research to \\ fact-check a text-based claim \cite{micallef2022true}\end{tabular}                 & \begin{tabular}[c]{@{}l@{}}Look at previously tagged fact-checks \cite{hassan_claimbuster_2017};\\ Go through official documents\end{tabular}                                                                                                                                                                                  \\ \midrule
Verification: Image analysis \cite{bellingcat_verification}   & \begin{tabular}[c]{@{}l@{}}Identify visual clues,\\ metadata, potential \\ manipulation/editing \cite{FirstDraft_verification}\end{tabular}                                      & \begin{tabular}[c]{@{}l@{}}Use Exif metadata\cite{exif}; \\ Reverse image search; \\ Look at shadows, weather during the time \\ at the location\end{tabular}                                                                                                                                        \\ \midrule
Verification: Image geolocation \cite{verification_handbook} & \begin{tabular}[c]{@{}l@{}}Identify where on Earth \\ a photo or video was taken \cite{Bellingcat_guide}\end{tabular}                                                     & \begin{tabular}[c]{@{}l@{}}Reverse image search; \\ Navigate through satellite imagery \\ (Google earth and Google street view)\end{tabular}   \\ \bottomrule
\label{tab:tasks}
\end{tabular}}
\end{table*}

\subsection{Design Arguments}
\label{subsec:design-arguments}
To scale and speed up OSINT investigations carried out by journalists, fact-checkers, law enforcement officers and human rights investigators, our work sought to develop a collaborative crowdsourcing framework that: (a) trained the crowd in OSINT analysis skills both in terms of tasks and the ethical aspects; (b) divided real-life expert investigations into constituent discovery and verification tasks; and (c) enabled synchronous collaboration between experts and a trained crowd. OSINT Research Studios (ORS) provides a sociotechnical approach that orchestrates training and deployment of a crowd for performing efficient and ethical investigations involving OSINT.

ORS addresses the design challenges for crowdsourcing complex OSINT tasks in the following ways:

\subsubsection{Delegation of complex OSINT tasks}
To cater to the needs of analysis of open source information across multiple domains, ORS uses the OSINT framework \cite{williams_defining_2018} to divide larger investigations into tasks. Prior work \cite{williams_defining_2018,belghith_compete_2022} divided OSINT investigations into four steps through the OSINT cycle: (1) discover relevant information; (2) verify its provenance and look at the veracity of claims around that content; (3) archive them for future reference; and (4) report on the findings of the investigation. ORS dives deeper and further divides the discovery and verification phases into five parallelizable macrotasks. Each of these tasks is associated with distinct OSINT tools and techniques. These tasks can contribute to rapid, focused investigations as they do not require extensive background knowledge from the crowd and generate results quickly. We provide details about these tasks in Table~\ref{tab:tasks}. Work on journalistic processes of information discovery and verification by Brands et al. \cite{brands2018social} and practitioner resources from  First Draft News \cite{FirstDraft_discovery, FirstDraft_verification}, Bellingcat \cite{Bellingcat_guide}, and Poynter \cite{verification_handbook} guide the design of these tasks. Micallef et al. \cite{micallef2022true} include most of these resources in their review of computational tools used by fact-checkers.

The five independent tasks serve two purposes in ORS: they (1) form the basis for training the crowd; and (2) break down expert investigations into one or more tasks. These tasks are designed to generate rapid, focused output that can augment ongoing expert investigations. As the crowd is trained to perform these tasks efficiently, they \newt{should be} able to complete them when prompted by the experts. Following the ethos of the field of OSINT, the tasks discourage over-reliance on custom tools, and training is based on techniques for working directly with data sources. The contexts vary widely based on the domain and the particular investigation, but are circumscribed by OSINT analysis. Therefore, these tasks map the expert investigations to the crowd's \newt{skill set} and provide a way to garner the required support for investigations.

\paragraph{Iteration:} We iterated on making the tasks easy for experts to relate to and use. Initially, we had two discovery tasks for local and global events. However, we realized that the crowd applied the same tools and techniques to both, and experts did not need to specify the scope beforehand. They could delegate the task as discovery and the scope would be determined by the available information. 

\subsubsection{Safety, privacy, and ethical considerations}
\label{sec:safety-considerations}
ORS elevates expert supervision and provides the crowd with a solid understanding of the ethical considerations of investigations. ORS operationalizes the ethos of OSINT, including prioritizing transparency and avoiding the use of subterfuge, in the following ways: 

\begin{itemize}
    \item {Prioritizing transparency:} ORS encourages each piece of information submitted to be archived to counter the ephemeral nature of the information. Source identification and documenting all relevant details are also part of the process.
    \item{Avoiding the use of subterfuge:} ORS prohibits hacking and, given a classroom setting, the honor code is applied to curb any such attempts.
    \item{Limiting investigations to passive reconnaissance:} ORS promotes the use of sock-puppet accounts that help to anonymize the identity of the investigator. This is to ensure that ongoing investigations are not interfered with and investigators are not putting themselves at risk, in case organized crime groups apply countermeasures. Crowdworkers are provided with access to virtual machines and VPNs to access unsecured information without jeopardizing their computing systems.
    \item{Fostering accountability:} Having a dedicated, accountable crowd of known participants (i.e., the students) also helps preserve privacy and reduce the risks of leaks and sabotage.
\end{itemize}

\newtext{As a part of the training process, we ensured that crowdworkers were well-versed in the guidelines and ethical considerations of OSINT investigations. We used real-life scenarios and case studies to help the crowd understand the potential risks and challenges. We also informed the crowd about the potential harm to individuals and communities and the responsibility they held in conducting investigations ethically.}

\newtext{ORS enables experts to stay in control of the investigation. We helped experts outline the scope, objectives, and boundaries of the OSINT investigation during the planning phase by providing examples of previous investigations and explaining the scope of the tasks. We also encouraged experts to balance providing context and disclosing sensitive information to protect the integrity of their investigation.}

\newtext{During each investigation, we established mechanisms for feedback and accountability. We encourage crowdworkers to report any issues or concerns and plan to address them promptly. We ensured that experienced investigators were actively involved in guiding and overseeing the investigation. Investigators utilized their expertise to validate information, make informed decisions, and navigate complex situations. They provided feedback to crowdworkers to ensure conclusions and findings are based on accurate, reliable, and verified information. We also conducted a post-investigation review with each expert to evaluate the process and outcomes.}

\paragraph{Iteration:} We iterated on the crowd submission forms to enforce the requirement of archiving any information put on the spreadsheet for transparency. A required field was added to provide archived links for investigated social media content. \newtext{We also added a notes field to promote further analysis from the crowd and highlight interesting findings for the expert.}

\subsubsection{Organizational overheads of synchronous collaboration} 
ORS employs a piggyback prototyping \cite{grevet2015piggyback} approach to develop a workflow for synchronous collaboration between the expert leading the session and the crowd. The goal was to make the crowd submissions accessible to the expert and present all relevant details in one place. Google Forms are used for specifying the requirements for the information requested. These submissions were collated in a Google Sheet. This sheet was accessible to all the participants of the session including the experts, the crowd, and the researchers. Experts provided qualitative feedback for the submissions on the sheet. This piggybacking approach proved beneficial for three reasons: (1) the crowd and experts had previous experience with the interfaces \cite{brands2018social}, (2) the established applications are robust, and (3) data can be easily exported or read from external applications for further processing. 

Another major issue in such crowdsourced settings is duplication of effort \cite{venkatagiri2023cosint}. ORS tries to circumvent this issue through these two aspects of team dynamics. First, the crowd works in teams, and each team is assigned an exclusive task. The division of labor can be on the basis of social media platforms, location, subtopics, and/or OSINT subtasks. For example, if one team is searching for information in France, the other teams are looking at countries other than France. Second, each team is responsible for submitting unique entries. This requires collaboration within the team and being aware of what other team members are working on.

\paragraph{Iteration:} We initially created a team leader role who was responsible for dividing up and delegating tasks as well as communicating directly with the expert. However, the team leader requirement was relaxed based on crowd feedback that teams did not find that role useful. \newt{Without the formal leader role, we dedicated 5 minutes at the start of each session for teams to discuss their strategy.}

\subsubsection{Maintaining the Quality of Investigations}
Previous studies have found that feedback from experts and self-evaluation both improve the quality of crowdwork \cite{dow2012shepherding}. ORS \newt{seeks to leverage these feedback benefits} as follows. First, experts are asked to look through the submissions during the session and provide qualitative feedback that can help crowdworkers improve their performance. Second, the crowd rates their own submissions based on three measures. For a piece of content to fulfill the requirements, it must be specific, verifiable, and relevant. 

Experts are also encouraged to communicate verbally with the crowd and make interventions to direct the crowd in productive directions. For example, if the expert finds the information coming in to be of a certain type which is not helpful, they can qualify their requirements further and help the crowd generate more appropriate information. Experts can also move teams from one task to another based on progress across tasks at any time during the session.

\paragraph{Iteration:} Based on feedback from the crowd, we solicited more involvement from experts. We asked experts to provide feedback quickly. We also prompted experts to ``think aloud'' (in the virtual meeting) any information that they found helpful \newtext{and discuss any high-level feedback that could help the crowd improve further during their debrief at the end of the session.}
\section{Findings}
\label{sec:findings}
\subsection{Role of OSINT Macrotasks in the Crowdsourcing Setup}

\subsubsection{Experts found OSINT macrotasks to be relevant} Experts mentioned that they spend a lot of time on the 5 OSINT macrotasks in their typical investigations. We asked experts to rate the tasks on a scale of 1 (not useful) to 3 (very useful) based on how relevant they are to their investigations. Overall the tasks received an average rating of 2.76 out of 3. The verification tasks were rated more favorably compared to the discovery task. 

\paragraph{Discovery}
The discovery task received an average rating of 2.4 out of 3. E2 mentioned that discovery is their first step in any investigation and any help there would be beneficial. E5 stated that the task is important in news discovery. E1 and E3 found it to be less relevant as  the discovery task can become overwhelming and collecting too much unverifiable information is counterproductive. Experts also thought that the task was suitable for crowds. E5 thought that the crowd has ``far more sort of computer or digital literacy'' \newt{than he does} to perform well-directed advanced searches and mine information from a wide range of social media platforms.

\paragraph{Verification}
Verification tasks had an aggregate rating of 2.85 out of 3. Each sub-task was found to be relevant to the verification process that is integral to OSINT investigations and the work practices of the investigators. 

Source analysis enabled experts to dig deeper into interesting leads and possibly reach out to them for further investigation. E1 thought that the crowd would be especially strong in this task, and he could imagine delegating this work of finding background information on creators of social media posts as a part of his workflow.

Fact-checking was thought to be a form of deep research and termed “fairly straightforward.” E1 and E4 thought that this task could be performed well by a group of trained people. E1 mentioned that it is an essential step and he would “still feel the need to do it [himself].” But assistance from the crowd would be helpful to speed up the process and potentially find evidence that might get missed.

Experts found image analysis to be a hard task, especially detecting manipulation, but thought it would be useful to have members of the crowd look at it independently and come to a conclusion.

Geolocation was a part of all the investigators’ work practices and received a maximum usefulness rating of 3. Experts explained the importance of the task in identifying recycled information. For example, it is the next step for most evidence in human rights investigations, placing the content in a particular geographical location. E4 had previous experience with crowdsourcing geolocation tasks and thought it worked very well.

\subsubsection{\newt{Crowd was confident about applying OSINT skills that they were trained on}} 
\newt{Crowdworkers felt that they understood the requirements clearly and could work towards the solution. CW6 recalled, “The submissions that I gave kind of knew exactly like what kind of information they're looking for and how much detail they wanted it to like go into. And what kind of information wouldn't be too useful to submit to the Google Doc.” CW1 mentioned that the practice sessions were useful to learn the skills, especially geolocation. Practice using tools like reverse image search, Google Street View, and looking up the language of signs were all useful during expert sessions. He felt it was “like a game but [investigators] use that in real life.”} 

\newt{The crowd had a positive experience being able to apply the learned skills. They appreciated how the tools and techniques learned during training could be applied to impactful real-life investigations. For example, CW2 mentioned, “...taking all of the techniques that we learned in class and going in seeing people who actually use those on their day to day it was really interesting.”}

\subsection{Session Planning}

\subsubsection{Investigations were decomposed into OSINT macrotasks} 
We worked with experts to break down their ongoing investigations into prompts that each apply one of these five tasks described in Table~\ref{tab:tasks}: discovery, source analysis, image analysis, fact-checking, and geolocation. Experts came up with the prompts for students after we described the scope, tools and techniques involved and examples of previous practice and expert investigations relating to each task. In all 5 study sessions, experts chose a mix of discovery and verification tasks for the crowd as described in Table \ref{tab:sessions}. The tasks were framed as questions that looked for detailed answers and supporting links and documents. For example, in the session with E1, five different questions were asked based on a viral Instagram video clip of Dr. Fauci speaking at an event:
\begin{itemize}
    \item Is this a legitimate video, not one doctored to make it look like Dr. Fauci speaking?
    \item What is the context of the video --- where/when/why did Dr. Fauci speak?
    \item What is the context of Dr. Fauci's remarks? He says something like, "you take an infectious agent and you introduce it into a population," making it seem like he is behind the HIV/AIDS epidemic, but what was he addressing with his remarks? What are some of the other related facts for the epidemic mentioned?
    \item What is the background of the Instagram user?
    \item Have any news articles or fact-checks been published about this particular video, or about Dr. Fauci's remarks in the video? Where else has this video been used?
\end{itemize}

\newt{Experts divided larger investigations into tasks for crowd teams across several dimensions, including social media platforms, location, subtopics, and OSINT subtasks. For example, in session 4, teams were looking into anti-vaccine protests across countries in Europe, and each team had a specific country to look at. This form of task assignment helped in the non-duplication of effort across teams. The investigator assigned teams to the scoped tasks randomly, as the crowdworkers were assumed to have the same skill level.}

\subsubsection{Experts sifted through information around topic before session} Experts found planning before the session essential to the process. Investigators mentioned that they spent substantial time trying to acclimatize themselves to the information surrounding the discourse of the topics. For example, E3 said, “I gave [the crowd] a bunch of information on the location of the first test” and thought “that was very helpful for them that allowed them to provide geospatial data on trails.” Investigators thought sifting through information around the topic could help ensure the right level of difficulty for the sessions. Discussing how his preparation would change if he were to do it again, E1 said, “I would definitely do my own research” ahead of time. The preparation helped experts respond faster and more reliably to the submissions. 

\subsubsection{Experts had different foci based on quantity and quality of submissions} 

\paragraph{Focus on quantity} \newt{E2 and E4} looked to gather a bunch of information that could then serve as a starting point for their investigation and future reporting. For example, E4 prioritized quantity and mentioned that his organization's investigation was at a stage where they would be more concerned with quantity over quality. He reasoned that for “taking this data and turning it into a project, I think we want to err on the side of letting us decide [later] what is useful or not.” He wanted the crowd to submit all relevant and interesting content without second-guessing. 
\paragraph{Focus on quality} \newt{E1, E3 and E5} looked for verified information and had a focus on the quality of submissions. They wanted additional relevant information for the discovered content. This information would come from follow-up verification tasks that identify visual elements like buildings, \newt{cars,} flags, and groups involved.  

\subsection{Collaboration Within Crowd Teams}

\subsubsection{Team formation} Team formation was initially based on proximity in terms of seating. Teams generally stuck together once they were formed during the first study session. There were very few changes to most of the teams across the sessions, typically caused by members being absent or leaving an existing team.
 
\subsubsection{Division of work within teams} Within their teams, the crowdworkers divided up the work keeping in mind the requirements set by the investigators. The team members worked individually after choosing non-overlapping search spaces for online information. Sometimes each member had a preferred social media platform to investigate and they tried to divide the work equitably. Most of the coordination was to make sure that there were no duplicate posts from the same team. CW12 explained, “if one of us found something, we let the other person know so we don't find the same thing twice.” They also got a sense of which social media platforms had more relevant content and pivoted their searches based on content discovered by other team members.

Only two teams reported having particular strategies in place to divide up the work which did not change over time. CW12 described this common pattern for the division of work, paraphrased as follows. First, get together and divide up the work individually, for example take up different social media platforms like Twitter, Reddit, Facebook, etc. Second, double check posts for duplicate when they find something relevant, especially if multiple members were working on the same platform. Third, add relevant verification details. Fourth, repeat this for all sessions, as they had the same team throughout. Other teams had members working individually on tasks without a particular strategy, but following a similar workflow.

\subsubsection{Benefits of teamwork} The major benefit of teamwork was observed when getting relevant information for the prompts was difficult. For example CW8 mentioned that he relied on his team when he felt it was “just kind of daunting” to take up one platform and “find all the information” individually. As discovery got harder, their team wanted to move from one platform to another as a team. The members of the team bounced off ideas with each other when they hit a dead end with one platform.

Team members discussed the feedback that they received from the investigators. They generally found the feedback to be helpful and wanted all the members to be aware of the pointers or corrections provided. The teams communicated verbally and talked about the posts they were working on to ensure the others knew. We also observed communication within teams when they got together to figure out the next steps after getting stuck with their individual investigations. Some of the crowdworkers wished for better communication as a team and but thought that they could not successfully enforce any organizational structure. No leadership roles were established within the crowd teams.

\subsection{Collaboration Between Experts and the Crowd}

\subsubsection{Collaboration through expert feedback} \newt{The primary mode of collaboration between experts and the crowd involved the submission of tasks by the crowd and the subsequent feedback provided by the experts.} Across all the sessions, investigators provided substantial feedback on the spreadsheets containing crowd submissions. Out of 196 submissions across the sessions, 112 submissions received long-form feedback, generally 1--2 sentences. \newt{E2} mentioned that they did not leave any comments if they thought that the submissions “didn't have any relevant information that would help us one way or the other for sure.” The feedback was used for two main purposes: getting the crowd to dig deeper into the discovered content and course-correcting the investigations. 

\paragraph{Feedback asking crowd to dig deeper} Experts pointed out specific parts of submissions and asked them follow-up questions to bolster the evidence. Some examples from written feedback on spreadsheets include, “Is there any quick verification you can do to confirm that this is actually Brussels or at least Belgium?”, “22 minutes of driving video.  This is the kind of stuff that could be really helpful.  How do we know when the video was shot?  Spreadsheet says 10/9 but youtube says it was posted on 10/15.” In one case, a crowd team submitted another social media post made by the same poster and indicated that the source had a history of spreading unverified information. The investigator asked the team to find more such historical posts that could be potential misinformation. 

\paragraph{Feedback asking crowd to course-correct} Experts reiterated the requirements if they found submitted information to be less relevant. Feedback on the sheet clarified the date ranges, helped submissions to focus on primary sources and called out the lack of details for the investigations. Some examples include, “Posted on the Oct 5th but the page says the footage was shot on 9/25.  Outside our range”, “Good. But best if we can find non-mass media sources from individuals on the ground”, “Hard to tell if this is a covid protest-- getting more footage from the protest on this date would help to verify.”

\subsubsection{Experts made effective interventions} Experts could make announcements during the session to influence the crowd as a whole. The crowd responded well to high level feedback and tailored their investigation to the needs of the expert. E3 asked for more details about the source and location of the media submissions and the crowd provided that. E5 prompted the crowd to discover information about deer killings in and around a city in the US. He found that the crowd took some time, but pivoted based on his feedback of avoiding trophy pictures. He thought that the delay was reasonable and attributed the latency to rabbit holes that investigators might have gotten into and the time needed to reformulate the searches.

\subsubsection{Crowdworkers found feedback from the investigators to be helpful} Feedback helped the crowd in navigating the information space around the topic. The crowdworkers felt that it was important to receive quick feedback from experts to improve their performance. CW6 recalled session 4 and said, “[E4] gave feedback right away, which is, which is a big deal for us because we're able to understand Okay, this is exactly what [the expert] is looking for \ldots now we can kind of start tailoring our searches to that.” Feedback from the experts helped the crowd improve their subsequent submissions. Quick feedback motivated the crowd to stay engaged and keep looking for new information. On the other hand, slow feedback from some of the experts made CW10 feel that his submissions were less useful.

\subsubsection{Self-evaluation}
\paragraph{Experts speculated about the usefulness of ORS}
Investigators did not use the self-evaluation scores from the crowd during the sessions. However, they thought it could be useful for sifting through information like in the case of geolocation, where the ratings could reflect their confidence in the result. One of the experts mentioned that having them can benefit longer running investigations with a huge number of submissions. E3 and E5 mentioned that they found self-evaluation ratings from students to be \newt{"fairly accurate." Likewise, there tended to be positive expert feedback on submissions with high self-evaluation scores, whereas submissions that did not meet requirements as pointed out by experts had low self-evaluation scores in the relevance and verifiability measures.} Others like E4, who had a hard time figuring out the self-evaluation part of the sheet, mentioned that those measures were not clearly communicated to him.

\paragraph {It helped the crowd to reflect but did not improve quality of submissions} The crowdworkers had to rate each of their form submissions based on the three metrics of specificity, relevance and verifiability across all the expert sessions. The crowd thought self-evaluations reinforced important aspects of what the expert investigators were looking for into the submissions. CW13 felt that “it was really useful to evaluate ourselves just to kind of keep up the quality in our submissions and make sure the experts that were kind of like looking over our submissions actually got something out of like coming to the sessions.” Crowdworkers did not feel that self-evaluations improved their performance, as it was after the submission that they added the scores. But they were able to reflect on their submissions and sometimes admitted that their results were lacking. \newt{The average scores on a scale of 0--2 across the three measures were high, and ranged between 1.67 and 1.93 across sessions. Submissions with all 2s ranged from 57.7\% to 81.5\% of total submissions during sessions.} There was no noticeable improvement or significant changes in these scores over time.

\subsection{\newt{Factors Affecting the Crowd's Performance During the Investigations}}

\newt{Each team was assigned a particular prompt and they submitted information to cater to that. The number of Google Form submissions varied widely across the sessions based on the difficulty and type of task. For example, discovery tasks generated more submissions, whereas verification tasks required more information and research which reduced the quantity. The highest number of submissions (71) were made during session 4, which was about documenting anti-vaccine protests across Europe. Other, more geographically specific and time-bound investigations generated leads ranging from 25 to 38 in number. Expert feedback, applicability of OSINT skills, difficulty and context of tasks, and the topic of investigation all may have influenced their performance.}

\subsubsection{\newt{Applicability of OSINT skills}} \newt{Some of the crowdworkers felt that their performance was highly dependent on the expert prompt and how well relevant information could be mined using OSINT tools and techniques. CW5 talked specifically about session 3 which he thought was the hardest because information was difficult to find and the task was not suited for their skills. He found sessions to be easier when they featured the use of geolocation and searching on Twitter and Facebook. Another compounding factor that negatively impacted the performance of the crowd was explained by CW6 as the lack of clarity about the requirement: “... if [experts] were a little too broad and the subject matter expert didn't actively kind of talk about what they were looking for, as far as topics for mis or disinformation, it was kind of hard to hit the mark in some instances.”} 

\subsubsection{\newt{Difficulty of task}} \newt{The crowd mentioned that the difficulty of tasks impacted productivity during sessions with experts. They wanted the tasks to be challenging but where they could make progress and have submissions to show for the session. CW12 mentioned that his favorite expert session, “...was a good balance between not being too easy or a little too hard, so I thought that was a good like middle ground where we had enough to work with \ldots there's like meat on the bone to work with.” High difficulty significantly contributed to the least favorite sessions for the crowdworkers. This was because the crowd did not feel that they could contribute to the investigations meaningfully and some reached a dead end even before the end of the session. But a few of the crowdworkers enjoyed a challenging thread of investigation and were fueled by competition while trying to gather information on a hard task.}

\subsubsection{\newt{Context for task}} \newt{Crowdworkers felt more engaged and had a positive experience if they understood the context for the investigations and got how the information they submitted could be used effectively. This also helped them provide more relevant information and dig deeper into interesting pieces of information. CW3 described one scenario where this was not the case: “I was able to find the picture and geo-located it, which was fun, but I just wasn't you know overly sure of how it was actually helping,” so this made the session his least favorite. CW10 found that the expert in the first session kept reiterating that the crowd submissions matched his own findings on the topic; this made him feel that he did not contribute to the investigation. The crowd was motivated by the context around the investigation and needed a clear objective to be specified right at the beginning of the session.} 

\subsubsection{\newt{Topic of investigation}} \newt{CW4 disliked finding COVID vaccine misinformation because it was “kind of boring” and “in the news all the time, I read about [it] all the time.” The crowd responded well to topics that had close physical proximity; for example, discovering information about a tornado that hit the town where the crowd's university was located, or verifying claims about a local leader of a nearby city. CW2 wanted the topic to have "...a balance between like it'd be interesting topic for us and, like an important topic to do."}

\subsection{Reflections on Synchronous Collaborative Crowdsourcing Setup}

\subsubsection {Collaborative workflow during session}

\paragraph{Information access}
The proposed model of working matched how experts themselves organize some of the tasks in terms of the use of Google Docs and Sheets. Investigators were able to access the information seamlessly and stay on top of it. E5 mentioned,  “I liked the ease of the process. You know sort of I had all the information at my fingertips.” Talking about the experience of monitoring submissions, E1 thought a spreadsheet was efficient for him to provide quick feedback.

\paragraph{Hybrid setting of sessions}
Investigators liked the hybrid setup of getting them to join remotely and work with two groups of colocated crowds. This was explained by E4 when he mentioned the shift to remote work which was exacerbated by the pandemic and how this model gets at that problem. He thought working on investigations is “not the same as doing group activities on zoom. It just doesn't work.” This model was helpful for him to collaborate with 30 participants on Zoom and perform OSINT analysis. However, he mentioned that the same level of engagement and work atmosphere that can be achieved by having the investigators and the crowd in the same physical space cannot be emulated in such a hybrid setup.

During hybrid sessions where experts joined remotely, experts spent most of their time going through the submissions and providing feedback. In the only session where the investigator was present in the classroom, he spent the majority of their time helping the students think through the issues while walking around, and the remaining time on evaluating online submissions.

\paragraph{Crowd liked the setting but wanted to learn more from experts} The crowdworkers appreciated the remote involvement of expert investigators and how they were able to collaborate through forms and sheets. The crowd felt they understood the requirements of the ongoing investigation based on the brief of the investigation and how the tasks were carved out. Having more detailed requirements like date ranges and locations on the task helped the crowd to engage more efficiently with the available information. However, some of them wanted lessons and demonstrations from experts. For example, CW5 wished, “If experts could talk a little bit about what tools they would use for some things, just give us a little more insight on how they would solve it if they were you know in our seat.” Such discussions could enrich their skill set and possibly help them discover new and efficient methods.

\subsubsection{Experts had varied experiences while keeping up with crowd submissions} 4 of the 5 investigators mentioned that they were able to keep up with the information that was coming in and were able to provide timely feedback. E5 talked about his professional experience in such a role and thought he was able to put his ability to peruse information to use and “didn't need to catch up.”

Experts pointed out how submissions picked up pace during the later part of the session as the crowd got a hang of the topic. E4 talked about the difficulty of providing feedback to each submission with a quick turnaround time. He explained this problem, “...when I'm working by myself, I'm working at the speed of me, right, so I've got the link and I'm going to archive it. But if I've got seven groups of people submitting things at the same time. Then suddenly, not only am I working, but I have to work faster because I've got all this data coming in, but then I'm also not in a single thread in my head.” This issue was more prominent in session 4 due to the high volume of submissions (71 form submissions). He also shared his thought about the time limit for such deployments. He thought these sessions could be extended to 2 hours without much negative impact. However, for a normal work day which is around 8 hours, the setup would get unsustainably taxing for the expert.

\subsubsection{Lack of interaction}

Experts mentioned a lack of interaction with the crowd; the teams did not reach out to the experts with any clarifications or concerns. E4 did not have any interaction other than providing feedback on submissions. He thought of this interaction as a trade-off and said “I'm not sure how you would do that and still get done what you get done during the class.” Experts felt that the sessions could be more interactive through questions and verbal feedback, but acknowledged the challenges with the size of the crowd and time constraints. E4 thought having an ice-breaker pre-session with at least one member of the team members, possibly the team leader, could be helpful for the crowd to reach out to them.

Some of the crowdworkers also felt the lack of interaction during investigations. They mentioned possible advantages of having the experts join in person as that could add more communication channels and help gather quick verbal feedback.

\subsection{Reflections on the Quality and Quantity of Submissions}

\subsubsection{Experts reflected positively on the efficiency of crowd} Investigators appreciated the contributions of the crowd investigations. E2 described the challenge with their task as “...people are actually probably posting stuff all day long and there are probably thousands of entries every day, you probably be super overwhelmed …we just don't have that kind of manpower.” He mentioned that the results from crowd investigation were important to them and the information would provide “more avenues of investigation because now [they] can potentially go back to some of these users and ask them for more data.” Reiterating the lack of manpower and describing how the session can speed up their work, E5 mentioned, “we have some great talented freelancers occasionally at the [local newspaper], but it is mostly just me …so having you know people who can help in this sort of process, particularly in some of these cases, you know investigative work it's time-consuming and looking through things would be super useful.”

E1 speculated about the efficiency of the crowd in terms of speed and accuracy, “...for the most part, [the information submitted] was very strong, particularly how quickly they were able to respond. I would say that they pretty much grasped what I was asking for and provided good answers.” E2 added, “The type of work, you guys did in an hour would take us you know, certainly all day with one person doing it, if not longer.” The investigators found the information to be good in terms of both quality and quantity. They also mentioned particular pieces of information that they found to be very promising and could lead to breakthroughs in their investigations. E3 recalled two such posts and said, “...that was something that I had been searching for but couldn't find so that was really great.” Experts could successfully crowdsource parts of their ongoing investigations to gather rapid, focused results.

\subsubsection{Experts found trained crowd more suitable than the general public}

Crowdsourced investigations with the general public have been plagued by issues of sabotage, low output, and leaks \cite{allen2022birds, nhan2017digilantism, trottier2017digital}. E4 talked about the challenges faced with their investigation of the US Capitol insurrection on January 6, 2021 which involved collecting pictures of the riot from mostly Twitter followers of an OSINT organization. He mentioned that they got a lot of “junk links” which was probably to slow them down. They had to be mindful of bad actors who want to “feed garbage.” They used virtual machines to access links from unknown social media platforms. E5 mentioned that it might be risky to involve an online crowd, for example, by asking the Twitter followers of the journalist to find out more about a lead. They do not want to tip off competition and are mindful of the sensitive nature of the information that they are trying to present.

E2 and E4 talked about the comparative advantage of such a trained crowd. Compared to the leads from the general public through investigations that seek the same information from the general public, experts thought the current crowd was able to perform significantly better in terms of the quality and quantity of relevant information generated. E2 described it as “...to finish with 30-plus entries and I'm very happy with that because we nearly doubled what we had from the public in a really short span.” To be precise, there were 29 submissions during the session compared to 15 leads that the expert received through an open call to the public.

Investigators appreciated the crowd’s effort to archive and document the steps of their investigation as it is fundamental to the transparency of OSINT analysis. The archived links helped investigators to have access to the social media posts readily and use it for future reference. The crowd added relevant verification details for the discovered information and added archived links to peripheral information, map coordinates for geolocation, and found original sources for recycled information. Investigators trusted the process and thought they were more confident about the crowd’s abilities than they started out with.

E5 thought that working with a crowd like the one during the session reduces the chances of a leak or creating potential misinformation about the subject of the investigation. Regarding the issues of safety and relevance, E4 felt that “with a group of students you know who are not deliberately trying to sabotage the investigation, those concerns are not there.”

\subsubsection{Experts identified weaknesses in crowd submissions}
Some of the expert expectations were not met during the sessions. The investigators reported that only a fraction of the submissions hit the exact target. Characterizing the low-quality submissions, investigators talked about the inconsistency among teams in filling out the required details about the social media links. For example, E3 talks about a particular example, “for the notes and the visual cues section, some people are, I think, correctly saying that this is in front of the Trump hotel, which is super useful. But other people would write things like 'signs' or like, 'a building,' which is less useful.” 

During interviews, the experts shared issues related to the specificity, relevance, and verifiability of crowd submissions. These issues led to the lower quality of a part of the submissions. E3 pointed out a lack of understanding of the context behind the investigation and limited time during sessions for information that missed the required details. Some information was found to be only “tangentially relevant.” Some submissions were news articles that were copied and pasted, without looking at the veracity of those sources. 

Investigators reflected on what they could have done differently to avoid this. E5 thought useful clarifications could be provided before starting off the crowd like specifying the type of content and social media platforms to be mined. The prompts and the little time that investigators had while introducing the topic was crucial. Interventions were successful, but investigators felt giving the form of content that would be most valuable at the beginning could have improved the quality of submissions. For example, E2 reflected, “I don't know if I'd mentioned the GoPro cameras or not, but I wish I would have, if I didn't I wish I would have said it.”

Some of the experts mentioned that the tasks were not fully completed. E2 said, “I wouldn't say that what we did today is the end of it because we're still asking the public to send this information to but I'm certain what you guys produced today will be helpful.” E3 stated that the information collected from their first task would “basically reduce the time that [he] would need in finding other subsequent information.” It was therefore hard to complete investigations based on the results of a single 75 minute session. 

\subsubsection{The crowd perceived that their performance improved over time} \newt{Crowdworkers thought that they got faster with their searches and application of tools.} CW3 listed the factors of “repetition of like practicing [techniques] over and over again” and “some decent feedback [from experts] where I'd be like, 'Okay, so I need to do like some of this more'” as helpful for improving their performance. Other crowdworkers mentioned factors like better teamwork as a result of getting to know their team members, being more independent and dividing up work efficiently. The crowd had a better understanding of how to discover content on heavily used social media platforms like Twitter, Facebook, Instagram and Reddit as they remained common across sessions. 

Crowdworkers mentioned improvement in both the quality and quantity of submissions over the sessions. In terms of quantity, they attributed faster searches to practice with the tools and techniques and increasing ease of navigating social media platforms for discovery. CW7 thought “everything got a lot easier” and improved on all major tasks including archival, discovery through tools like Hoaxy \cite{shao_hoaxy_2016} and verification including geolocation. Crowdworkers satisfied the major requirement of not generating duplicate submissions and there were no duplicates after the first session (session 1 had 2 duplicate entries among 38 submissions). CW9 talked about how he got more comfortable with this requirement by checking for duplicates and putting “a different spin on [submission], make sure it was coming from a different angle.” 
 
More importantly, the crowdworkers thought that their quality of submissions improved. CW6 said, “And maybe not so much the quantity, I mean I guess it went up from the beginning, just because things were quicker knowing how to do things but definitely, was able to I felt like I found one post that was like very relevant to the topic every time.” Combined with ongoing training and a better understanding of the requirements, the crowd was able to provide relevant details for the submissions as the sessions went on. They talked about making more refined searches and leveraging previously successful strategies for verification.

However, CW8 found the sessions to be repetitive in terms of the tasks and for him “performance over the different expert sessions didn't really change or improve or anything like that.”

\subsection{\newt{Future Engagement}}

\newt{All experts reported that it was an enjoyable experience for them. E3 felt that “...this is really promising and interesting and I think it was fruitful. We did this for an hour and I think it was fun.” E5 said, “I mean when [course instructor] said, you know, like five minutes up, I was like, oh wow we're already done, so time flies when you're having fun.”}

\newt{Investigators thought the positive experience of working collaboratively could translate into longer term engagement with such a trained crowd in the future. They talked about how such a crowdsourcing framework can be incorporated into their regular work. E1 thought this could be a way to break away from the solo activity of investigating online information and delegate work to the crowd reliably. E1 thought identifying “areas of expertise” of the crowd can be helpful as those could be leveraged while planning a new investigation. E1 mentioned that for him, “there's always something available, that is not super time-sensitive and those would be the ones that would allow for me to have the pre-meeting, and then develop questions for the students, so I think it can be replicated.” E1 also talked about how the crowd could perform peripheral tasks like looking at the background of the person/bot who made the post while he “might be digging into more of the central questions about the fact check.”} 

\newt{Investigators suggested longer and deeper projects as a good fit for the trained crowd, as they were able to perform these rapid and focused investigations. That would allow the crowd to take up investigations with larger scope and hone their skills further.} 
\section{Discussion}

\begin{table*}[]
\caption{Addressing the challenges faced by investigators for crowdsourced OSINT investigations through ORS. \\ (Section numbers are provided in parentheses in Expert After column to refer back to the Findings.)
}
\resizebox{2\columnwidth}{!}{
\begin{tabular}{@{}llll@{}}
\toprule
\textbf{\newt{Design Challenge}} & \textbf{Expert Before} & \textbf{\newt{Design Argument}} & \textbf{Expert After} \\ \midrule
\begin{tabular}[c]{@{}l@{}}Delegation of diverse, \\ complex OSINT tasks \\ in ways that crowds \\ can meaningfully help\end{tabular} &
  \begin{tabular}[c]{@{}l@{}}Experts avoid \\ crowdsourcing complex \\ work altogether or \\ delegate very narrow \\ microtasks.\end{tabular} &
  \begin{tabular}[c]{@{}l@{}}We help experts to delegate complex \\ tasks to trained crowdworkers using \\ the OSINT framework they already know.\end{tabular} &
  \begin{tabular}[c]{@{}l@{}}Experts found the individual tasks to \\ be relevant to their work practices (5.1.1). \\ Experts got valuable information and \\ leads to augment their ongoing \\ investigations from the crowd (5.7.1).\end{tabular} \\ \midrule
\begin{tabular}[c]{@{}l@{}}Safety, privacy,\\ and ethical considerations \\ for investigations \\ performed by crowds\end{tabular} &
  \begin{tabular}[c]{@{}l@{}}Experts can’t delegate \\ work to any online \\ crowd due to privacy \\ and ethical concerns\end{tabular} &
  \begin{tabular}[c]{@{}l@{}}Experts are in control over the information \\ and benefit from the crowd’s training in \\ transparency and privacy. \\ The ethos of the field are imparted to the \\ crowd and upheld during sessions.\end{tabular} &
  \begin{tabular}[c]{@{}l@{}}Experts had better experience  \\ compared to a novice online crowd (5.7.2). \\ They were optimistic about \\future engagement with the setup (5.8).\end{tabular} \\ \midrule
\begin{tabular}[c]{@{}l@{}}Logistical challenges \\of synchronous \\ collaboration\end{tabular} &
  \begin{tabular}[c]{@{}l@{}}Experts don’t work \\ synchronously with \\ crowds to generate \\ useful leads\end{tabular} &
  \begin{tabular}[c]{@{}l@{}}We run multiple sessions to bring the \\ crowd and expert together and provide \\ technical support using Google Forms \\ and sheets. The crowd works in teams \\ and submits relevant information without \\ duplication of effort and are driven by \\ quick feedback.\end{tabular} &
  \begin{tabular}[c]{@{}l@{}}Experts found the information from \\ crowd investigation to be easily \\ accessible (5.6.1). The setup was \\ convenient to provide feedback for the \\submissions (5.4.1).\end{tabular} \\ \midrule
\begin{tabular}[c]{@{}l@{}}Control over\\ the quality and \\ direction of \\investigation\end{tabular} &
  \begin{tabular}[c]{@{}l@{}}Experts have limited \\ communication with \\ the crowd and cannot \\ provide feedback to \\ improve crowd \\ performance\end{tabular} &
  \begin{tabular}[c]{@{}l@{}} Experts provide detailed qualitative \\feedback to crowd  submissions. \\ Experts use interventions to address \\any unfavorable patterns in the \\ submissions.\end{tabular} &
  \begin{tabular}[c]{@{}l@{}}Experts influenced the crowd \\ investigation positively and drilled \\ down on interesting leads (5.4.2, 5.7.4). \\ Expert guidance improved the overall \\ performance of the crowd for \\ the tasks (5.4.3).\end{tabular} \\ \bottomrule
\label{tab:ORS_overview}
\end{tabular}}
\end{table*}

We designed and evaluated OSINT Research Studios (ORS), a sociotechnical framework that enables collaboration between investigators and a trained crowd. Through ORS, we address design challenges in crowdsourced OSINT investigations including the delegation of diverse, complex OSINT tasks; safety, privacy, and ethical considerations; organizational overheads of synchronous collaboration; and maintaining the quality of investigation. \newt{Table~\ref{tab:ORS_overview} summarizes how the challenges faced by investigators while performing crowdsourced OSINT investigations are addressed through ORS.}

The OSINT lab course was a semester-long deployment of ORS, where the first half involved training a group of 30 undergraduate students. During the latter half, this trained crowd collaborated with professionals, including a journalist, a fact-checker, law enforcement investigator, and a human rights analyst. The crowd performed time-boxed and highly targeted investigations based on prompts from the expert, that alluded to one or more information discovery and verification tasks. Experts said that the results from these investigations were useful for solving parts of their broader investigations, find new leads for subsequent investigation and validate some of their own findings, thereby helping scale up and speed up OSINT investigations. \newt{We revisit how the challenges were addressed and identify opportunities for future research.}

\subsection{Delegation of Diverse and Complex OSINT Tasks}

\subsubsection{Effectiveness of macrotasks} 
ORS focused on developing crowd expertise in five macrotasks --- discovery, source analysis, image analysis, fact-checking, and geolocation. \newt{The tasks helped decompose crowdsourced OSINT investigations and gather results from the crowd, providing a structure that has been lacking in previous investigations \cite{cochrane2022citizen, venkatagiri2021crowdsolve}.} Similar to CrowdForge \cite{kittur2011crowdforge}, we found that high-level initial decomposition of complex work by experts is effective in helping crowdworkers complete assigned tasks. \newt{ORS contributes to an expanding body of research that demonstrates how crowds can effectively tackle complex tasks, given they possess adequate motivation, support, training, and autonomy \cite{venkatagiri2023cosint, venkatagiri2021crowdsolve, harris2019joining, retelny2017noworkflow}.} Our exploration of training based on how to approach different tasks without overdependency on tools matched Wang et al. \cite{wang_exploring_2018}’s results of training based on analytical thinking skills. In both cases, crowdworkers developed domain expertise and applied their knowledge to solve complex tasks. All expert tasks leveraged the crowd's capabilities to discover of information from multiple online platforms and identify provenance and location information required for verification. The crowd's results were viewed as potential leads and valued by experts for their speed and quantity. 

\subsubsection{Role of training in ORS}
\newt{In this work, we argue that training can enable crowdworkers to augment investigations carried out by journalists, fact-checkers, human rights investigators, and law enforcement officers.} Doroudi et al. \cite{doroudi_toward_2016} mention challenges like the unavailability and unwillingness of experts for training in a crowdsourcing framework. Experts might sometimes lack understanding about how to ensure the successful completion of tasks, which makes it harder for them to train. In the context of OSINT, these challenges are met by the field's unique characteristics. Belghith et al. \cite{belghith_compete_2022} described OSINT as a community of practice with legitimate peripheral participation \cite{lave_situated_1991}. This involves training novice practitioners through low-risk tasks as they grow into the roles of experts in the community. Experts in the community participate in this model and strike a balance between practice and training endeavors. In terms of skills, OSINT has a wide range of tasks and applications and experts tend to be generalists. Experts learn from each other by sharing how they performed challenging investigations. Training is also available through online certification \cite{SANS,NICCS} and MOOC programs \cite{Udemy} for skills related to investigation of people, online information and websites \cite{SANS}. \newt{Training involving multiple OSINT skills is time-consuming, but just-in-time training and developing shorter modules that can enable novice crowdworkers to contribute to particular investigations can be an effective alternative, as shown here.} With additional available resources and further modification, trained crowds can be employed by experts outside of our classroom study setting.

\subsubsection{Extending ORS to other domains}
\newt{In the current work, investigations involved discovering social media posts and verifying their content. OSINT investigations ranged from fact-checking videos, documenting human rights violations, finding traces of homicide suspects, and investigating the whereabouts of public servants (presented in Table 1).} Given the parallelizable nature of the tasks that we chose, this setup can be scaled up to contribute to rapid response scenarios that are crucial to fight misinformation \cite{wardle2017information} and respond to crisis events \cite{dailey2014crowdsourcerers}. \newt{ORS has the potential for seamless adaptation across domains beyond investigations of online information. For instance, it can be employed to facilitate the coordination of physical search-and-rescue operations for missing individuals or animals \cite{tracelabs_2022_twitter, white2014digital}, as well as to evaluate the extent of damage following both natural and man-made disasters \cite{bittner2016turning}.} The parallelizable and targeted characteristics combined with the inherent adversarial nature of investigations also makes the ORS framework applicable to the domains of cybersecurity \cite{esteban_borges_securitytrails_2019} and finance \cite{rouach_competitive_2001, ryan_hunt_thirty-seven_2012}.

\subsubsection{Extending ORS to other types of crowds}
To scale up the ORS framework further, future work can consider different crowds outside of a classroom setting. Organizations like Bellingcat \cite{bellingcat}, Trace Labs~\cite{cox2018tracelabs}, and the Syrian Archive~\cite{syrian_archive} engage thousands of volunteers in their efforts to collect and verify open source information. \newt{In order to incorporate individuals possessing relevant skills and diverse backgrounds, volunteering efforts should collaborate closely with such established communities of practice \cite{wenger1998cop} in OSINT.} However, there are open challenges involved with engaging workers of different skill levels, motivating them for sustained participation, having the right mechanism for quality control and aggregating results \cite{correia2023designing,venkatagiri2022supporting}. 

\subsection{Safety, Privacy, and Ethical Considerations}
 
Based on experts' responses, our findings show that the crowd could conduct OSINT macrotasks and contribute safely and meaningfully to experts' investigation. We achieved this by operationalizing the OSINT ethos: prioritizing transparency, avoiding subterfuge, and limiting investigations to passive reconnaissance \cite{belghith_compete_2022}. Training was essential for the crowd and experts to implement strategies ensuring ethical investigations. Detailed results and descriptions of the steps involved made the investigations transparent and reproducible. The use of virtual machines, sock puppet accounts, and VPNs helped the crowd mitigate any risks of retaliation and safeguard their systems. We found these measures to be essential for investigators' safety and sufficient for the tasks. The student crowd's accountability ensured private investigations, free from information leaks and vigilante behavior. \strike{The experts could also balanced providing context and disclosing sensitive information to safeguard their investigation while still generating useful results.}

\subsubsection{\newtext{Safeguarding against harms caused by crowd inaccuracy}}

\newtext{One major concern with crowdsourced investigations is the accuracy of the information generated and its implications for results of an investigation. In our work, the investigations are scoped to tasks meant to generate leads for the experts to follow up on. Experts play an important role in providing the right context that enables the crowd to generate useful results. For example, in study session 2, E2 (law enforcement officer) was working on finding digital traces of a homicide suspect. Instead of revealing the suspect's identity and asking the crowd to research them, E2 tasked the crowd with collecting social media images and videos of a particular mountain range within certain dates that contained any people or vehicles. This ensured that the crowd could augment the investigation without risks of misidentification and leaking sensitive information. Similarly, GroundTruth used a diagram-drawing technique to allow investigators to crowdsource image geolocation tasks without sharing the target photo or video, which might contain confidential details or disturbing imagery~\cite{venkatagiri2019groundtruth}. We advocate for pervasive expert oversight, including quality assurance processes where experts review and verify the information gathered by the crowd. Given that crowd submissions may not always meet the requirements set by experts, systems should provide experts the control to make conclusions and findings based on accurate, reliable, and verified information.}

\newtext{Experts must often make make critical decisions, especially within sensitive or high-risk situations. They can leverage their experience and judgment to avoid potential pitfalls and ensure the investigation stays on track. Investigators can conduct a comprehensive risk assessment before initiating the crowdsourced investigation to identify potential risks, vulnerabilities, and threats. Researchers and domain experts should engage in scenario-planning to anticipate and prepare for possible challenges and develop strategies to mitigate them. Collaboration with legal authorities can help to clearly define what is permissible and what is not, ensuring alignment with legal and ethical standards.}

\subsubsection{\newtext{Protecting crowd investigators and students from harm}}

\newtext{Operational Security (OPSEC) in the context of OSINT investigations refers to the process and strategies used to protect sensitive information, ensure personal and organizational security, and maintain the effectiveness and integrity of the investigation. As mentioned in Section~\ref{sec:safety-considerations}, we apply key OPSEC practices in our deployment of ORS to ensure crowdworkers can conduct research without revealing their identity or affiliation. These include using Virtual Private Networks (VPNs), secure communication platforms for messaging and email, and virtual machines. Crowdworkers used sock puppet accounts to limit the exposure of personal and sensitive information on social media platforms.}

\newt{In our work, we were careful about the appropriateness of investigations in terms of the topic and the general nature of information surrounding it to make it more suitable for the classroom setting. To further broaden the scope of investigations, future work should consider potential risks, such as issues related to secondary trauma and exposure to sensitive content and triggers \cite{cochrane2022citizen}. Secondary trauma refers to the emotional stress experienced by individuals as a result of witnessing or being exposed to traumatic events indirectly. To address this, it is crucial to provide psychological support and resources for investigators, including access to counseling services and regular debriefings \cite{williams_defining_2018,steiger2021wellbeing}. Training programs designed to help crowdworkers deal with secondary trauma through self-care, establishing boundaries, seeking peer support, and recognizing warning signs can be beneficial in managing the emotional demands of their work and maintaining their mental well-being \cite{SANS,cochrane2022citizen,williams_defining_2018}.}

\newtext{When recruiting OSINT crowds in a classroom context, as we did, it is especially important to consider the unique needs of students. Experiential learning is valued in higher education because it provides authentic, real-world learning experiences~\cite{kolb_experiential_2017}. However, in investigative or adversarial contexts, it can also expose students to some risk. For example, some universities teach courses where students conduct real-world OSINT investigations in human rights~\cite{HRC_Berkeley} and cybersecurity contexts~\cite{MITCybersecurityClinic}. This work may require students to engage with disturbing or illegal material and investigate bad actors. But it can also be personally meaningful and societally impactful, and prepares students for successful careers as professionals in these fields. By employing the safeguards above, OSINT students can learn investigative work in a relatively safer environment compared to those requiring direct interaction with persons of interest. Nevertheless, it is important for instructors to communicate these risks and trade-offs to students and to help experts provide appropriate levels of exposure during investigations that match students' developing skill levels.}

\subsection{Organizational Overheads of Synchronous Collaboration}

\subsubsection{Session planning}
Experts played an active role in breaking down ongoing investigation(s) and came up with prompts that suited the crowd's expertise. Interestingly, the three major factors for the overall crowd experience - topic, difficulty, and context of investigation (section 5.5) - were heavily influenced by planning. Experts were able to guide the crowd better by doing prior research and providing actionable feedback. Based on crowd feedback, session planning can be improved by specifying how crowd contributions fit in with the larger investigation, thereby setting clear expectations about the type and level of details for submissions. \newt{Future research can look at systems that conduct surveys among the crowd, enabling them to communicate their motivations to the expert \cite{pintrich2004conceptual}. Based on this feedback, the expert can make appropriate modifications to their tasks.}

\newt{The crowd collaborated with each expert for 75 minutes. This setup helped scope out tasks related to discovery and quick verification of social media content. Future research can explore longer running and more in-depth OSINT investigations as seen in investigative journalism and human rights advocacy programs \cite{muller_gatekeeper_2021, HRC_Berkeley}. Such collaboration and training will enable the workers to dive deeper, apply advanced skills, and learn new ones while solving complex OSINT tasks. Based on prior work showing the benefits of competition in crowdsourcing \cite{venkatagiri2023cosint, belghith_compete_2022}, gamifying the collaborative process can make the sessions more productive.}

\newt{The skill level of workers was assumed to be the same as they go through the same training and start with no background in OSINT investigations. There were also 8 teams participating in the sessions throughout. This allowed the process of assigning tasks to crowd teams to be random and manual. To scale up collaboration and accommodate crowdworkers of varying skill levels, AI-mediated crowdsourcing shows promise in automating task assignment and skill assessment \cite{correia2023designing}. Future efforts can focus on efficient task allocation and result aggregation based on expert need \cite{suzuki_atelier:_2016,kittur2011crowdforge}, particularly for large-scale deployments involving participants from crowd marketplaces like Amazon Mechanical Turk.}

 \subsubsection{Teamwork and communication}
The crowd took up tasks in teams of three to four members. Crowdworkers worked individually after dividing up work based on different social media platforms, geographical locations, and OSINT sub-tasks. Communication within teams grew during difficult tasks as they discussed individual findings and feedback from experts to decide on the next steps. Participants felt that there was a lack of interaction between crowd and the expert due to a hybrid setup (with remote experts collaborating synchronously with co-located crowds distributed across two locations), which on the other hand, enabled having experts from across the country. Participants acknowledged that increasing interaction also comes with the cost of slowing down investigations. \newt{ Hybrid intelligence involving crowd-AI interaction \cite{correia2023designing} can be explored to provide relevant context while maintaining the fast and focused nature of investigations.}

\subsection{Maintaining the Quality of Investigations}
The major form of collaboration between the experts and the crowd was through crowd submissions and expert feedback on them. Feedback helped the crowd direct their efforts in the right direction and improve on their performance. Experts suggested, based on their prior experience, that the trained crowd performed better than the working with general public. Experts wanted to engage with the crowd in the future based on other investigative tasks like source analysis, discovery, and geolocation. ORS also addressed our design goal to limit duplication of effort, which has been an issue across crowdsourcing solutions \cite{belghith_compete_2022,venkatagiri2021crowdsolve}. Adding flexibility to the expert-led crowdsourcing framework \cite{venkatagiri2021crowdsolve, venkatagiri2019groundtruth}, a trained crowd quickly responded to feedback and changed course to meet the requirements on a wider range of OSINT tasks. Based on crowd feedback, training strategies like solving more tasks and setting goals (explored through practice sessions), and expert feedback (explored through expert sessions) were found effective as seen in \cite{zhu2014reviewing,dow2012shepherding}. However, self-evaluations did not have an impact on our model as compared to prior work \cite{dow2012shepherding}. Self-evaluation effectiveness can be improved through automated quality checks by flagging unverified sources and missing archival links. To improve the quality of submissions further, peer review, which is effective as an adversarial training strategy \cite{suzuki_atelier:_2016}, can be implemented to elicit feedback from other teams. 

\subsubsection{\newt{Automating expert feedback}}

\newt{Leveraging insights from expert feedback in the current study, future crowdsourced OSINT investigations could automate specific feedback mechanisms using large language models (LLMs) \cite{cao2023leveraging}. Automating specific feedback saves time and effort for both experts and crowdworkers, as the system can provide tailored feedback to crowdworkers \cite{tekian2017qualitative} while allowing experts to focus on nuanced assessments or more complex tasks. While automation benefits efficiency, a balanced approach is needed to consider experts' unique perspective \cite{kim2020toward}. There is a need for combining automated and human feedback to ensure comprehensive evaluation and boost the productivity of OSINT investigations.}

\section{Conclusion}

In our work, we supported the need to scale up and speed up OSINT investigations across multiple domains. We addressed the practical challenges of crowdsourcing OSINT investigations through crowd training and synchronous collaboration. Training was based on the technical and ethical aspects of OSINT and contributed to successful completion of a wide range of tasks. Collaboration was centered around feedback that experts said improved the overall quality of their investigations. Taking a design-based research (DBR) approach, we iteratively designed OSINT Research Studios (ORS), a sociotechnical system that facilitated rapid and focused OSINT investigations. Through the OSINT lab course, we had a semester long deployment of ORS including evaluation sessions with investigators from the domains of journalism, fact-checking, law enforcement and human rights investigation to evaluate the system. Experts found the sessions to be useful, and mentioned strengths like speed, safety, high quality and quantity of submissions across tasks, and the crowd's adaptability to feedback. The crowd enjoyed working with experts and successfully applied their OSINT skills. In conclusion, ORS enabled ethical and effective crowdsourced OSINT investigations.

\bibliographystyle{ACM-Reference-Format}
\bibliography{OSINTResearchStudios_arXiv/main}


\begin{thebibliography}{139}


\ifx \showCODEN    \undefined \def \showCODEN     #1{\unskip}     \fi
\ifx \showDOI      \undefined \def \showDOI       #1{#1}\fi
\ifx \showISBNx    \undefined \def \showISBNx     #1{\unskip}     \fi
\ifx \showISBNxiii \undefined \def \showISBNxiii  #1{\unskip}     \fi
\ifx \showISSN     \undefined \def \showISSN      #1{\unskip}     \fi
\ifx \showLCCN     \undefined \def \showLCCN      #1{\unskip}     \fi
\ifx \shownote     \undefined \def \shownote      #1{#1}          \fi
\ifx \showarticletitle \undefined \def \showarticletitle #1{#1}   \fi
\ifx \showURL      \undefined \def \showURL       {\relax}        \fi
\providecommand\bibfield[2]{#2}
\providecommand\bibinfo[2]{#2}
\providecommand\natexlab[1]{#1}
\providecommand\showeprint[2][]{arXiv:#2}

\bibitem[bel(2015)]%
        {bellingcat2015calltoarms}
 \bibinfo{year}{2015}\natexlab{}.
\newblock \bibinfo{title}{A {Call} to {Arms}: {Open} {Source} {Intelligence} and {Evidence} {Based} {Policymaking}}.
\newblock
\newblock
\urldef\tempurl%
\url{https://www.bellingcat.com/resources/articles/2015/01/20/a-call-to-arms-open-source-intelligence-and-evidence-based-policymaking/}
\showURL{%
\tempurl}


\bibitem[HRC(2021)]%
        {HRC_Berkeley}
 \bibinfo{year}{2021}\natexlab{}.
\newblock \bibinfo{booktitle}{\emph{J298 OSINT Seminar --- Open Source Investigations}}.
\newblock
\urldef\tempurl%
\url{https://journalism.berkeley.edu/course-section/j298-human-rights-center-seminar-f21/}
\showURL{%
\tempurl}


\bibitem[noa(2021)]%
        {noauthor_whatsapp_2021}
 \bibinfo{year}{2021}\natexlab{}.
\newblock \bibinfo{title}{{WhatsApp} can be a black box of misinformation, but {Maldita} may have opened a window}.
\newblock
\newblock
\urldef\tempurl%
\url{https://www.poynter.org/fact-checking/2021/whatsapp-can-be-a-black-box-of-misinformation-but-maldita-may-have-opened-a-window/}
\showURL{%
\tempurl}


\bibitem[hun(2022)]%
        {hunchly}
 \bibinfo{year}{2022}\natexlab{}.
\newblock \bibinfo{booktitle}{\emph{Hunchly - OSINT Software for Cybersecurity, Law Enforcement, Journalists, Private Investigators, and more}}.
\newblock
\urldef\tempurl%
\url{https://www.hunch.ly/}
\showURL{%
\tempurl}


\bibitem[noa(2022)]%
        {noauthor_russia_2022}
 \bibinfo{year}{2022}\natexlab{}.
\newblock \bibinfo{title}{Russia is losing so much equipment in {Ukraine} that weapons monitors can’t keep up}.
\newblock
\newblock
\urldef\tempurl%
\url{https://www.independent.co.uk/news/world/europe/russia-ukraine-military-equipment-losses-b2049613.html}
\showURL{%
\tempurl}
\newblock
\shownote{Section: News}.


\bibitem[tra(2022)]%
        {tracelabs_2022_twitter}
 \bibinfo{year}{2022}\natexlab{}.
\newblock \bibinfo{title}{{TraceLabs} {Twitter} {Account}}.
\newblock
\newblock
\urldef\tempurl%
\url{https://twitter.com/tracelabs/status/1558831625986777088}
\showURL{%
\tempurl}


\bibitem[bel(2023)]%
        {bellingcat}
 \bibinfo{year}{2023}\natexlab{}.
\newblock \bibinfo{booktitle}{\emph{About Bellingcat}}.
\newblock
\urldef\tempurl%
\url{https://www.bellingcat.com/about/}
\showURL{%
\tempurl}


\bibitem[arc(2023)]%
        {archive_is}
 \bibinfo{year}{2023}\natexlab{}.
\newblock \bibinfo{title}{archive.is}.
\newblock
\newblock
\urldef\tempurl%
\url{https://archive.is/}
\showURL{%
\tempurl}


\bibitem[bot(2023)]%
        {bot_sentinel}
 \bibinfo{year}{2023}\natexlab{}.
\newblock \bibinfo{booktitle}{\emph{Bot {Sentinel} - {Dashboard}}}.
\newblock
\urldef\tempurl%
\url{https://botsentinel.com/}
\showURL{%
\tempurl}


\bibitem[NIC(2023)]%
        {NICCS}
 \bibinfo{year}{2023}\natexlab{}.
\newblock \bibinfo{booktitle}{\emph{Certified in {Open} {Source} {Intelligence} ({C}{\textbar}{OSINT}) from {McAfee} {Institute} {\textbar} {NICCS}}}.
\newblock
\urldef\tempurl%
\url{https://niccs.cisa.gov/education-training/catalog/mcafee-institute/certified-open-source-intelligence-cosint}
\showURL{%
\tempurl}


\bibitem[che(2023)]%
        {checkdesk}
 \bibinfo{year}{2023}\natexlab{}.
\newblock \bibinfo{title}{Check (@checkdesk) / Twitter}.
\newblock
\newblock
\urldef\tempurl%
\url{https://twitter.com/checkdesk}
\showURL{%
\tempurl}


\bibitem[exi(2023)]%
        {exif}
 \bibinfo{year}{2023}\natexlab{}.
\newblock \bibinfo{booktitle}{\emph{{EXIF} {Data} {Viewer}}}.
\newblock
\urldef\tempurl%
\url{https://exifdata.com/}
\showURL{%
\tempurl}


\bibitem[goo(2023)]%
        {google_image}
 \bibinfo{year}{2023}\natexlab{}.
\newblock \bibinfo{title}{Google {Images}}.
\newblock
\newblock
\urldef\tempurl%
\url{https://images.google.com/}
\showURL{%
\tempurl}


\bibitem[new(2023)]%
        {news}
 \bibinfo{year}{2023}\natexlab{}.
\newblock \bibinfo{title}{Join us in pushing back on misinformation.}
\newblock
\newblock
\urldef\tempurl%
\url{https://our.news/}
\showURL{%
\tempurl}


\bibitem[Fir(2023a)]%
        {FirstDraft_discovery}
 \bibinfo{year}{2023}\natexlab{a}.
\newblock \bibinfo{title}{Newsgathering and {Monitoring} on the {Social} {Web}}.
\newblock
\newblock
\urldef\tempurl%
\url{https://firstdraftnews.org:443/long-form-article/newsgathering-and-monitoring-on-the-social-web/}
\showURL{%
\tempurl}


\bibitem[Ude(2023)]%
        {Udemy}
 \bibinfo{year}{2023}\natexlab{}.
\newblock \bibinfo{booktitle}{\emph{{OSINT}: {Open}-{Source} {Intelligence}}}.
\newblock
\urldef\tempurl%
\url{https://www.udemy.com/course/osint-open-source-intelligence/}
\showURL{%
\tempurl}


\bibitem[SAN(2023)]%
        {SANS}
 \bibinfo{year}{2023}\natexlab{}.
\newblock \bibinfo{booktitle}{\emph{Practical {Open}-{Source} {Intelligence} ({OSINT}) {Training} --- {SANS} {SEC497}}}.
\newblock
\urldef\tempurl%
\url{https://www.sans.org/cyber-security-courses/practical-open-source-intelligence/}
\showURL{%
\tempurl}


\bibitem[Tra(2023)]%
        {TraceLabs_search}
 \bibinfo{year}{2023}\natexlab{}.
\newblock \bibinfo{title}{Search {Party} {Rules}}.
\newblock
\newblock
\urldef\tempurl%
\url{https://www.tracelabs.org/about/search-party-rules}
\showURL{%
\tempurl}


\bibitem[syr(2023)]%
        {syrian_archive}
 \bibinfo{year}{2023}\natexlab{}.
\newblock \bibinfo{title}{Syrian {Archive} {\textbar} {Syrian} {Archive}}.
\newblock
\newblock
\urldef\tempurl%
\url{https://syrianarchive.org/}
\showURL{%
\tempurl}


\bibitem[tin(2023)]%
        {tineye}
 \bibinfo{year}{2023}\natexlab{}.
\newblock \bibinfo{booktitle}{\emph{{TinEye} {Reverse} {Image} {Search}}}.
\newblock
\urldef\tempurl%
\url{https://tineye.com/}
\showURL{%
\tempurl}


\bibitem[Fir(2023b)]%
        {FirstDraft_verification}
 \bibinfo{year}{2023}\natexlab{b}.
\newblock \bibinfo{booktitle}{\emph{Verifying {Online} {Information}}}.
\newblock
\urldef\tempurl%
\url{https://firstdraftnews.org:443/long-form-article/verifying-online-information/}
\showURL{%
\tempurl}


\bibitem[yan(2023)]%
        {yandex}
 \bibinfo{year}{2023}\natexlab{}.
\newblock \bibinfo{booktitle}{\emph{Yandex}}.
\newblock
\urldef\tempurl%
\url{https://yandex.com/}
\showURL{%
\tempurl}


\bibitem[Abdullah et~al\mbox{.}(2021)]%
        {abdullah_osint_2021}
\bibfield{author}{\bibinfo{person}{Alwan Abdullah}, \bibinfo{person}{Shams~A. Laghari}, \bibinfo{person}{Ashish Jaisan}, {and} \bibinfo{person}{Shankar Karuppayah}.} \bibinfo{year}{2021}\natexlab{}.
\newblock \showarticletitle{{OSINT} {Explorer}: {A} {Tool} {Recommender} {Framework} for {OSINT} {Sources}}. In \bibinfo{booktitle}{\emph{Advances in {Cyber} {Security}}} \emph{(\bibinfo{series}{Communications in {Computer} and {Information} {Science}})}, \bibfield{editor}{\bibinfo{person}{Nibras Abdullah}, \bibinfo{person}{Selvakumar Manickam}, {and} \bibinfo{person}{Mohammed Anbar}} (Eds.). \bibinfo{publisher}{Springer}, \bibinfo{address}{Singapore}, \bibinfo{pages}{389--400}.
\newblock
\showISBNx{9789811680595}
\urldef\tempurl%
\url{https://doi.org/10.1007/978-981-16-8059-5_24}
\showDOI{\tempurl}


\bibitem[Agapie et~al\mbox{.}(2015)]%
        {agapie2015crowdsourcing}
\bibfield{author}{\bibinfo{person}{Elena Agapie}, \bibinfo{person}{Jaime Teevan}, {and} \bibinfo{person}{Andr{\'e}s Monroy-Hern{\'a}ndez}.} \bibinfo{year}{2015}\natexlab{}.
\newblock \showarticletitle{Crowdsourcing in the field: A case study using local crowds for event reporting}. In \bibinfo{booktitle}{\emph{Third AAAI Conference on Human Computation and Crowdsourcing}}. \bibinfo{publisher}{Association for the Advancement of Artificial Intelligence}, \bibinfo{pages}{11}.
\newblock
\urldef\tempurl%
\url{https://www.microsoft.com/en-us/research/publication/crowdsourcing-in-the-field-a-case-study-using-local-crowds-for-event-reporting/}
\showURL{%
\tempurl}


\bibitem[Alcaidinho et~al\mbox{.}(2017)]%
        {alcaidinho_mobile_2017}
\bibfield{author}{\bibinfo{person}{Joelle Alcaidinho}, \bibinfo{person}{Larry Freil}, \bibinfo{person}{Taylor Kelly}, \bibinfo{person}{Kayla Marland}, \bibinfo{person}{Chunhui Wu}, \bibinfo{person}{Bradley Wittenbrook}, \bibinfo{person}{Giancarlo Valentin}, {and} \bibinfo{person}{Melody Jackson}.} \bibinfo{year}{2017}\natexlab{}.
\newblock \showarticletitle{Mobile {Collaboration} for {Human} and {Canine} {Police} {Explosive} {Detection} {Teams}}. In \bibinfo{booktitle}{\emph{Proceedings of the 2017 {ACM} {Conference} on {Computer} {Supported} {Cooperative} {Work} and {Social} {Computing}}} \emph{(\bibinfo{series}{{CSCW} '17})}. \bibinfo{publisher}{Association for Computing Machinery}, \bibinfo{address}{New York, NY, USA}, \bibinfo{pages}{925--933}.
\newblock
\showISBNx{978-1-4503-4335-0}
\urldef\tempurl%
\url{https://doi.org/10.1145/2998181.2998271}
\showDOI{\tempurl}


\bibitem[Alharthi et~al\mbox{.}(2021)]%
        {alharthi_2021}
\bibfield{author}{\bibinfo{person}{Sultan~A. Alharthi}, \bibinfo{person}{Nicolas~James LaLone}, \bibinfo{person}{Hitesh~Nidhi Sharma}, \bibinfo{person}{Igor Dolgov}, {and} \bibinfo{person}{Z~O. Toups}.} \bibinfo{year}{2021}\natexlab{}.
\newblock \showarticletitle{An Activity Theory Analysis of Search and; Rescue Collective Sensemaking and Planning Practices}. In \bibinfo{booktitle}{\emph{Proceedings of the 2021 CHI Conference on Human Factors in Computing Systems}} (Yokohama, Japan) \emph{(\bibinfo{series}{CHI '21})}. \bibinfo{publisher}{Association for Computing Machinery}, \bibinfo{address}{New York, NY, USA}, Article \bibinfo{articleno}{146}, \bibinfo{numpages}{20}~pages.
\newblock
\showISBNx{9781450380966}
\urldef\tempurl%
\url{https://doi.org/10.1145/3411764.3445272}
\showDOI{\tempurl}


\bibitem[Aliprandi et~al\mbox{.}(2014)]%
        {aliprandi_caper_nodate}
\bibfield{author}{\bibinfo{person}{Carlo Aliprandi}, \bibinfo{person}{Juan~Arraiza Irujo}, \bibinfo{person}{Montse Cuadros}, \bibinfo{person}{Sebastian Maier}, \bibinfo{person}{Felipe Melero}, {and} \bibinfo{person}{Matteo Raffaelli}.} \bibinfo{year}{2014}\natexlab{}.
\newblock \showarticletitle{{CAPER}: {Collaborative} {Information}, {Acquisition}, {Processing}, {Exploitation} and {Reporting} for the {Prevention} of {Organised} {Crime}}.
\newblock \bibinfo{journal}{\emph{Communications in Computer and Information Science}} (\bibinfo{year}{2014}), \bibinfo{pages}{6}.
\newblock


\bibitem[Allen et~al\mbox{.}(2021)]%
        {allen2021scaling}
\bibfield{author}{\bibinfo{person}{Jennifer Allen}, \bibinfo{person}{Antonio~A Arechar}, \bibinfo{person}{Gordon Pennycook}, {and} \bibinfo{person}{David~G Rand}.} \bibinfo{year}{2021}\natexlab{}.
\newblock \showarticletitle{Scaling up fact-checking using the wisdom of crowds}.
\newblock \bibinfo{journal}{\emph{Science advances}} \bibinfo{volume}{7}, \bibinfo{number}{36} (\bibinfo{year}{2021}), \bibinfo{pages}{eabf4393}.
\newblock


\bibitem[Allen et~al\mbox{.}(2022)]%
        {allen2022birds}
\bibfield{author}{\bibinfo{person}{Jennifer Allen}, \bibinfo{person}{Cameron Martel}, {and} \bibinfo{person}{David~G Rand}.} \bibinfo{year}{2022}\natexlab{}.
\newblock \showarticletitle{Birds of a feather don’t fact-check each other: Partisanship and the evaluation of news in Twitter’s Birdwatch crowdsourced fact-checking program}. In \bibinfo{booktitle}{\emph{CHI Conference on Human Factors in Computing Systems}}. \bibinfo{publisher}{PsyArXiv}, \bibinfo{pages}{1--19}.
\newblock
\urldef\tempurl%
\url{https://doi.org/10.31234/osf.io/57e3q}
\showDOI{\tempurl}


\bibitem[Alvarado~Garcia et~al\mbox{.}(2021)]%
        {garcia2021data}
\bibfield{author}{\bibinfo{person}{Adriana Alvarado~Garcia}, \bibinfo{person}{Matthew~J. Britton}, \bibinfo{person}{Dhairya~Manish Doshi}, \bibinfo{person}{Munmun De~Choudhury}, {and} \bibinfo{person}{Christopher~A. Le~Dantec}.} \bibinfo{year}{2021}\natexlab{}.
\newblock \showarticletitle{Data Migrations: Exploring the Use of Social Media Data as Evidence for Human Rights Advocacy}.
\newblock \bibinfo{journal}{\emph{Proc. ACM Hum.-Comput. Interact.}} \bibinfo{volume}{4}, \bibinfo{number}{CSCW3}, Article \bibinfo{articleno}{268} (\bibinfo{date}{jan} \bibinfo{year}{2021}), \bibinfo{numpages}{25}~pages.
\newblock
\urldef\tempurl%
\url{https://doi.org/10.1145/3434177}
\showDOI{\tempurl}


\bibitem[Alvarado~Garcia and Le~Dantec(2018)]%
        {garcia2018quotidian}
\bibfield{author}{\bibinfo{person}{Adriana Alvarado~Garcia} {and} \bibinfo{person}{Christopher~A. Le~Dantec}.} \bibinfo{year}{2018}\natexlab{}.
\newblock \showarticletitle{Quotidian Report: Grassroots Data Practices to Address Public Safety}.
\newblock \bibinfo{journal}{\emph{Proc. ACM Hum.-Comput. Interact.}} \bibinfo{volume}{2}, \bibinfo{number}{CSCW}, Article \bibinfo{articleno}{17} (\bibinfo{date}{nov} \bibinfo{year}{2018}), \bibinfo{numpages}{18}~pages.
\newblock
\urldef\tempurl%
\url{https://doi.org/10.1145/3274286}
\showDOI{\tempurl}


\bibitem[{Amnesty International}(2020)]%
        {amnesty_international_syria_2020}
\bibfield{author}{\bibinfo{person}{{Amnesty International}}.} \bibinfo{year}{2020}\natexlab{}.
\newblock \bibinfo{booktitle}{\emph{Syria: '{Nowhere} is {Safe} for {Us}': {Unlawful} {Attacks} and {Mass} {Displacement} in {North}-{West} {Syria}}}.
\newblock \bibinfo{type}{{T}echnical {R}eport}.
\newblock
\urldef\tempurl%
\url{https://www.amnesty.org/en/documents/document/?indexNumber=MDE24%2f2089%2f2020&language=en}
\showURL{%
\tempurl}


\bibitem[Arif et~al\mbox{.}(2017)]%
        {arif_closer_2017}
\bibfield{author}{\bibinfo{person}{Ahmer Arif}, \bibinfo{person}{John~J. Robinson}, \bibinfo{person}{Stephanie~A. Stanek}, \bibinfo{person}{Elodie~S. Fichet}, \bibinfo{person}{Paul Townsend}, \bibinfo{person}{Zena Worku}, {and} \bibinfo{person}{Kate Starbird}.} \bibinfo{year}{2017}\natexlab{}.
\newblock \showarticletitle{A {Closer} {Look} at the {Self}-{Correcting} {Crowd}: {Examining} {Corrections} in {Online} {Rumors}}. In \bibinfo{booktitle}{\emph{Proceedings of the 2017 {ACM} {Conference} on {Computer} {Supported} {Cooperative} {Work} and {Social} {Computing}}} \emph{(\bibinfo{series}{{CSCW} '17})}. \bibinfo{publisher}{ACM}, \bibinfo{address}{Portland Oregon USA}, \bibinfo{pages}{155--168}.
\newblock
\showISBNx{978-1-4503-4335-0}
\urldef\tempurl%
\url{https://doi.org/10.1145/2998181.2998294}
\showDOI{\tempurl}


\bibitem[Aula and Russell(2008)]%
        {aula2008complex}
\bibfield{author}{\bibinfo{person}{Anne Aula} {and} \bibinfo{person}{Daniel~M Russell}.} \bibinfo{year}{2008}\natexlab{}.
\newblock \showarticletitle{Complex and exploratory web search}. In \bibinfo{booktitle}{\emph{Information Seeking Support Systems Workshop (ISSS 2008), Chapel Hill, NC, USA}}. Citeseer.
\newblock


\bibitem[Beckett(2017)]%
        {beckett2017wikitribune}
\bibfield{author}{\bibinfo{person}{Charlie Beckett}.} \bibinfo{year}{2017}\natexlab{}.
\newblock \showarticletitle{Wikitribune: can crowd-sourced journalism solve the crisis of trust in news?}
\newblock \bibinfo{journal}{\emph{POLIS: journalism and society at the LSE}} (\bibinfo{year}{2017}).
\newblock


\bibitem[Belghith et~al\mbox{.}(2022)]%
        {belghith_compete_2022}
\bibfield{author}{\bibinfo{person}{Yasmine Belghith}, \bibinfo{person}{Sukrit Venkatagiri}, {and} \bibinfo{person}{Kurt Luther}.} \bibinfo{year}{2022}\natexlab{}.
\newblock \showarticletitle{Compete, {Collaborate}, {Investigate}: {Exploring} the {Social} {Structures} of {Open} {Source} {Intelligence} {Investigations}}. In \bibinfo{booktitle}{\emph{Proceedings of the 2022 {CHI} {Conference} on {Human} {Factors} in {Computing} {Systems}}} \emph{(\bibinfo{series}{{CHI} '22})}. \bibinfo{publisher}{Association for Computing Machinery}, \bibinfo{address}{New York, NY, USA}, \bibinfo{pages}{1--18}.
\newblock
\showISBNx{978-1-4503-9157-3}
\urldef\tempurl%
\url{https://doi.org/10.1145/3491102.3517526}
\showDOI{\tempurl}


\bibitem[{Bellingcat Investigation Team}(2015)]%
        {bellingcat_investigation_team_diversifying_2015}
\bibfield{author}{\bibinfo{person}{{Bellingcat Investigation Team}}.} \bibinfo{year}{2015}\natexlab{}.
\newblock \bibinfo{title}{Diversifying {OSINT}: {Women} {Experts}}.
\newblock
\newblock
\urldef\tempurl%
\url{https://www.bellingcat.com/resources/articles/2015/12/08/women-in-osint-diversifying-the-field/}
\showURL{%
\tempurl}


\bibitem[Bernstein et~al\mbox{.}(2010)]%
        {bernstein2010soylent}
\bibfield{author}{\bibinfo{person}{Michael~S Bernstein}, \bibinfo{person}{Greg Little}, \bibinfo{person}{Robert~C Miller}, \bibinfo{person}{Bj{\"o}rn Hartmann}, \bibinfo{person}{Mark~S Ackerman}, \bibinfo{person}{David~R Karger}, \bibinfo{person}{David Crowell}, {and} \bibinfo{person}{Katrina Panovich}.} \bibinfo{year}{2010}\natexlab{}.
\newblock \showarticletitle{Soylent: a word processor with a crowd inside}. In \bibinfo{booktitle}{\emph{Proceedings of the 23nd annual ACM symposium on User interface software and technology}}. ACM, \bibinfo{pages}{313--322}.
\newblock


\bibitem[Bittner et~al\mbox{.}(2016)]%
        {bittner2016turning}
\bibfield{author}{\bibinfo{person}{Christian Bittner}, \bibinfo{person}{Boris Michel}, {and} \bibinfo{person}{Cate Turk}.} \bibinfo{year}{2016}\natexlab{}.
\newblock \showarticletitle{Turning the spotlight on the crowd: Examining the participatory ethics and practices of crisis mapping}.
\newblock \bibinfo{journal}{\emph{ACME: An International Journal for Critical Geographies}} \bibinfo{volume}{15}, \bibinfo{number}{1} (\bibinfo{year}{2016}), \bibinfo{pages}{207--229}.
\newblock


\bibitem[Brands et~al\mbox{.}(2018)]%
        {brands2018social}
\bibfield{author}{\bibinfo{person}{Bert~Jan Brands}, \bibinfo{person}{Todd Graham}, {and} \bibinfo{person}{Marcel Broersma}.} \bibinfo{year}{2018}\natexlab{}.
\newblock \showarticletitle{Social media sourcing practices: How Dutch newspapers use tweets in political news coverage}.
\newblock \bibinfo{journal}{\emph{Managing democracy in the digital age: Internet regulation, social media use, and online civic engagement}} (\bibinfo{year}{2018}), \bibinfo{pages}{159--178}.
\newblock


\bibitem[Braun and Clarke(2006)]%
        {braun2006thematic}
\bibfield{author}{\bibinfo{person}{Virginia Braun} {and} \bibinfo{person}{Victoria Clarke}.} \bibinfo{year}{2006}\natexlab{}.
\newblock \showarticletitle{Using thematic analysis in psychology}.
\newblock \bibinfo{journal}{\emph{Qualitative research in psychology}} \bibinfo{volume}{3}, \bibinfo{number}{2} (\bibinfo{date}{Jan.} \bibinfo{year}{2006}), \bibinfo{pages}{77--101}.
\newblock
\showISSN{1478-0887}
\urldef\tempurl%
\url{https://doi.org/10.1191/1478088706qp063oa}
\showDOI{\tempurl}
\newblock
\shownote{Publisher: Routledge \_eprint: https://www.tandfonline.com/doi/pdf/10.1191/1478088706qp063oa}.


\bibitem[Broughton et~al\mbox{.}(2014)]%
        {y13}
\bibfield{author}{\bibinfo{person}{Andrea Broughton}, \bibinfo{person}{Beth Foley}, \bibinfo{person}{Stefanie Ledermaier}, {and} \bibinfo{person}{Annette Cox}.} \bibinfo{year}{2014}\natexlab{}.
\newblock \showarticletitle{The use of social media in the recruitment process. (2014)}.
\newblock   \bibinfo{volume}{81} (\bibinfo{year}{2014}).
\newblock


\bibitem[Cao(2023)]%
        {cao2023leveraging}
\bibfield{author}{\bibinfo{person}{Chen Cao}.} \bibinfo{year}{2023}\natexlab{}.
\newblock \showarticletitle{Leveraging Large Language Model and Story-Based Gamification in Intelligent Tutoring System to Scaffold Introductory Programming Courses: A Design-Based Research Study}.
\newblock \bibinfo{journal}{\emph{arXiv preprint arXiv:2302.12834}} (\bibinfo{year}{2023}).
\newblock


\bibitem[Carneiro et~al\mbox{.}(2019)]%
        {carneiro2019deb8}
\bibfield{author}{\bibinfo{person}{Guilherme Carneiro}, \bibinfo{person}{Miguel Nacenta}, \bibinfo{person}{Alice Toniolo}, \bibinfo{person}{Gonzalo Mendez}, {and} \bibinfo{person}{Aaron~J Quigley}.} \bibinfo{year}{2019}\natexlab{}.
\newblock \showarticletitle{Deb8: A tool for collaborative analysis of video}. In \bibinfo{booktitle}{\emph{Proceedings of the 2019 ACM International Conference on Interactive Experiences for TV and Online Video}}. \bibinfo{pages}{47--58}.
\newblock


\bibitem[Center(2021)]%
        {splc_2021_jan6}
\bibfield{author}{\bibinfo{person}{Southern Poverty~Law Center}.} \bibinfo{year}{2021}\natexlab{}.
\newblock \bibinfo{title}{The {Road} to {Jan}. 6: {A} {Year} of {Extremist} {Mobilization}}.
\newblock
\newblock
\urldef\tempurl%
\url{https://www.splcenter.org/news/2021/12/30/road-jan-6-year-extremist-mobilization}
\showURL{%
\tempurl}


\bibitem[Chan et~al\mbox{.}(2018)]%
        {chan_solvent:_2018}
\bibfield{author}{\bibinfo{person}{Joel Chan}, \bibinfo{person}{Joseph~Chee Chang}, \bibinfo{person}{Tom Hope}, \bibinfo{person}{Dafna Shahaf}, {and} \bibinfo{person}{Aniket Kittur}.} \bibinfo{year}{2018}\natexlab{}.
\newblock \showarticletitle{{SOLVENT}: A Mixed Initiative System for Finding Analogies Between Research Papers}.
\newblock   \bibinfo{volume}{2} (\bibinfo{year}{2018}), \bibinfo{pages}{31:1--31:21}.
\newblock
Issue {CSCW}.
\showISSN{2573-0142}
\urldef\tempurl%
\url{https://doi.org/10.1145/3274300}
\showDOI{\tempurl}


\bibitem[Cobb et~al\mbox{.}(2003)]%
        {cobb2003design}
\bibfield{author}{\bibinfo{person}{Paul Cobb}, \bibinfo{person}{Jere Confrey}, \bibinfo{person}{Andrea DiSessa}, \bibinfo{person}{Richard Lehrer}, {and} \bibinfo{person}{Leona Schauble}.} \bibinfo{year}{2003}\natexlab{}.
\newblock \showarticletitle{Design experiments in educational research}.
\newblock \bibinfo{journal}{\emph{Educational researcher}} \bibinfo{volume}{32}, \bibinfo{number}{1} (\bibinfo{year}{2003}), \bibinfo{pages}{9--13}.
\newblock


\bibitem[Cochrane(2022)]%
        {cochrane2022citizen}
\bibfield{author}{\bibinfo{person}{Josie Cochrane}.} \bibinfo{year}{2022}\natexlab{}.
\newblock \bibinfo{title}{Citizen OSINT Analysts: Motivations of Open-Source Intelligence Volunteers}.
\newblock
\newblock


\bibitem[Correia et~al\mbox{.}(2023)]%
        {correia2023designing}
\bibfield{author}{\bibinfo{person}{Ant{\'o}nio Correia}, \bibinfo{person}{Andrea Grover}, \bibinfo{person}{Daniel Schneider}, \bibinfo{person}{Ana~Paula Pimentel}, \bibinfo{person}{Ramon Chaves}, \bibinfo{person}{Marcos~Antonio De~Almeida}, {and} \bibinfo{person}{Benjamim Fonseca}.} \bibinfo{year}{2023}\natexlab{}.
\newblock \showarticletitle{Designing for Hybrid Intelligence: A Taxonomy and Survey of Crowd-Machine Interaction}.
\newblock \bibinfo{journal}{\emph{Applied Sciences}} \bibinfo{volume}{13}, \bibinfo{number}{4} (\bibinfo{year}{2023}), \bibinfo{pages}{2198}.
\newblock


\bibitem[Cox(2018)]%
        {cox2018tracelabs}
\bibfield{author}{\bibinfo{person}{Joseph Cox}.} \bibinfo{year}{2018}\natexlab{}.
\newblock \showarticletitle{The Hackers Hunting Down Missing People: Nonprofit TraceLabs ran DEF CON’s first crowdsourced event for tracking missing people through public information}.
\newblock \bibinfo{journal}{\emph{Vice}} (\bibinfo{year}{2018}).
\newblock
\urldef\tempurl%
\url{https://www.vice.com/en_us/article/qvmm3x/hackers-hunting-missing-people-osint-defcon-tracelabs}
\showURL{%
\tempurl}


\bibitem[Dailey and Starbird(2014)]%
        {dailey2014crowdsourcerers}
\bibfield{author}{\bibinfo{person}{Dharma Dailey} {and} \bibinfo{person}{Kate Starbird}.} \bibinfo{year}{2014}\natexlab{}.
\newblock \showarticletitle{Journalists as Crowdsourcerers: Responding to Crisis by Reporting with a Crowd}.
\newblock \bibinfo{journal}{\emph{Computer Supported Cooperative Work (CSCW)}} \bibinfo{volume}{23}, \bibinfo{number}{4} (\bibinfo{date}{01 Dec} \bibinfo{year}{2014}), \bibinfo{pages}{445--481}.
\newblock
\showISSN{1573-7551}
\urldef\tempurl%
\url{https://doi.org/10.1007/s10606-014-9208-z}
\showDOI{\tempurl}


\bibitem[Dailey and Starbird(2015)]%
        {dailey_dispersants_2015}
\bibfield{author}{\bibinfo{person}{Dharma Dailey} {and} \bibinfo{person}{Kate Starbird}.} \bibinfo{year}{2015}\natexlab{}.
\newblock \showarticletitle{"It's Raining Dispersants": Collective Sensemaking of Complex Information in Crisis Contexts}. In \bibinfo{booktitle}{\emph{Proceedings of the 18th ACM Conference Companion on Computer Supported Cooperative Work and; Social Computing}} (Vancouver, BC, Canada) \emph{(\bibinfo{series}{CSCW'15 Companion})}. \bibinfo{publisher}{Association for Computing Machinery}, \bibinfo{address}{New York, NY, USA}, \bibinfo{pages}{155–158}.
\newblock
\showISBNx{9781450329460}
\urldef\tempurl%
\url{https://doi.org/10.1145/2685553.2698995}
\showDOI{\tempurl}


\bibitem[Diakopoulos et~al\mbox{.}(2021)]%
        {diakopoulos_towards_2021}
\bibfield{author}{\bibinfo{person}{Nicholas Diakopoulos}, \bibinfo{person}{Daniel Trielli}, {and} \bibinfo{person}{Grace Lee}.} \bibinfo{year}{2021}\natexlab{}.
\newblock \showarticletitle{Towards {Understanding} and {Supporting} {Journalistic} {Practices} {Using} {Semi}-{Automated} {News} {Discovery} {Tools}}.
\newblock \bibinfo{journal}{\emph{Proceedings of the ACM on Human-Computer Interaction}} \bibinfo{volume}{5}, \bibinfo{number}{CSCW2} (\bibinfo{date}{Oct.} \bibinfo{year}{2021}), \bibinfo{pages}{406:1--406:30}.
\newblock
\urldef\tempurl%
\url{https://doi.org/10.1145/3479550}
\showDOI{\tempurl}


\bibitem[Doroudi et~al\mbox{.}(2016)]%
        {doroudi_toward_2016}
\bibfield{author}{\bibinfo{person}{Shayan Doroudi}, \bibinfo{person}{Ece Kamar}, \bibinfo{person}{Emma Brunskill}, {and} \bibinfo{person}{Eric Horvitz}.} \bibinfo{year}{2016}\natexlab{}.
\newblock \showarticletitle{Toward a {Learning} {Science} for {Complex} {Crowdsourcing} {Tasks}}. In \bibinfo{booktitle}{\emph{Proceedings of the 2016 {CHI} {Conference} on {Human} {Factors} in {Computing} {Systems}}} \emph{(\bibinfo{series}{{CHI} '16})}. \bibinfo{publisher}{Association for Computing Machinery}, \bibinfo{address}{New York, NY, USA}, \bibinfo{pages}{2623--2634}.
\newblock
\showISBNx{978-1-4503-3362-7}
\urldef\tempurl%
\url{https://doi.org/10.1145/2858036.2858268}
\showDOI{\tempurl}


\bibitem[Dow et~al\mbox{.}(2012)]%
        {dow2012shepherding}
\bibfield{author}{\bibinfo{person}{Steven Dow}, \bibinfo{person}{Anand Kulkarni}, \bibinfo{person}{Scott Klemmer}, {and} \bibinfo{person}{Bj\"{o}rn Hartmann}.} \bibinfo{year}{2012}\natexlab{}.
\newblock \showarticletitle{Shepherding the Crowd Yields Better Work}. In \bibinfo{booktitle}{\emph{Proceedings of the ACM 2012 Conference on Computer Supported Cooperative Work}} (Seattle, Washington, USA) \emph{(\bibinfo{series}{CSCW ’12})}. \bibinfo{publisher}{Association for Computing Machinery}, \bibinfo{address}{New York, NY, USA}, \bibinfo{pages}{1013–1022}.
\newblock
\showISBNx{9781450310864}
\urldef\tempurl%
\url{https://doi.org/10.1145/2145204.2145355}
\showDOI{\tempurl}


\bibitem[Dubberley et~al\mbox{.}(2020)]%
        {noauthor_digital_2020}
\bibfield{editor}{\bibinfo{person}{Sam Dubberley}, \bibinfo{person}{Alexa Koenig}, {and} \bibinfo{person}{Daragh Murray}} (Eds.). \bibinfo{year}{2020}\natexlab{}.
\newblock \bibinfo{booktitle}{\emph{Digital {Witness}: {Using} {Open} {Source} {Information} for {Human} {Rights} {Investigation}, {Documentation}, and {Accountability}}}.
\newblock \bibinfo{publisher}{Oxford University Press}, \bibinfo{address}{Oxford, New York}.
\newblock
\showISBNx{978-0-19-883607-0}


\bibitem[Easterday et~al\mbox{.}(2014)]%
        {easterday2014design}
\bibfield{author}{\bibinfo{person}{Matthew~W Easterday}, \bibinfo{person}{Daniel~Rees Lewis}, {and} \bibinfo{person}{Elizabeth~M Gerber}.} \bibinfo{year}{2014}\natexlab{}.
\newblock \showarticletitle{Design-based research process: Problems, phases, and applications}.
\newblock \bibinfo{publisher}{Boulder, CO: International Society of the Learning Sciences}.
\newblock


\bibitem[Erete(2015)]%
        {erete_engaging_2015}
\bibfield{author}{\bibinfo{person}{Sheena~L. Erete}.} \bibinfo{year}{2015}\natexlab{}.
\newblock \showarticletitle{Engaging {Around} {Neighborhood} {Issues}: {How} {Online} {Communication} {Affects} {Offline} {Behavior}}. In \bibinfo{booktitle}{\emph{Proceedings of the 18th {ACM} {Conference} on {Computer} {Supported} {Cooperative} {Work} \& {Social} {Computing}}} \emph{(\bibinfo{series}{{CSCW} '15})}. \bibinfo{publisher}{Association for Computing Machinery}, \bibinfo{address}{New York, NY, USA}, \bibinfo{pages}{1590--1601}.
\newblock
\showISBNx{978-1-4503-2922-4}
\urldef\tempurl%
\url{https://doi.org/10.1145/2675133.2675182}
\showDOI{\tempurl}


\bibitem[{Esteban Borges}(2019)]%
        {esteban_borges_securitytrails_2019}
\bibfield{author}{\bibinfo{person}{{Esteban Borges}}.} \bibinfo{year}{2019}\natexlab{}.
\newblock \bibinfo{title}{{SecurityTrails} {\textbar} {OSINT} {Framework}: {The} {Perfect} {Cybersecurity} {Intel} {Gathering} {Tool}}.
\newblock
\newblock
\urldef\tempurl%
\url{https://securitytrails.com/blog/osint-framework}
\showURL{%
\tempurl}


\bibitem[Fiorella(2021)]%
        {bellingcat_verification}
\bibfield{author}{\bibinfo{person}{Giancarlo Fiorella}.} \bibinfo{year}{2021}\natexlab{}.
\newblock \bibinfo{booktitle}{\emph{First {Steps} to {Getting} {Started} in {Open} {Source} {Research}}}.
\newblock
\urldef\tempurl%
\url{https://www.bellingcat.com/resources/2021/11/09/first-steps-to-getting-started-in-open-source-research/}
\showURL{%
\tempurl}


\bibitem[Fisher et~al\mbox{.}(2012)]%
        {fisher2012distributed}
\bibfield{author}{\bibinfo{person}{Kristie Fisher}, \bibinfo{person}{Scott Counts}, {and} \bibinfo{person}{Aniket Kittur}.} \bibinfo{year}{2012}\natexlab{}.
\newblock \showarticletitle{Distributed Sensemaking: Improving Sensemaking by Leveraging the Efforts of Previous Users}. In \bibinfo{booktitle}{\emph{Proceedings of the {SIGCHI} Conference on Human Factors in Computing Systems}} (New York, {NY}, {USA}) \emph{(\bibinfo{series}{{CHI} '12})}. \bibinfo{publisher}{{ACM}}, \bibinfo{address}{New York, NY, USA}, \bibinfo{pages}{247--256}.
\newblock
\showISBNx{978-1-4503-1015-4}
\urldef\tempurl%
\url{https://doi.org/10.1145/2207676.2207711}
\showDOI{\tempurl}


\bibitem[Fletcher et~al\mbox{.}(2018)]%
        {fletcher2018measuring}
\bibfield{author}{\bibinfo{person}{Richard Fletcher}, \bibinfo{person}{Alessio Cornia}, \bibinfo{person}{Lucas Graves}, {and} \bibinfo{person}{Rasmus~Kleis Nielsen}.} \bibinfo{year}{2018}\natexlab{}.
\newblock \showarticletitle{Measuring the reach of" fake news" and online disinformation in Europe}.
\newblock \bibinfo{journal}{\emph{Australasian Policing}} \bibinfo{volume}{10}, \bibinfo{number}{2} (\bibinfo{year}{2018}).
\newblock


\bibitem[Flores-Saviaga et~al\mbox{.}(2022)]%
        {flores2022datavoidant}
\bibfield{author}{\bibinfo{person}{Claudia Flores-Saviaga}, \bibinfo{person}{Shangbin Feng}, {and} \bibinfo{person}{Saiph Savage}.} \bibinfo{year}{2022}\natexlab{}.
\newblock \showarticletitle{Datavoidant: An AI System for Addressing Political Data Voids on Social Media}.
\newblock \bibinfo{journal}{\emph{Proceedings of the ACM on Human-Computer Interaction}} \bibinfo{volume}{6}, \bibinfo{number}{CSCW2} (\bibinfo{year}{2022}), \bibinfo{pages}{1--29}.
\newblock


\bibitem[Ghioni et~al\mbox{.}(2022)]%
        {ghioni2022open}
\bibfield{author}{\bibinfo{person}{Riccardo Ghioni}, \bibinfo{person}{Mariarosaria Taddeo}, {and} \bibinfo{person}{Luciano Floridi}.} \bibinfo{year}{2022}\natexlab{}.
\newblock \showarticletitle{Open Source Intelligence and AI: a Systematic Review of the GELSI Literature}.
\newblock \bibinfo{journal}{\emph{AI \& Society}} (\bibinfo{year}{2022}).
\newblock


\bibitem[Glassman and Kang(2012)]%
        {glassman2012intelligence}
\bibfield{author}{\bibinfo{person}{Michael Glassman} {and} \bibinfo{person}{Min~Ju Kang}.} \bibinfo{year}{2012}\natexlab{}.
\newblock \showarticletitle{Intelligence in the internet age: The emergence and evolution of Open Source Intelligence (OSINT)}.
\newblock \bibinfo{journal}{\emph{Computers in Human Behavior}} \bibinfo{volume}{28}, \bibinfo{number}{2} (\bibinfo{date}{March} \bibinfo{year}{2012}), \bibinfo{pages}{673--682}.
\newblock
\showISSN{0747-5632}
\urldef\tempurl%
\url{https://doi.org/10.1016/j.chb.2011.11.014}
\showDOI{\tempurl}


\bibitem[Godel et~al\mbox{.}(2021)]%
        {godel2021moderating}
\bibfield{author}{\bibinfo{person}{William Godel}, \bibinfo{person}{Zeve Sanderson}, \bibinfo{person}{Kevin Aslett}, \bibinfo{person}{Jonathan Nagler}, \bibinfo{person}{Richard Bonneau}, \bibinfo{person}{Nathaniel Persily}, {and} \bibinfo{person}{Joshua~A Tucker}.} \bibinfo{year}{2021}\natexlab{}.
\newblock \showarticletitle{Moderating with the mob: Evaluating the efficacy of real-time crowdsourced fact-checking}.
\newblock \bibinfo{journal}{\emph{Journal of Online Trust and Safety}} \bibinfo{volume}{1}, \bibinfo{number}{1} (\bibinfo{year}{2021}).
\newblock


\bibitem[Gong et~al\mbox{.}(2019)]%
        {gong_social_2019}
\bibfield{author}{\bibinfo{person}{Miaomiao Gong}, \bibinfo{person}{Yuling Sun}, {and} \bibinfo{person}{Liang He}.} \bibinfo{year}{2019}\natexlab{}.
\newblock \showarticletitle{A {Social} {Network} {Engaged} {Crowdsourcing} {Framework} for {Expert} {Tasks}}. In \bibinfo{booktitle}{\emph{2019 {IEEE} 23rd {International} {Conference} on {Computer} {Supported} {Cooperative} {Work} in {Design} ({CSCWD})}}. IEEE, \bibinfo{pages}{249--254}.
\newblock
\urldef\tempurl%
\url{https://doi.org/10.1109/CSCWD.2019.8791923}
\showDOI{\tempurl}


\bibitem[Grevet and Gilbert(2015)]%
        {grevet2015piggyback}
\bibfield{author}{\bibinfo{person}{Catherine Grevet} {and} \bibinfo{person}{Eric Gilbert}.} \bibinfo{year}{2015}\natexlab{}.
\newblock \showarticletitle{Piggyback prototyping: Using existing, large-scale social computing systems to prototype new ones}. In \bibinfo{booktitle}{\emph{Proceedings of the 33rd Annual ACM Conference on Human Factors in Computing Systems}}. \bibinfo{pages}{4047--4056}.
\newblock


\bibitem[Gutierrez and Torpey(2015)]%
        {gutierrez2015digital}
\bibfield{author}{\bibinfo{person}{Pablo Gutierrez} {and} \bibinfo{person}{Paul Torpey}.} \bibinfo{year}{2015}\natexlab{}.
\newblock \showarticletitle{How Digital Detectives Say They Proved Ukraine Attacks Came from Russia}.
\newblock \bibinfo{journal}{\emph{The Guardian}} (\bibinfo{year}{2015}).
\newblock


\bibitem[Hahn et~al\mbox{.}(2016)]%
        {hahn2016knowledge}
\bibfield{author}{\bibinfo{person}{Nathan Hahn}, \bibinfo{person}{Joseph Chang}, \bibinfo{person}{Ji~Eun Kim}, {and} \bibinfo{person}{Aniket Kittur}.} \bibinfo{year}{2016}\natexlab{}.
\newblock \showarticletitle{The Knowledge Accelerator: Big Picture Thinking in Small Pieces}. In \bibinfo{booktitle}{\emph{Proceedings of the 2016 {CHI} Conference on Human Factors in Computing Systems}} (New York, {NY}, {USA}) \emph{(\bibinfo{series}{{CHI} '16})}. \bibinfo{publisher}{{ACM}}, \bibinfo{pages}{2258--2270}.
\newblock
\showISBNx{978-1-4503-3362-7}
\urldef\tempurl%
\url{http://doi.acm.org/10.1145/2858036.2858364}
\showURL{%
\tempurl}


\bibitem[Hanham and Shin(2020)]%
        {osintethics2020hanham}
\bibfield{author}{\bibinfo{person}{Melissa Hanham} {and} \bibinfo{person}{Jaewoo Shin}.} \bibinfo{year}{2020}\natexlab{}.
\newblock \showarticletitle{Ethics in the Age of OSINT Innocence}.
\newblock \bibinfo{journal}{\emph{Stanley Center for Peace and Security}} (\bibinfo{date}{May} \bibinfo{year}{2020}), \bibinfo{pages}{6}.
\newblock
\urldef\tempurl%
\url{https://stanleycenter.org/publications/ethics-osint-innocence/}
\showURL{%
\tempurl}


\bibitem[Harris et~al\mbox{.}(2019)]%
        {harris2019joining}
\bibfield{author}{\bibinfo{person}{Alexa~M Harris}, \bibinfo{person}{Diego G{\'o}mez-Zar{\'a}}, \bibinfo{person}{Leslie~A DeChurch}, {and} \bibinfo{person}{Noshir~S Contractor}.} \bibinfo{year}{2019}\natexlab{}.
\newblock \showarticletitle{Joining together online: the trajectory of CSCW scholarship on group formation}.
\newblock \bibinfo{journal}{\emph{Proceedings of the ACM on Human-Computer Interaction}} \bibinfo{volume}{3}, \bibinfo{number}{CSCW} (\bibinfo{year}{2019}), \bibinfo{pages}{1--27}.
\newblock


\bibitem[Hassan et~al\mbox{.}(2017)]%
        {hassan_claimbuster_2017}
\bibfield{author}{\bibinfo{person}{Naeemul Hassan}, \bibinfo{person}{Gensheng Zhang}, \bibinfo{person}{Fatma Arslan}, \bibinfo{person}{Josue Caraballo}, \bibinfo{person}{Damian Jimenez}, \bibinfo{person}{Siddhant Gawsane}, \bibinfo{person}{Shohedul Hasan}, \bibinfo{person}{Minumol Joseph}, \bibinfo{person}{Aaditya Kulkarni}, \bibinfo{person}{Anil~Kumar Nayak}, \bibinfo{person}{Vikas Sable}, \bibinfo{person}{Chengkai Li}, {and} \bibinfo{person}{Mark Tremayne}.} \bibinfo{year}{2017}\natexlab{}.
\newblock \showarticletitle{{ClaimBuster}: the first-ever end-to-end fact-checking system}.
\newblock \bibinfo{journal}{\emph{Proceedings of the VLDB Endowment}} \bibinfo{volume}{10}, \bibinfo{number}{12} (\bibinfo{date}{Aug.} \bibinfo{year}{2017}), \bibinfo{pages}{1945--1948}.
\newblock
\showISSN{2150-8097}
\urldef\tempurl%
\url{https://doi.org/10.14778/3137765.3137815}
\showDOI{\tempurl}


\bibitem[Higgins(2021a)]%
        {Bellingcat_guide}
\bibfield{author}{\bibinfo{person}{Annique~Mossou Higgins, Ross}.} \bibinfo{year}{2021}\natexlab{a}.
\newblock \bibinfo{booktitle}{\emph{A {Beginner}'s {Guide} to {Social} {Media} {Verification}}}.
\newblock
\urldef\tempurl%
\url{https://www.bellingcat.com/resources/2021/11/01/a-beginners-guide-to-social-media-verification/}
\showURL{%
\tempurl}


\bibitem[Higgins(2021b)]%
        {higgins2021we}
\bibfield{author}{\bibinfo{person}{Eliot Higgins}.} \bibinfo{year}{2021}\natexlab{b}.
\newblock \bibinfo{booktitle}{\emph{We Are Bellingcat : An Intelligence Agency for the People}}.
\newblock \bibinfo{publisher}{Bloomsbury Publishing}, \bibinfo{address}{London}.
\newblock
\showISBNx{978-1-5266-1575-6}


\bibitem[Hswe et~al\mbox{.}(2009)]%
        {hswe_web_2009}
\bibfield{author}{\bibinfo{person}{Patricia Hswe}, \bibinfo{person}{Joanne Kaczmarek}, \bibinfo{person}{Leah Houser}, {and} \bibinfo{person}{Janet Eke}.} \bibinfo{year}{2009}\natexlab{}.
\newblock \showarticletitle{The {Web} {Archives} {Workbench} ({WAW}) {Tool} {Suite}: {Taking} an {Archival} {Approach} to the {Preservation} of {Web} {Content}}.
\newblock \bibinfo{journal}{\emph{Library Trends}} \bibinfo{volume}{57}, \bibinfo{number}{3} (\bibinfo{year}{2009}), \bibinfo{pages}{442--460}.
\newblock
\showISSN{1559-0682}
\urldef\tempurl%
\url{https://doi.org/10.1353/lib.0.0046}
\showDOI{\tempurl}
\newblock
\shownote{Publisher: Johns Hopkins University Press}.


\bibitem[Huang et~al\mbox{.}(2015)]%
        {huang_connected_2015}
\bibfield{author}{\bibinfo{person}{Y.~Linlin Huang}, \bibinfo{person}{Kate Starbird}, \bibinfo{person}{Mania Orand}, \bibinfo{person}{Stephanie~A. Stanek}, {and} \bibinfo{person}{Heather~T. Pedersen}.} \bibinfo{year}{2015}\natexlab{}.
\newblock \showarticletitle{Connected {Through} {Crisis}: {Emotional} {Proximity} and the {Spread} of {Misinformation} {Online}}. In \bibinfo{booktitle}{\emph{Proceedings of the 18th {ACM} {Conference} on {Computer} {Supported} {Cooperative} {Work} \& {Social} {Computing}}} \emph{(\bibinfo{series}{{CSCW} '15})}. \bibinfo{publisher}{Association for Computing Machinery}, \bibinfo{address}{New York, NY, USA}, \bibinfo{pages}{969--980}.
\newblock
\showISBNx{978-1-4503-2922-4}
\urldef\tempurl%
\url{https://doi.org/10.1145/2675133.2675202}
\showDOI{\tempurl}


\bibitem[Hulnick(2002)]%
        {hulnick_downside_2002}
\bibfield{author}{\bibinfo{person}{Arthur~S. Hulnick}.} \bibinfo{year}{2002}\natexlab{}.
\newblock \showarticletitle{The {Downside} of {Open} {Source} {Intelligence}}.
\newblock \bibinfo{journal}{\emph{International Journal of Intelligence and CounterIntelligence}} \bibinfo{volume}{15}, \bibinfo{number}{4} (\bibinfo{date}{Nov.} \bibinfo{year}{2002}), \bibinfo{pages}{565--579}.
\newblock
\showISSN{0885-0607, 1521-0561}
\urldef\tempurl%
\url{https://doi.org/10.1080/08850600290101767}
\showDOI{\tempurl}


\bibitem[Iorga et~al\mbox{.}(2021)]%
        {iorga2021early}
\bibfield{author}{\bibinfo{person}{Denis Iorga}, \bibinfo{person}{Octavian Grigorescu}, \bibinfo{person}{Mihai Predoiu}, \bibinfo{person}{Cristian Sandescu}, \bibinfo{person}{Mihai Dascalu}, {and} \bibinfo{person}{Razvan Rughinis}.} \bibinfo{year}{2021}\natexlab{}.
\newblock \showarticletitle{Early Usability Evaluation to Enhance User Interfaces-A Use Case on the Yggdrasil Cybersecurity Mockup.}. In \bibinfo{booktitle}{\emph{RoCHI}}. \bibinfo{pages}{103--110}.
\newblock


\bibitem[Kermode et~al\mbox{.}(2020)]%
        {kermode_objects_2020}
\bibfield{author}{\bibinfo{person}{Lachlan Kermode}, \bibinfo{person}{Jan Freyberg}, \bibinfo{person}{Alican Akturk}, \bibinfo{person}{Robert Trafford}, \bibinfo{person}{Denis Kochetkov}, \bibinfo{person}{Rafael Pardinas}, \bibinfo{person}{Eyal Weizman}, {and} \bibinfo{person}{Julien Cornebise}.} \bibinfo{year}{2020}\natexlab{}.
\newblock \showarticletitle{Objects of violence: synthetic data for practical {ML} in human rights investigations}.
\newblock \bibinfo{journal}{\emph{arXiv:2004.01030 [cs]}} (\bibinfo{date}{April} \bibinfo{year}{2020}), \bibinfo{pages}{12}.
\newblock
\urldef\tempurl%
\url{http://arxiv.org/abs/2004.01030}
\showURL{%
\tempurl}
\newblock
\shownote{arXiv: 2004.01030}.


\bibitem[Kim et~al\mbox{.}(2020)]%
        {kim2020toward}
\bibfield{author}{\bibinfo{person}{Sung-Kyung Kim}, \bibinfo{person}{Eun-Tae Jang}, {and} \bibinfo{person}{Ki-Woong Park}.} \bibinfo{year}{2020}\natexlab{}.
\newblock \showarticletitle{Toward a fine-grained evaluation of the Pwnable CTF}. In \bibinfo{booktitle}{\emph{Information Security Applications: 21st International Conference, WISA 2020, Jeju Island, South Korea, August 26--28, 2020, Revised Selected Papers}}. Springer, \bibinfo{pages}{179--190}.
\newblock


\bibitem[Kim et~al\mbox{.}(2018)]%
        {kim2018hit}
\bibfield{author}{\bibinfo{person}{Yongsung Kim}, \bibinfo{person}{Darren Gergle}, {and} \bibinfo{person}{Haoqi Zhang}.} \bibinfo{year}{2018}\natexlab{}.
\newblock \showarticletitle{Hit-or-wait: Coordinating opportunistic low-effort contributions to achieve global outcomes in on-the-go crowdsourcing}. In \bibinfo{booktitle}{\emph{Proceedings of the 2018 CHI Conference on Human Factors in Computing Systems}}. \bibinfo{pages}{1--12}.
\newblock


\bibitem[Kittur et~al\mbox{.}(2014)]%
        {kittur_standing_2014}
\bibfield{author}{\bibinfo{person}{Aniket Kittur}, \bibinfo{person}{Andrew~M. Peters}, \bibinfo{person}{Abdigani Diriye}, {and} \bibinfo{person}{Michael Bove}.} \bibinfo{year}{2014}\natexlab{}.
\newblock \showarticletitle{Standing on the Schemas of Giants: Socially Augmented Information Foraging}. In \bibinfo{booktitle}{\emph{Proceedings of the 17th {ACM} Conference on Computer Supported Cooperative Work \& Social Computing}} (New York, {NY}, {USA}) \emph{(\bibinfo{series}{{CSCW} '14})}. \bibinfo{publisher}{{ACM}}, \bibinfo{pages}{999--1010}.
\newblock
\showISBNx{978-1-4503-2540-0}
\urldef\tempurl%
\url{https://doi.org/10.1145/2531602.2531644}
\showDOI{\tempurl}


\bibitem[Kittur et~al\mbox{.}(2011)]%
        {kittur2011crowdforge}
\bibfield{author}{\bibinfo{person}{Aniket Kittur}, \bibinfo{person}{Boris Smus}, \bibinfo{person}{Susheel Khamkar}, {and} \bibinfo{person}{Robert~E. Kraut}.} \bibinfo{year}{2011}\natexlab{}.
\newblock \showarticletitle{{CrowdForge}: crowdsourcing complex work}. In \bibinfo{booktitle}{\emph{Proceedings of the 24th annual {ACM} symposium on User interface software and technology}} (New York, {NY}, {USA}) \emph{(\bibinfo{series}{{UIST} '11})}. \bibinfo{publisher}{{ACM}}, \bibinfo{pages}{43--52}.
\newblock
\showISBNx{978-1-4503-0716-1}
\urldef\tempurl%
\url{https://doi.org/10.1145/2047196.2047202}
\showDOI{\tempurl}


\bibitem[Kolb and Kolb(2017)]%
        {kolb_experiential_2017}
\bibfield{author}{\bibinfo{person}{Alice Kolb} {and} \bibinfo{person}{David Kolb}.} \bibinfo{year}{2017}\natexlab{}.
\newblock \showarticletitle{Experiential {Learning} {Theory} as a {Guide} for {Experiential} {Educators} in {Higher} {Education}}.
\newblock \bibinfo{journal}{\emph{Experiential Learning \& Teaching in Higher Education}} \bibinfo{volume}{1}, \bibinfo{number}{1} (\bibinfo{date}{June} \bibinfo{year}{2017}), \bibinfo{pages}{7--44}.
\newblock
\showISSN{2474-3410 (print)}
\urldef\tempurl%
\url{https://nsuworks.nova.edu/elthe/vol1/iss1/7}
\showURL{%
\tempurl}


\bibitem[Kornfield(2021)]%
        {kornfield2021wrong}
\bibfield{author}{\bibinfo{person}{Meryl Kornfield}.} \bibinfo{year}{2021}\natexlab{}.
\newblock \bibinfo{booktitle}{\emph{The wrong {ID}: {Retired} firefighter, comedian and {Chuck} {Norris} falsely accused of being {Capitol} rioters}}.
\newblock Washington Post.
\newblock
\showISSN{0190-8286}
\urldef\tempurl%
\url{https://www.washingtonpost.com/technology/2021/01/16/sleuths-falsely-identify-rioters/}
\showURL{%
\tempurl}


\bibitem[Lave and Wenger(1991)]%
        {lave_situated_1991}
\bibfield{author}{\bibinfo{person}{Jean Lave} {and} \bibinfo{person}{Etienne Wenger}.} \bibinfo{year}{1991}\natexlab{}.
\newblock \bibinfo{booktitle}{\emph{Situated {Learning}: {Legitimate} {Peripheral} {Participation}}}.
\newblock \bibinfo{publisher}{Cambridge University Press}, \bibinfo{address}{Cambridge, UK}.
\newblock
\showISBNx{978-0-521-42374-8}
\newblock
\shownote{Google-Books-ID: CAVIOrW3vYAC}.


\bibitem[Li et~al\mbox{.}(2018)]%
        {li2018crowdia}
\bibfield{author}{\bibinfo{person}{Tianyi Li}, \bibinfo{person}{Kurt Luther}, {and} \bibinfo{person}{Chris North}.} \bibinfo{year}{2018}\natexlab{}.
\newblock \showarticletitle{{CrowdIA}: Solving Mysteries with Crowdsourced Sensemaking}.
\newblock   \bibinfo{volume}{2} (\bibinfo{year}{2018}), \bibinfo{pages}{105:1--105:29}.
\newblock
Issue {CSCW}.
\showISSN{2573-0142}
\urldef\tempurl%
\url{https://doi.org/10.1145/3274374}
\showDOI{\tempurl}


\bibitem[Li et~al\mbox{.}(2019)]%
        {li2019dropping}
\bibfield{author}{\bibinfo{person}{Tianyi Li}, \bibinfo{person}{Chandler~J Manns}, \bibinfo{person}{Chris North}, {and} \bibinfo{person}{Kurt Luther}.} \bibinfo{year}{2019}\natexlab{}.
\newblock \showarticletitle{Dropping the baton? Understanding errors and bottlenecks in a crowdsourced sensemaking pipeline}.
\newblock \bibinfo{journal}{\emph{Proceedings of the ACM on Human-Computer Interaction}} \bibinfo{volume}{3}, \bibinfo{number}{CSCW} (\bibinfo{year}{2019}), \bibinfo{pages}{1--26}.
\newblock


\bibitem[Matatov et~al\mbox{.}(2018)]%
        {Matatov2018DejaVu}
\bibfield{author}{\bibinfo{person}{Hana Matatov}, \bibinfo{person}{Adina Bechhofer}, \bibinfo{person}{Lora Aroyo}, \bibinfo{person}{Ofra Amir}, {and} \bibinfo{person}{Mor Naaman}.} \bibinfo{year}{2018}\natexlab{}.
\newblock \showarticletitle{DejaVu: A System for Journalists to Collaboratively Address Visual Misinformation}.
\newblock


\bibitem[McClure~Haughey et~al\mbox{.}(2020)]%
        {haughey2020misinformationbeat}
\bibfield{author}{\bibinfo{person}{Melinda McClure~Haughey}, \bibinfo{person}{Meena~Devii Muralikumar}, \bibinfo{person}{Cameron~A. Wood}, {and} \bibinfo{person}{Kate Starbird}.} \bibinfo{year}{2020}\natexlab{}.
\newblock \showarticletitle{On the Misinformation Beat: Understanding the Work of Investigative Journalists Reporting on Problematic Information Online}.
\newblock \bibinfo{journal}{\emph{Proc. ACM Hum.-Comput. Interact.}} \bibinfo{volume}{4}, \bibinfo{number}{CSCW2}, Article \bibinfo{articleno}{133} (\bibinfo{date}{Oct.} \bibinfo{year}{2020}), \bibinfo{numpages}{22}~pages.
\newblock
\urldef\tempurl%
\url{https://doi.org/10.1145/3415204}
\showDOI{\tempurl}


\bibitem[McKeown et~al\mbox{.}(2014)]%
        {mckeown_investigating_2014}
\bibfield{author}{\bibinfo{person}{Sean McKeown}, \bibinfo{person}{David Maxwell}, \bibinfo{person}{Leif Azzopardi}, {and} \bibinfo{person}{William~Bradley Glisson}.} \bibinfo{year}{2014}\natexlab{}.
\newblock \showarticletitle{Investigating people: a qualitative analysis of the search behaviours of open-source intelligence analysts}. In \bibinfo{booktitle}{\emph{Proceedings of the 5th {Information} {Interaction} in {Context} {Symposium}}} \emph{(\bibinfo{series}{{IIiX} '14})}. \bibinfo{publisher}{Association for Computing Machinery}, \bibinfo{address}{New York, NY, USA}, \bibinfo{pages}{175--184}.
\newblock
\showISBNx{978-1-4503-2976-7}
\urldef\tempurl%
\url{https://doi.org/10.1145/2637002.2637023}
\showDOI{\tempurl}


\bibitem[Mercado(2004)]%
        {y48}
\bibfield{author}{\bibinfo{person}{Stephen~C. Mercado}.} \bibinfo{year}{2004}\natexlab{}.
\newblock \bibinfo{booktitle}{\emph{Sailing the sea of OSINT in the information age. Technical Report}}.
\newblock \bibinfo{publisher}{dataset}, \bibinfo{address}{American Psychological Association. type}.
\newblock
\urldef\tempurl%
\url{https://doi.org/10.1037/e741272011-005}
\showURL{%
\tempurl}


\bibitem[Metaxas and Finn(2017)]%
        {metaxas2017infamous}
\bibfield{author}{\bibinfo{person}{Panagiotis Metaxas} {and} \bibinfo{person}{Samantha~T Finn}.} \bibinfo{year}{2017}\natexlab{}.
\newblock \showarticletitle{The infamous\# Pizzagate conspiracy theory: Insight from a TwitterTrails investigation}.
\newblock  (\bibinfo{year}{2017}).
\newblock


\bibitem[Micallef et~al\mbox{.}(2022)]%
        {micallef2022true}
\bibfield{author}{\bibinfo{person}{Nicholas Micallef}, \bibinfo{person}{Vivienne Armacost}, \bibinfo{person}{Nasir Memon}, {and} \bibinfo{person}{Sameer Patil}.} \bibinfo{year}{2022}\natexlab{}.
\newblock \showarticletitle{True or False: Studying the Work Practices of Professional Fact-Checkers}.
\newblock \bibinfo{journal}{\emph{Proceedings of the ACM on Human-Computer Interaction}} \bibinfo{volume}{6}, \bibinfo{number}{CSCW1} (\bibinfo{year}{2022}), \bibinfo{pages}{1--44}.
\newblock


\bibitem[Mitra et~al\mbox{.}(2015)]%
        {mitra2015comparing}
\bibfield{author}{\bibinfo{person}{Tanushree Mitra}, \bibinfo{person}{Clayton~J Hutto}, {and} \bibinfo{person}{Eric Gilbert}.} \bibinfo{year}{2015}\natexlab{}.
\newblock \showarticletitle{Comparing person-and process-centric strategies for obtaining quality data on amazon mechanical turk}. In \bibinfo{booktitle}{\emph{Proceedings of the 33rd Annual ACM Conference on Human Factors in Computing Systems}}. \bibinfo{pages}{1345--1354}.
\newblock


\bibitem[Müller and Wiik(2021)]%
        {muller_gatekeeper_2021}
\bibfield{author}{\bibinfo{person}{Nina~C. Müller} {and} \bibinfo{person}{Jenny Wiik}.} \bibinfo{year}{2021}\natexlab{}.
\newblock \showarticletitle{From {Gatekeeper} to {Gate}-opener: {Open}-{Source} {Spaces} in {Investigative} {Journalism}}.
\newblock \bibinfo{journal}{\emph{Journalism Practice}} \bibinfo{volume}{0}, \bibinfo{number}{0} (\bibinfo{date}{May} \bibinfo{year}{2021}), \bibinfo{pages}{1--20}.
\newblock
\showISSN{1751-2786}
\urldef\tempurl%
\url{https://doi.org/10.1080/17512786.2021.1919543}
\showDOI{\tempurl}
\newblock
\shownote{Publisher: Routledge \_eprint: https://doi.org/10.1080/17512786.2021.1919543}.


\bibitem[Nhan et~al\mbox{.}(2017)]%
        {nhan2017digilantism}
\bibfield{author}{\bibinfo{person}{Johnny Nhan}, \bibinfo{person}{Laura Huey}, {and} \bibinfo{person}{Ryan Broll}.} \bibinfo{year}{2017}\natexlab{}.
\newblock \showarticletitle{Digilantism: An analysis of crowdsourcing and the Boston marathon bombings}.
\newblock \bibinfo{journal}{\emph{The British journal of criminology}} \bibinfo{volume}{57}, \bibinfo{number}{2} (\bibinfo{date}{March} \bibinfo{year}{2017}), \bibinfo{pages}{341--361}.
\newblock
\showISSN{0007-0955}
\urldef\tempurl%
\url{https://doi.org/10.1093/bjc/azv118}
\showDOI{\tempurl}


\bibitem[Noronha et~al\mbox{.}(2011)]%
        {noronha2011platemate}
\bibfield{author}{\bibinfo{person}{Jon Noronha}, \bibinfo{person}{Eric Hysen}, \bibinfo{person}{Haoqi Zhang}, {and} \bibinfo{person}{Krzysztof~Z Gajos}.} \bibinfo{year}{2011}\natexlab{}.
\newblock \showarticletitle{Platemate: crowdsourcing nutritional analysis from food photographs}. In \bibinfo{booktitle}{\emph{Proceedings of the 24th annual ACM symposium on User interface software and technology}}. \bibinfo{pages}{1--12}.
\newblock


\bibitem[of~Homeland~Security(2010)]%
        {department_of_homeland_security_ufouoles_2010}
\bibfield{author}{\bibinfo{person}{Department of Homeland~Security}.} \bibinfo{year}{2010}\natexlab{}.
\newblock \bibinfo{title}{({U}//{FOUO}//{LES}) {DHS} {Terrorist} {Use} of {Social} {Networking} {Facebook} {Case} {Study} {\textbar} {Public} {Intelligence}}.
\newblock
\newblock
\urldef\tempurl%
\url{https://publicintelligence.net/ufouoles-dhs-terrorist-use-of-social-networking-facebook-case-study/}
\showURL{%
\tempurl}


\bibitem[of~Technology(2023)]%
        {MITCybersecurityClinic}
\bibfield{author}{\bibinfo{person}{Massachusetts~Institute of Technology}.} \bibinfo{year}{2023}\natexlab{}.
\newblock \bibinfo{title}{Urban Cyber Defense: Cybersecurity Clinic}.
\newblock
\newblock
\urldef\tempurl%
\url{http://urbancyberdefense.mit.edu/cybersecurityclinic}
\showURL{%
\tempurl}
\newblock
\shownote{Accessed: 2023-10-02}.


\bibitem[Oleson et~al\mbox{.}(2013)]%
        {oleson2013evaluating}
\bibfield{author}{\bibinfo{person}{David Oleson}, \bibinfo{person}{Alexander Sorokin}, \bibinfo{person}{Greg Laughlin}, \bibinfo{person}{Vaughn Hester}, \bibinfo{person}{John Le}, \bibinfo{person}{Christopher~R Van~Pelt}, {and} \bibinfo{person}{Lukas~A Biewald}.} \bibinfo{year}{2013}\natexlab{}.
\newblock \bibinfo{title}{Evaluating a worker in performing crowd sourced tasks and providing in-task training through programmatically generated test tasks}.
\newblock
\newblock
\newblock
\shownote{US Patent 8,554,605}.


\bibitem[Papoutsaki et~al\mbox{.}(2015)]%
        {papoutsaki_crowdsourcing_2015}
\bibfield{author}{\bibinfo{person}{Alexandra Papoutsaki}, \bibinfo{person}{Hua Guo}, \bibinfo{person}{Danae Metaxa-Kakavouli}, \bibinfo{person}{Connor Gramazio}, \bibinfo{person}{Jeff Rasley}, \bibinfo{person}{Wenting Xie}, \bibinfo{person}{Guan Wang}, {and} \bibinfo{person}{Jeff Huang}.} \bibinfo{year}{2015}\natexlab{}.
\newblock \showarticletitle{Crowdsourcing from Scratch: A Pragmatic Experiment in Data Collection by Novice Requesters}. In \bibinfo{booktitle}{\emph{Third {AAAI} Conference on Human Computation and Crowdsourcing}}.
\newblock
\urldef\tempurl%
\url{https://www.aaai.org/ocs/index.php/HCOMP/HCOMP15/paper/view/11582}
\showURL{%
\tempurl}


\bibitem[Pennycook and Rand(2019)]%
        {pennycook_fighting_2019}
\bibfield{author}{\bibinfo{person}{Gordon Pennycook} {and} \bibinfo{person}{David~G. Rand}.} \bibinfo{year}{2019}\natexlab{}.
\newblock \showarticletitle{Fighting misinformation on social media using crowdsourced judgments of news source quality}.
\newblock \bibinfo{journal}{\emph{Proceedings of the National Academy of Sciences}} \bibinfo{volume}{116}, \bibinfo{number}{7} (\bibinfo{date}{Feb.} \bibinfo{year}{2019}), \bibinfo{pages}{2521--2526}.
\newblock
\showISSN{0027-8424, 1091-6490}
\urldef\tempurl%
\url{https://doi.org/10.1073/pnas.1806781116}
\showDOI{\tempurl}
\newblock
\shownote{Publisher: National Academy of Sciences Section: Social Sciences}.


\bibitem[Pintrich(2004)]%
        {pintrich2004conceptual}
\bibfield{author}{\bibinfo{person}{Paul~R Pintrich}.} \bibinfo{year}{2004}\natexlab{}.
\newblock \showarticletitle{A conceptual framework for assessing motivation and self-regulated learning in college students}.
\newblock \bibinfo{journal}{\emph{Educational psychology review}} \bibinfo{volume}{16}, \bibinfo{number}{4} (\bibinfo{year}{2004}), \bibinfo{pages}{385--407}.
\newblock


\bibitem[Plomp et~al\mbox{.}(2013)]%
        {plomp2013educational}
\bibfield{author}{\bibinfo{person}{Tjeerd Plomp} {et~al\mbox{.}}} \bibinfo{year}{2013}\natexlab{}.
\newblock \showarticletitle{Educational design research: An introduction}.
\newblock \bibinfo{journal}{\emph{Educational design research}} (\bibinfo{year}{2013}), \bibinfo{pages}{11--50}.
\newblock


\bibitem[Poelman et~al\mbox{.}(2012)]%
        {poelman_as_2012}
\bibfield{author}{\bibinfo{person}{Ronald Poelman}, \bibinfo{person}{Oytun Akman}, \bibinfo{person}{Stephan Lukosch}, {and} \bibinfo{person}{Pieter Jonker}.} \bibinfo{year}{2012}\natexlab{}.
\newblock \showarticletitle{As if being there: mediated reality for crime scene investigation}. In \bibinfo{booktitle}{\emph{Proceedings of the {ACM} 2012 conference on {Computer} {Supported} {Cooperative} {Work}}} \emph{(\bibinfo{series}{{CSCW} '12})}. \bibinfo{publisher}{Association for Computing Machinery}, \bibinfo{address}{New York, NY, USA}, \bibinfo{pages}{1267--1276}.
\newblock
\showISBNx{978-1-4503-1086-4}
\urldef\tempurl%
\url{https://doi.org/10.1145/2145204.2145394}
\showDOI{\tempurl}


\bibitem[Rechkemmer and Yin(2020)]%
        {rechkemmer_motivating_2020}
\bibfield{author}{\bibinfo{person}{Amy Rechkemmer} {and} \bibinfo{person}{Ming Yin}.} \bibinfo{year}{2020}\natexlab{}.
\newblock \showarticletitle{Motivating {Novice} {Crowd} {Workers} through {Goal} {Setting}: {An} {Investigation} into the {Effects} on {Complex} {Crowdsourcing} {Task} {Training}}.
\newblock \bibinfo{journal}{\emph{Proceedings of the AAAI Conference on Human Computation and Crowdsourcing}}  \bibinfo{volume}{8} (\bibinfo{date}{Oct.} \bibinfo{year}{2020}), \bibinfo{pages}{122--131}.
\newblock
\showISSN{2769-1349}
\urldef\tempurl%
\url{https://doi.org/10.1609/hcomp.v8i1.7470}
\showDOI{\tempurl}


\bibitem[Retelny et~al\mbox{.}(2017)]%
        {retelny2017noworkflow}
\bibfield{author}{\bibinfo{person}{Daniela Retelny}, \bibinfo{person}{Michael~S. Bernstein}, {and} \bibinfo{person}{Melissa~A. Valentine}.} \bibinfo{year}{2017}\natexlab{}.
\newblock \showarticletitle{No Workflow Can Ever Be Enough: How Crowdsourcing Workflows Constrain Complex Work}.
\newblock \bibinfo{journal}{\emph{Proc. ACM Hum.-Comput. Interact.}} \bibinfo{volume}{1}, \bibinfo{number}{CSCW}, Article \bibinfo{articleno}{89} (\bibinfo{date}{Dec.} \bibinfo{year}{2017}), \bibinfo{numpages}{23}~pages.
\newblock
\urldef\tempurl%
\url{https://doi.org/10.1145/3134724}
\showDOI{\tempurl}


\bibitem[Rouach and Santi(2001)]%
        {rouach_competitive_2001}
\bibfield{author}{\bibinfo{person}{Daniel Rouach} {and} \bibinfo{person}{Patrice Santi}.} \bibinfo{year}{2001}\natexlab{}.
\newblock \showarticletitle{Competitive {Intelligence} {Adds} {Value}:: {Five} {Intelligence} {Attitudes}}.
\newblock \bibinfo{journal}{\emph{European Management Journal}} \bibinfo{volume}{19}, \bibinfo{number}{5} (\bibinfo{date}{Oct.} \bibinfo{year}{2001}), \bibinfo{pages}{552--559}.
\newblock
\showISSN{0263-2373}
\urldef\tempurl%
\url{https://doi.org/10.1016/S0263-2373(01)00069-X}
\showDOI{\tempurl}


\bibitem[{Ryan Hunt}(2012)]%
        {ryan_hunt_thirty-seven_2012}
\bibfield{author}{\bibinfo{person}{{Ryan Hunt}}.} \bibinfo{year}{2012}\natexlab{}.
\newblock \bibinfo{title}{Thirty-{Seven} {Percent} of {Companies} {Use} {Social} {Networks} to {Research} {Potential} {Job} {Candidates}, {According} to {New} {CareerBuilder} {Survey} - {Apr} 18, 2012}.
\newblock
\newblock
\urldef\tempurl%
\url{http://press.careerbuilder.com/2012-04-18-Thirty-Seven-Percent-of-Companies-Use-Social-Networks-to-Research-Potential-Job-Candidates-According-to-New-CareerBuilder-Survey}
\showURL{%
\tempurl}


\bibitem[Saeed et~al\mbox{.}(2022)]%
        {saeed2022crowdsourced}
\bibfield{author}{\bibinfo{person}{Mohammed Saeed}, \bibinfo{person}{Nicolas Traub}, \bibinfo{person}{Maelle Nicolas}, \bibinfo{person}{Gianluca Demartini}, {and} \bibinfo{person}{Paolo Papotti}.} \bibinfo{year}{2022}\natexlab{}.
\newblock \showarticletitle{Crowdsourced Fact-Checking at Twitter: How Does the Crowd Compare With Experts?}. In \bibinfo{booktitle}{\emph{Proceedings of the 31st ACM International Conference on Information \& Knowledge Management}}. \bibinfo{pages}{1736--1746}.
\newblock


\bibitem[Schumacher-Matos(2012)]%
        {schumacher-matos_election_2012}
\bibfield{author}{\bibinfo{person}{Edward Schumacher-Matos}.} \bibinfo{year}{2012}\natexlab{}.
\newblock \showarticletitle{Election 1: {Fact} {Checking} {The} {NPR} {Fact} {Checkers}}.
\newblock \bibinfo{journal}{\emph{NPR}} (\bibinfo{date}{Oct.} \bibinfo{year}{2012}).
\newblock
\urldef\tempurl%
\url{https://www.npr.org/sections/publiceditor/2012/10/28/161839145/election-1-fact-checking-the-npr-fact-checkers}
\showURL{%
\tempurl}


\bibitem[Sethi and Rangaraju(2018)]%
        {sethi2018extinguishing}
\bibfield{author}{\bibinfo{person}{Ricky Sethi} {and} \bibinfo{person}{Raghuram Rangaraju}.} \bibinfo{year}{2018}\natexlab{}.
\newblock \showarticletitle{Extinguishing the backfire effect: using emotions in online social collaborative argumentation for fact checking}. In \bibinfo{booktitle}{\emph{2018 IEEE International conference on web services (ICWS)}}. IEEE, \bibinfo{pages}{363--366}.
\newblock


\bibitem[Shao et~al\mbox{.}(2016)]%
        {shao_hoaxy_2016}
\bibfield{author}{\bibinfo{person}{Chengcheng Shao}, \bibinfo{person}{Giovanni~Luca Ciampaglia}, \bibinfo{person}{Alessandro Flammini}, {and} \bibinfo{person}{Filippo Menczer}.} \bibinfo{year}{2016}\natexlab{}.
\newblock \showarticletitle{Hoaxy: {A} {Platform} for {Tracking} {Online} {Misinformation}}. In \bibinfo{booktitle}{\emph{Proceedings of the 25th {International} {Conference} {Companion} on {World} {Wide} {Web}}} \emph{(\bibinfo{series}{{WWW} '16 {Companion}})}. \bibinfo{publisher}{International World Wide Web Conferences Steering Committee}, \bibinfo{address}{Republic and Canton of Geneva, CHE}, \bibinfo{pages}{745--750}.
\newblock
\showISBNx{978-1-4503-4144-8}
\urldef\tempurl%
\url{https://doi.org/10.1145/2872518.2890098}
\showDOI{\tempurl}


\bibitem[Silverman(2013)]%
        {verification_handbook}
\bibfield{author}{\bibinfo{person}{Craig Silverman}.} \bibinfo{year}{2013}\natexlab{}.
\newblock \bibinfo{title}{Verification Handbook: A Definitive Guide to Verifying Digital Content for Emergency Coverage}.
\newblock
\newblock
\urldef\tempurl%
\url{http://verificationhandbook.com/}
\showURL{%
\tempurl}


\bibitem[Steiger et~al\mbox{.}(2021)]%
        {steiger2021wellbeing}
\bibfield{author}{\bibinfo{person}{Miriah Steiger}, \bibinfo{person}{Timir~J Bharucha}, \bibinfo{person}{Sukrit Venkatagiri}, \bibinfo{person}{Martin~J. Riedl}, {and} \bibinfo{person}{Matthew Lease}.} \bibinfo{year}{2021}\natexlab{}.
\newblock \showarticletitle{The Psychological Well-Being of Content Moderators: The Emotional Labor of Commercial Moderation and Avenues for Improving Support}. In \bibinfo{booktitle}{\emph{Proceedings of the 2021 CHI Conference on Human Factors in Computing Systems}} (Yokohama, Japan) \emph{(\bibinfo{series}{CHI '21})}. \bibinfo{publisher}{Association for Computing Machinery}, \bibinfo{address}{New York, NY, USA}, Article \bibinfo{articleno}{341}, \bibinfo{numpages}{14}~pages.
\newblock
\showISBNx{9781450380966}
\urldef\tempurl%
\url{https://doi.org/10.1145/3411764.3445092}
\showDOI{\tempurl}


\bibitem[Suzuki et~al\mbox{.}(2016)]%
        {suzuki_atelier:_2016}
\bibfield{author}{\bibinfo{person}{Ryo Suzuki}, \bibinfo{person}{Niloufar Salehi}, \bibinfo{person}{Michelle~S. Lam}, \bibinfo{person}{Juan~C. Marroquin}, {and} \bibinfo{person}{Michael~S. Bernstein}.} \bibinfo{year}{2016}\natexlab{}.
\newblock \showarticletitle{Atelier: {Repurposing} {Expert} {Crowdsourcing} {Tasks} {As} {Micro}-internships}. In \bibinfo{booktitle}{\emph{Proceedings of the 2016 {CHI} {Conference} on {Human} {Factors} in {Computing} {Systems}}} \emph{(\bibinfo{series}{{CHI} '16})}. \bibinfo{publisher}{ACM}, \bibinfo{address}{New York, NY, USA}, \bibinfo{pages}{2645--2656}.
\newblock
\showISBNx{978-1-4503-3362-7}
\urldef\tempurl%
\url{https://doi.org/10.1145/2858036.2858121}
\showDOI{\tempurl}


\bibitem[Tausczik and Boons(2018)]%
        {tausczik_distributed_2018}
\bibfield{author}{\bibinfo{person}{Yla Tausczik} {and} \bibinfo{person}{Mark Boons}.} \bibinfo{year}{2018}\natexlab{}.
\newblock \showarticletitle{Distributed Knowledge in Crowds: Crowd Performance on Hidden Profile Tasks}. In \bibinfo{booktitle}{\emph{Twelfth International {AAAI} Conference on Web and Social Media}}.
\newblock
\urldef\tempurl%
\url{https://aaai.org/ocs/index.php/ICWSM/ICWSM18/paper/view/17817}
\showURL{%
\tempurl}


\bibitem[Tekian et~al\mbox{.}(2017)]%
        {tekian2017qualitative}
\bibfield{author}{\bibinfo{person}{Ara Tekian}, \bibinfo{person}{Christopher~J Watling}, \bibinfo{person}{Trudie~E Roberts}, \bibinfo{person}{Yvonne Steinert}, {and} \bibinfo{person}{John Norcini}.} \bibinfo{year}{2017}\natexlab{}.
\newblock \showarticletitle{Qualitative and quantitative feedback in the context of competency-based education}.
\newblock \bibinfo{journal}{\emph{Medical teacher}} \bibinfo{volume}{39}, \bibinfo{number}{12} (\bibinfo{year}{2017}), \bibinfo{pages}{1245--1249}.
\newblock


\bibitem[Trottier(2017)]%
        {trottier2017digital}
\bibfield{author}{\bibinfo{person}{Daniel Trottier}.} \bibinfo{year}{2017}\natexlab{}.
\newblock \showarticletitle{Digital vigilantism as weaponisation of visibility}.
\newblock \bibinfo{journal}{\emph{Philosophy \& Technology}} \bibinfo{volume}{30}, \bibinfo{number}{1} (\bibinfo{date}{March} \bibinfo{year}{2017}), \bibinfo{pages}{55--72}.
\newblock
\showISSN{2210-5441}
\urldef\tempurl%
\url{https://doi.org/10.1007/s13347-016-0216-4}
\showDOI{\tempurl}


\bibitem[Varma(2010)]%
        {varma2010so}
\bibfield{author}{\bibinfo{person}{Roli Varma}.} \bibinfo{year}{2010}\natexlab{}.
\newblock \showarticletitle{Why so few women enroll in computing? Gender and ethnic differences in students' perception}.
\newblock \bibinfo{journal}{\emph{Computer Science Education}} \bibinfo{volume}{20}, \bibinfo{number}{4} (\bibinfo{year}{2010}), \bibinfo{pages}{301--316}.
\newblock


\bibitem[Venkatagiri(2022)]%
        {venkatagiri2022supporting}
\bibfield{author}{\bibinfo{person}{Sukrit Venkatagiri}.} \bibinfo{year}{2022}\natexlab{}.
\newblock \emph{\bibinfo{title}{Supporting and Transforming High-Stakes Investigations with Expert-Led Crowdsourcing}}.
\newblock \bibinfo{thesistype}{Ph.\,D. Dissertation}. \bibinfo{school}{Virginia Tech}.
\newblock


\bibitem[Venkatagiri et~al\mbox{.}(2021)]%
        {venkatagiri2021crowdsolve}
\bibfield{author}{\bibinfo{person}{Sukrit Venkatagiri}, \bibinfo{person}{Aakash Gautam}, {and} \bibinfo{person}{Kurt Luther}.} \bibinfo{year}{2021}\natexlab{}.
\newblock \showarticletitle{CrowdSolve: Managing Tensions in an Expert-Led Crowdsourced Investigation}.
\newblock \bibinfo{journal}{\emph{Proc. ACM Hum.-Comput. Interact.}} \bibinfo{volume}{5}, \bibinfo{number}{CSCW1}, Article \bibinfo{articleno}{118} (\bibinfo{date}{April} \bibinfo{year}{2021}), \bibinfo{numpages}{30}~pages.
\newblock
\urldef\tempurl%
\url{https://doi.org/10.1145/3449192}
\showDOI{\tempurl}


\bibitem[Venkatagiri et~al\mbox{.}(2023)]%
        {venkatagiri2023cosint}
\bibfield{author}{\bibinfo{person}{Sukrit Venkatagiri}, \bibinfo{person}{Anirban Mukhopadhyay}, \bibinfo{person}{David Hicks}, \bibinfo{person}{Aaron Brantly}, {and} \bibinfo{person}{Kurt Luther}.} \bibinfo{year}{2023}\natexlab{}.
\newblock \showarticletitle{CoSINT: Designing a Collaborative Capture the Flag Competition to Investigate Misinformation}. In \bibinfo{booktitle}{\emph{Proceedings of the 2023 ACM Designing Interactive Systems Conference}} (Pittsburgh, PA, USA) \emph{(\bibinfo{series}{DIS '23})}. \bibinfo{publisher}{Association for Computing Machinery}, \bibinfo{address}{New York, NY, USA}, \bibinfo{pages}{2551–2572}.
\newblock
\showISBNx{9781450398930}
\urldef\tempurl%
\url{https://doi.org/10.1145/3563657.3595997}
\showDOI{\tempurl}


\bibitem[Venkatagiri et~al\mbox{.}(2019)]%
        {venkatagiri2019groundtruth}
\bibfield{author}{\bibinfo{person}{Sukrit Venkatagiri}, \bibinfo{person}{Jacob Thebault-Spieker}, \bibinfo{person}{Rachel Kohler}, \bibinfo{person}{John Purviance}, \bibinfo{person}{Rifat~Sabbir Mansur}, {and} \bibinfo{person}{Kurt Luther}.} \bibinfo{year}{2019}\natexlab{}.
\newblock \showarticletitle{GroundTruth: Augmenting Expert Image Geolocation with Crowdsourcing and Shared Representations}.
\newblock \bibinfo{journal}{\emph{Proc. ACM Hum.-Comput. Interact.}} \bibinfo{volume}{3}, \bibinfo{number}{CSCW}, Article \bibinfo{articleno}{107} (\bibinfo{date}{Nov.} \bibinfo{year}{2019}), \bibinfo{numpages}{30}~pages.
\newblock
\urldef\tempurl%
\url{https://doi.org/10.1145/3359209}
\showDOI{\tempurl}


\bibitem[Vogt et~al\mbox{.}(2011)]%
        {vogt2011colocated}
\bibfield{author}{\bibinfo{person}{Katherine Vogt}, \bibinfo{person}{Lauren Bradel}, \bibinfo{person}{Christopher Andrews}, \bibinfo{person}{Chris North}, \bibinfo{person}{Alex Endert}, {and} \bibinfo{person}{Duke Hutchings}.} \bibinfo{year}{2011}\natexlab{}.
\newblock \showarticletitle{Co-located Collaborative Sensemaking on a Large High-Resolution Display with Multiple Input Devices}. In \bibinfo{booktitle}{\emph{Human-Computer Interaction -- INTERACT 2011}}. \bibinfo{publisher}{Springer Berlin Heidelberg}, \bibinfo{address}{Berlin, Heidelberg}, \bibinfo{pages}{589--604}.
\newblock


\bibitem[Wang et~al\mbox{.}(2018)]%
        {wang_exploring_2018}
\bibfield{author}{\bibinfo{person}{Nai-Ching Wang}, \bibinfo{person}{David Hicks}, {and} \bibinfo{person}{Kurt Luther}.} \bibinfo{year}{2018}\natexlab{}.
\newblock \showarticletitle{Exploring {Trade}-{Offs} {Between} {Learning} and {Productivity} in {Crowdsourced} {History}}.
\newblock \bibinfo{journal}{\emph{Proc. ACM Hum.-Comput. Interact.}} \bibinfo{volume}{2}, \bibinfo{number}{CSCW} (\bibinfo{date}{Nov.} \bibinfo{year}{2018}), \bibinfo{pages}{178:1--178:24}.
\newblock
\showISSN{2573-0142}
\urldef\tempurl%
\url{https://doi.org/10.1145/3274447}
\showDOI{\tempurl}


\bibitem[Wardle(2014)]%
        {wardle2014verifying}
\bibfield{author}{\bibinfo{person}{Claire Wardle}.} \bibinfo{year}{2014}\natexlab{}.
\newblock \showarticletitle{Verifying user-generated content}.
\newblock \bibinfo{journal}{\emph{Verification handbook: A definitive guide to verifying digital content for emergency coverage}} (\bibinfo{year}{2014}), \bibinfo{pages}{24--33}.
\newblock


\bibitem[Wardle and Derakhshan(2017)]%
        {wardle2017information}
\bibfield{author}{\bibinfo{person}{Claire Wardle} {and} \bibinfo{person}{Hossein Derakhshan}.} \bibinfo{year}{2017}\natexlab{}.
\newblock \bibinfo{title}{Information disorder: Toward an interdisciplinary framework for research and policymaking}.
\newblock
\newblock


\bibitem[Wenger et~al\mbox{.}(1998)]%
        {wenger1998cop}
\bibfield{author}{\bibinfo{person}{Etienne Wenger} {et~al\mbox{.}}} \bibinfo{year}{1998}\natexlab{}.
\newblock \showarticletitle{Communities of practice: Learning as a social system}.
\newblock \bibinfo{journal}{\emph{Systems thinker}} \bibinfo{volume}{9}, \bibinfo{number}{5} (\bibinfo{year}{1998}), \bibinfo{pages}{2--3}.
\newblock


\bibitem[White et~al\mbox{.}(2014)]%
        {white2014digital}
\bibfield{author}{\bibinfo{person}{Joanne~I White}, \bibinfo{person}{Leysia Palen}, {and} \bibinfo{person}{Kenneth~M Anderson}.} \bibinfo{year}{2014}\natexlab{}.
\newblock \showarticletitle{Digital mobilization in disaster response: the work \& self-organization of on-line pet advocates in response to hurricane sandy}. In \bibinfo{booktitle}{\emph{Proceedings of the 17th ACM Conference on Computer Supported Cooperative Work \& Social Computing}}. ACM, \bibinfo{pages}{866--876}.
\newblock


\bibitem[Williams and Blum(2018a)]%
        {williams_defining_2018}
\bibfield{author}{\bibinfo{person}{Heather~J. Williams} {and} \bibinfo{person}{Ilana Blum}.} \bibinfo{year}{2018}\natexlab{a}.
\newblock \bibinfo{booktitle}{\emph{Defining {Second} {Generation} {Open} {Source} {Intelligence} ({OSINT}) for the {Defense} {Enterprise}}}.
\newblock \bibinfo{type}{{T}echnical {R}eport}. \bibinfo{institution}{RAND Corporation}.
\newblock
\urldef\tempurl%
\url{https://www.rand.org/pubs/research_reports/RR1964.html}
\showURL{%
\tempurl}


\bibitem[Williams and Blum(2018b)]%
        {y77}
\bibfield{author}{\bibinfo{person}{Heather~J. Williams} {and} \bibinfo{person}{Ilana Blum}.} \bibinfo{year}{2018}\natexlab{b}.
\newblock \bibinfo{booktitle}{\emph{Defining second generation open source intelligence (OSINT) for the defense enterprise. Technical Report}}.
\newblock \bibinfo{publisher}{RAND Corporation Santa Monica United States}.
\newblock


\bibitem[Williams et~al\mbox{.}(2016)]%
        {williams2016axis}
\bibfield{author}{\bibinfo{person}{Joseph~Jay Williams}, \bibinfo{person}{Juho Kim}, \bibinfo{person}{Anna Rafferty}, \bibinfo{person}{Samuel Maldonado}, \bibinfo{person}{Krzysztof~Z Gajos}, \bibinfo{person}{Walter~S Lasecki}, {and} \bibinfo{person}{Neil Heffernan}.} \bibinfo{year}{2016}\natexlab{}.
\newblock \showarticletitle{Axis: Generating explanations at scale with learnersourcing and machine learning}. In \bibinfo{booktitle}{\emph{Proceedings of the Third (2016) ACM Conference on Learning@ Scale}}. \bibinfo{pages}{379--388}.
\newblock


\bibitem[Xu et~al\mbox{.}(2015)]%
        {xu2015classroom}
\bibfield{author}{\bibinfo{person}{Anbang Xu}, \bibinfo{person}{Huaming Rao}, \bibinfo{person}{Steven~P Dow}, {and} \bibinfo{person}{Brian~P Bailey}.} \bibinfo{year}{2015}\natexlab{}.
\newblock \showarticletitle{A classroom study of using crowd feedback in the iterative design process}. In \bibinfo{booktitle}{\emph{Proceedings of the 18th ACM conference on computer supported cooperative work \& social computing}}. \bibinfo{pages}{1637--1648}.
\newblock


\bibitem[Zhang et~al\mbox{.}(2017)]%
        {zhang2017agile}
\bibfield{author}{\bibinfo{person}{Haoqi Zhang}, \bibinfo{person}{Matthew~W Easterday}, \bibinfo{person}{Elizabeth~M Gerber}, \bibinfo{person}{Daniel Rees~Lewis}, {and} \bibinfo{person}{Leesha Maliakal}.} \bibinfo{year}{2017}\natexlab{}.
\newblock \showarticletitle{Agile research studios: Orchestrating communities of practice to advance research training}. In \bibinfo{booktitle}{\emph{Proceedings of the 2017 ACM Conference on Computer Supported Cooperative Work and Social Computing}}. \bibinfo{pages}{220--232}.
\newblock


\bibitem[Zhu et~al\mbox{.}(2014)]%
        {zhu2014reviewing}
\bibfield{author}{\bibinfo{person}{Haiyi Zhu}, \bibinfo{person}{Steven~P Dow}, \bibinfo{person}{Robert~E Kraut}, {and} \bibinfo{person}{Aniket Kittur}.} \bibinfo{year}{2014}\natexlab{}.
\newblock \showarticletitle{Reviewing versus doing: Learning and performance in crowd assessment}. In \bibinfo{booktitle}{\emph{Proceedings of the 17th ACM conference on Computer supported cooperative work \& social computing}}. \bibinfo{pages}{1445--1455}.
\newblock


\bibitem[Ünver(2018)]%
        {unver_digital_2018}
\bibfield{author}{\bibinfo{person}{H.~Akın Ünver}.} \bibinfo{year}{2018}\natexlab{}.
\newblock \bibinfo{booktitle}{\emph{Digital {Open} {Source} {Intelligence} and {International} {Security}: {A} {Primer}}}.
\newblock \bibinfo{type}{{T}echnical {R}eport}. \bibinfo{institution}{Centre for Economics and Foreign Policy Studies}.
\newblock
\urldef\tempurl%
\url{https://www.jstor.org/stable/resrep21048}
\showURL{%
\tempurl}


\end{thebibliography}

\appendix

\section{Appendix}
\label{appendix}

\subsection{\newtext{Reflection Survey}}
\hspace{0.5cm}\underline{\textbf{Task}}

\begin{enumerate}
    \item What is your name? *
    \item What was your team name? *
    \item What task was assigned to your team? *
\end{enumerate}

\hspace{0.2cm}\underline{\textbf{Coordination with Expert}}

\begin{enumerate}
    \item How was your overall experience working with the expert investigator? *
    \item What did you like about the expert collaboration? What could have been better?
\end{enumerate}

\hspace{0.2cm}\underline{\textbf{Teamwork}}

\begin{enumerate}
    \item How successful was your team overall for the investigative task(s)? * \\
    On a scale of 1-5, where 1 is not successful and 5 is very successful. 
    \item How did your team work together (or not) on the task(s)? How effective was this?
    \item What did you specifically do to help your team solve the task(s)?
\end{enumerate}

\hspace{0.2cm}\underline{\textbf{Overall}}

\begin{enumerate}
    \item How confident were you about applying the practised skill in real investigations? * \\
    On a scale of 1-5, where 1 is not confident at all and 5 is very confident.
    \item How difficult were the task(s) overall? * \\
    On a scale of 1-5, where 1 is very easy at all and 5 is very difficult.
    \item How enjoyable was the session? \\
    On a scale of 1-5, where 1 is not enjoyable at all and 5 is very enjoyable.
\end{enumerate}

\begin{figure*}[]
\includegraphics[width=15cm]{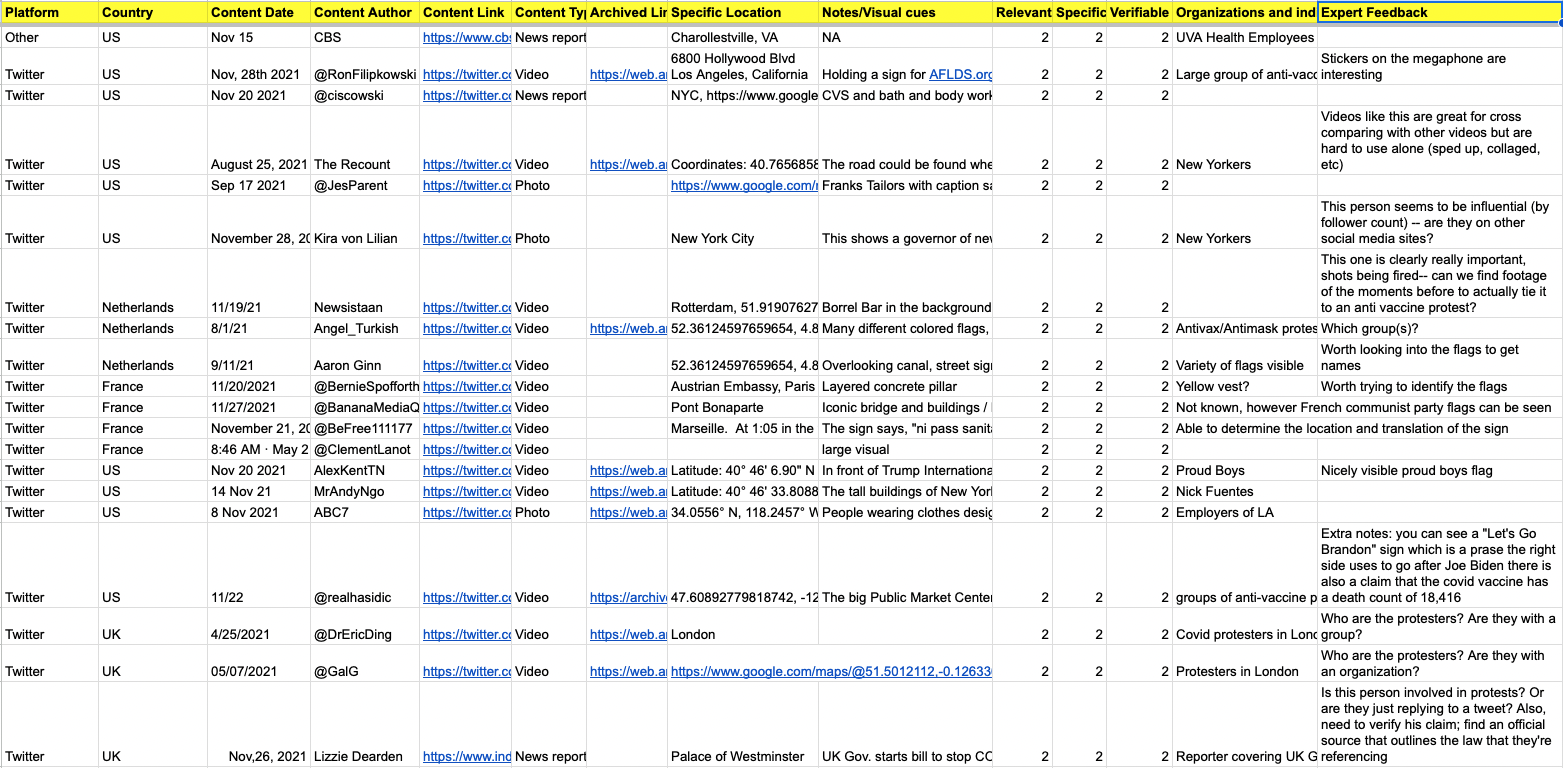}
\caption{\newtext{Snapshot of the spreadsheet containing crowd submissions and corresponding expert feedback for session 4. This session was led by an investigative journalist (E4). The goal of investigation was to identify discourse around anti-vaccine protests occurring throughout Europe as well as the groups involved. The investigation involved the discovery task and verification tasks like geolocation and source analysis.}}
\label{fig:crwod_submission}
\Description{}
\end{figure*}

\subsection{\newtext{Interview Guide for Crowdworkers}}

\hspace{1cm}\textbf{Focus group interviews with 3-4 students at a time.} [45 minutes]

\begin{enumerate}

    \item How did you form your teams and how did the relationship with teammates evolve over time? What were some issues that you faced as a part of the team?

    \item What changes did you notice over time in your performance during expert sessions? Quality/quantity?

    \item Can you tell me about your favorite task among the ones used during practice or expert sessions? [prompt - list of tasks slide]
    \begin{enumerate}
        \item What tools and techniques did you use for the task?
        \item Why did you like it? How is it different from other tasks?
    \end{enumerate}

    \item Can you tell me about your hardest task among the ones used during practice or expert sessions? [prompt - list of tasks slideshare]
    \begin{enumerate}
        \item What tools and techniques did you use for the task?
        \item Why did you find it to be hard? How is it different from other tasks?
        \item What helped you continue to work on the tasks when relevant information was hard to find?
    \end{enumerate}

    \item Can you tell me about your favorite expert session? What were the best parts of the session?

    \item Can you tell me about your least favorite expert session? What were the parts you did not like?

    \item Did the self-evaluation measures help you cover the requirements of the tasks? Why/why not?

    \item What else did you enjoy during this experience?
    \item  What could be improved during this experience?
\end{enumerate}

\subsection{\newtext{Interview guide for Experts}}

\hspace{1cm}\textbf{Post-session Individual interview with expert} [35-40 minutes]

\begin{enumerate}

    \item What techniques do you use for discovery and verification in your investigation?
    How do you collect the required information?
    \item How useful are the defined tasks (OSINT macrotasks) in your own investigations? Please rate each task. Follow-ups:
    \begin{itemize}
        \item Can the student crowd perform the task in the same way? Why/why not?
        \item How does this setup act differently?
    \end{itemize}
    \item What are your thoughts about the investigative tasks used in the session? Can you share examples of where you can incorporate these tasks into your work?
    \item Why did you choose the particular tasks from the list of tasks for working with the crowd?
    \item Only if not answered previously - When would you say that these tasks are successfully completed?

    \item What information did you decide to give the crowd before the session? Is there other information you wish you had given them?
    \item How did you plan to spend your time during the session, and what did you end up doing?
    \item How (if at all) did you interact/communicate with the crowd during the session? Any major communication issues?
    \item How well did the students respond to your interventions during the session?
    \item Any blockers in the collaborative setup?
    \item Overall, what did you think about the information submitted by the class in your investigation in terms of quality and quantity?
    \item Will you use it for your investigation? If yes, how? If not, why not?
    \item Do you have any experience with crowdsourcing for reporting? If yes, can you describe your experiences? If not, any particular reason?
    \item How would you describe the effectiveness of the students (compared to the general crowd)?
    \item What did you think about the crowd’s self-evaluation of the submissions?

    \item What else did you enjoy during this experience?

    \item What could be improved during this experience?

    \item Would you want to work with a crowd again this way in the future? Why / why not?

    \item Can you provide an example where you could have used the crowd in your investigation

\end{enumerate}

\end{document}